\documentclass[a4paper,11pt]{article}
\usepackage{jheppub} 
\makeatletter
\gdef\@fpheader{\textcolor{white}{preprint}}
\makeatother
\usepackage{lineno}
\usepackage{comment}

\title{\boldmath Regularized Master-Field Approximation for Large-$N$ Reduced Matrix Models}

\author{Reishi Maeta}
\affiliation{
Graduate School of Advanced Science and Engineering,\\
Hiroshima University,\\
Higashi-Hiroshima, Hiroshima 739-8526, Japan
}
\affiliation{
Department of Physics,\\
McGill University,\\
Montreal, QC H3A 2T8, Canada
}

\emailAdd{maeta-reishi@hiroshima-u.ac.jp}

\abstract{
We propose a numerical method based on the master field for large-$N$ reduced matrix models.
While the master field is originally an infinite-dimensional matrix, in this method it is regularized to a finite dimension, with the requirement that it satisfies the loop equations as much as possible.
This formulation can be directly implemented for numerical computation, and since there is no sign problem at the fundamental level, the method can be applied regardless of whether the model is of Euclidean or Minkowski type.
In numerical calculations for one- and two-matrix models, the exact solution is well reproduced in the Euclidean case, while perturbative results are well reproduced in the Minkowski case.
This demonstrates the effectiveness of the method and supports the idea that the matrix models studied in this paper admit a regularized master-field description.
}

\begin{document}
\maketitle
\flushbottom

\section{Introduction}
\label{sec:Introduction}

It was 't Hooft's insight that revealed that the gauge group rank $N$ can serve as a perturbative expansion parameter \cite{tHooft:1973alw}.
This observation led to a profound connection between gauge theory and string theory, and subsequently to developments such as dimensional reduction in the large-$N$ limit \cite{Eguchi:1982nm, GROSS1982440, Parisi:1982gp}, nonperturbative formulations of superstring theory via matrix models \cite{Banks:1996vh, Ishibashi:1996xs, Dijkgraaf:1997vv}, and the realization of the holographic principle through the AdS/CFT correspondence \cite{tHooft:1993dmi, Susskind:1994vu, Maldacena:1997re}.
Moreover, in the large-$N$ (planar) limit, planar diagrams dominate in the Feynman diagram expansion, and correlation functions factorize, leading to a certain classical description of the theory.
For this reason, matrix models have also been studied as a useful theoretical framework for understanding the structure of complex quantum systems \cite{Gross:1980he, Wadia:1980cp, Eguchi:1982nm, Gonzalez-Arroyo:1982hwr}.
More recently, it has also been proposed that matrix models in the large-$N$ limit can describe unitary gravitational interactions \cite{Ho:2025htr}, which provides a strong motivation to reconsider large-$N$ matrix models from a different perspective.

Despite the simplifications in the large-$N$ limit, it is still generally difficult to obtain exact analytic results.
In most cases, one resorts to approximations or numerical simulations, typically based on Monte Carlo methods, to compute expectation values \cite{Kabat:2000zv, Ambjorn:2000dx, Martin:2004un, Azuma:2004zq, Panero:2006bx, Hanada:2008gy, Gonzalez-Arroyo:2010omx, Hanada:2011fq, Kim:2011cr, Gonzalez-Arroyo:2014dua, Filev:2015hia}.
In recent years, bootstrap approaches based on positivity have also been developed \cite{Lin:2020mme, Kazakov:2021lel, Han:2020bkb, Lin:2023owt, Lin:2024vvg}, allowing rigorous upper and lower bounds on expectation values to be determined as an allowed region.

On the other hand, there are also challenges.
The standard approach to numerical computation in high-energy physics is the Monte Carlo method, which essentially performs numerical integration over weighted quantities, i.e., the evaluation of expectation values, but this can be carried out efficiently mainly when the weights are positive real numbers.
For this reason, there are fundamental difficulties in applying it to theories with real time $t$ and complex weights $e^{iS}$, as well as to some finite-temperature systems, and this is known as the notorious \textit{sign problem}.
Related to this, the matrix bootstrap method based on positivity is, in principle, not applicable to large-$N$ reduced matrix models with complex weights $e^{iS}$.
This is because positivity in large-$N$ reduced matrix models is defined under the assumption of the existence of a positive real weight $e^{-S}$. 
\footnote{The situation is different for matrix models with an explicit time coordinate $t$, namely matrix quantum mechanics.
In such theories, one can define a Hilbert space ${\cal H}$ and the corresponding state vectors $|\psi\rangle$, so that the positivity condition $\langle\psi|\psi\rangle\ge0$ is a physically meaningful constraint even when the matrix integral is weighted by $e^{iS}$.
For studies of time evolution in quantum mechanics based on this positivity, see \cite{Lawrence:2024mnj}.
By contrast, in large-$N$ reduced matrix models without a time coordinate $t$, no such Hilbert space exists a priori.
Instead, as will be discussed later, the bootstrap is formulated by employing the alternative notion of ``positivity,'' namely $\langle\mathrm{Tr}(M^{\dagger}M)\rangle\ge0$.
Unless otherwise stated, the term ``positivity'' will henceforth refer to the latter condition, namely $\langle\mathrm{Tr}(M^{\dagger}M)\rangle\ge0$.
For the sake of clarity, we also adopt the following terminology throughout this paper: a \emph{large-$N$ reduced matrix model} refers to a matrix model without explicit coordinate dependence such as $t$ or $x$, whereas \emph{matrix quantum mechanics} refers to a matrix model with an explicit time coordinate $t$.}

Against this background, various attempts have been made to improve both the Monte Carlo method and the matrix bootstrap method.
For the former, under the slogan of overcoming the sign problem, the complex Langevin method and the generalized Lefschetz thimble method have been developed and improved, and continue to be actively studied \cite{Parisi:1983mgm, Klauder:1983sp, Aarts:2009uq, Cristoforetti:2012su, Alexandru:2015sua}.
For applications of these methods to matrix models, see \cite{Nishimura:2019qal, Anagnostopoulos:2022dak, Chou:2025moy, Anagnostopoulos:2026qvz,  Anagnostopoulos:2026utg}.
On the other hand, for the matrix bootstrap method, several approaches that do not rely on the concept of positivity have been proposed.
Specifically, these include methods based on the cut structure of the resolvent \cite{Li:2024ggr}, methods that perform bootstrap approximations based on the eigenvalue distribution \cite{Maeta:2026oku}, and, for unitary matrix models, an asymptotic bootstrap method that exploits the behavior that the moments $a_{n}$ rapidly approach zero for $n \gg 1$ \cite{Berenstein:2025itw, Berenstein:2026wky}.

The present work significantly extends and improves the bootstrap approximation based on eigenvalues \cite{Maeta:2026oku}.
While that method was shown to work for Euclidean-type one-matrix models, where the weight of the matrix integral is given by $e^{-S}$, it had only conditional applicability to Minkowski-type theories, where the weight is given by $e^{iS}$, and its extension to multi-matrix models was not straightforward.
In contrast, the key feature of this work is the use of the large-$N$ master field, an infinite-dimensional matrix, as a natural extension of the eigenvalue distribution.

Although the existence of the master field had already been conjectured in the late 1970s \cite{Witten:1979pi}, it has not been clear how to construct it explicitly, or even whether such an object truly exists.
In this work, we assume its existence from the outset and regularize it in a finite-dimensional form, thereby formulating a concrete numerical method, which we call the regularized master-field approximation (RMA).
Essentially, this method reinterprets the \textit{large-$N$ master-field optimization} \cite{Jevicki:1982jj, Jevicki:1983hb, Koch:2021yeb, Mathaba:2023non} as a finite-dimensional regularization of the master field, while modifying part of the formulation so that it can be applied to a broader class of matrix models.
Applying this method to the Hermitian one- and two-matrix models, we find that for Euclidean-type models, both one- and two-matrix cases agree with known analytic and numerical results with high precision, and for the Minkowski-type one-matrix model, the results agree remarkably well with the formal solution as a working hypothesis.
Furthermore, the resulting approximate values are shown to be stable and insensitive to the details of the regularization scheme.
These results suggest that, at least under certain large-$N$ limits, an object analogous to the large-$N$ master field exists in one- and two-matrix models.

The organization of this paper is as follows.
In Section~\ref{sec:Method}, after briefly reviewing the conventional bootstrap approximation based on eigenvalue distributions, we discuss its usefulness and limitations.
We then introduce the large-$N$ master field as a concept that can overcome these limitations and explain how it can be implemented numerically as the RMA.
This section also includes a justification for finite-dimensional regularization of the master field and an explanation that the master field can be interpreted as a natural extension of the eigenvalue distribution.
In Section~\ref{sec:One-Matrix_Model}, as a first test case, we apply the RMA to Euclidean- and Minkowski-type one-matrix models, and in Section~\ref{sec:Two-Matrix_Model}, we extend it to two-matrix models.
Finally, in Section~\ref{sec:Summary_and_Discussion}, we summarize the method and numerical results and discuss future directions.
In the Appendices, we present the perturbative method based on the loop equations, together with supplementary details on the numerical analysis of the Euclidean one-matrix model.

\section{Method} \label{sec:Method}

\subsection{Bootstrap Approximation Based on Eigenvalue Distribution} \label{subsec:Bootstrap_Approximation_Based on_Eigenvalue_Distribution}

In \cite{Maeta:2026oku}, a bootstrap approximation was performed based on the existence of the eigenvalue distribution.
However, as already pointed out there, the eigenvalue distribution in multi-matrix models is not as powerful as in the one-matrix model.
To confirm this point, let us first review the eigenvalue distribution in the Euclidean-type one-matrix model.
\begin{equation} \label{eq:Action_of_One-matrix_Models}
\begin{aligned}
S_{\text{1-MM}}[\phi] & =N\text{Tr}(\frac{1}{2}\phi^{2}-\frac{g}{4}\phi^{4}),\\
Z_{E} & =\int d\phi e^{-S[\phi]}  =\int(\prod_{i=1}^{N}d\lambda_{i})\Delta^{2}(\lambda_{i})e^{-S(\lambda_{i})}.
\end{aligned}
\end{equation}
They are the action and partition function of the one-matrix model.
The mass $m$ has been normalized to unity in advance.
The rightmost expression for $Z_{E}$ is obtained by gauge fixing $\phi$ to $\phi=\text{diag}(\lambda_{1},...,\lambda_{N})$, and the square of the Vandermonde determinant $\Delta(\lambda_{i})$ appears as the corresponding Faddeev--Popov determinant.
Since there is only a single matrix, the eigenvalue distribution $\rho_{E}(\lambda)$ becomes a very powerful tool.
Specifically, the primary observables in this theory are defined by
\begin{equation}
w_{n}=\langle\text{tr}\phi^{n}\rangle_{E}=\langle\frac{1}{N}\text{Tr}\phi^{n}\rangle_{E},
\end{equation}
and once $\rho_{E}(\lambda)$ is known, $w_{n}$ for any $n\in\mathbb{N}=\{0,1,2,...\}$ can be computed as
\begin{equation}
w_{n}=\int_{\Gamma}d\lambda\lambda^{n}\rho_{E}(\lambda).
\end{equation}
Here, $\Gamma$ is the support of $\rho_{E}(\lambda)$, which corresponds to a finite interval on the real axis in the Euclidean-type one-matrix model.
If $\rho_{E}(\lambda)$ is regarded as a probability distribution, the right-hand side is nothing but the $n$th moment in the context of statistics, and thus $w_{n}$ is sometimes referred to as a moment.
We now recall the mathematical definition of $\rho_{E}(\lambda)$, which is formally given by
\begin{equation}
\begin{aligned}
\rho_{E}(\lambda) & \equiv \left\langle \frac{1}{N} \sum_{i=1}^{N} \delta(\lambda - \lambda_{i}) \right\rangle_{E} \\
& = \frac{1}{Z_{E}} \int \left( \prod_{i=1}^{N} d\lambda_{i} \right) \Delta^{2}(\lambda_{i}) \left\{ \frac{1}{N} \sum_{i=1}^{N} \delta(\lambda - \lambda_{i}) \right\} e^{-S(\lambda_{i})} ,
\end{aligned}
\end{equation}
that is, it is the expectation value of the spectral density $\frac{1}{N}\sum_{i=1}^{N}\delta(\lambda-\lambda_{i})$.
Note that this is the expectation value of the density itself, not the expectation values of the eigenvalues $\lambda_{i}$ (unless the $\mathbb{Z}_{2}$ symmetry of the theory is broken, one has $\langle\lambda_{i}\rangle=0$ for all $i=1,...,N$).
Here, $\lambda$ is merely a real parameter and is not intrinsically tied to the eigenvalues $\lambda_{i}$.
Therefore, the integration over $\lambda$, $\int_{\Gamma}d\lambda(\cdots)$, commutes with the matrix integral, and
\begin{equation}
\begin{aligned}
\int_{\Gamma}d\lambda\lambda^{n}\rho_{E}(\lambda) & =\frac{1}{Z_{E}}\int(\prod_{i=1}^{N}d\lambda_{i})\Delta^{2}(\lambda_{i})e^{-S(\lambda_{i})}\int_{\Gamma}d\lambda\lambda^{n}\{\frac{1}{N}\sum_{i=1}^{N}\delta(\lambda-\lambda_{i})\}\\
 & =\frac{1}{Z_{E}}\int(\prod_{i=1}^{N}d\lambda_{i})\Delta^{2}(\lambda_{i})(\frac{1}{N}\sum_{i=1}^{N}\lambda_{i}^{n})e^{-S(\lambda_{i})}\\
 & =\langle\text{tr}\phi^{n}\rangle_{E},
\end{aligned}
\end{equation}
indeed holds.
This is the definition and role of the eigenvalue distribution.

For theories involving more than one matrix, however, the eigenvalue distribution is unfortunately not so useful.
For example, consider the Euclidean-type multi-matrix model defined by
\begin{equation}
S_{E}=N\text{Tr}\{-\frac{h}{4}\delta^{\mu\rho}\delta^{\nu\lambda}[A_{\mu},A_{\nu}][A_{\rho},A_{\lambda}]+\frac{1}{2}\delta^{\mu\nu}A_{\mu}A_{\nu}+\frac{g}{4}\delta^{\mu\nu}A_{\mu}^{2}A_{\nu}^{2}\}.
\end{equation}
Here $\mu,\nu,\rho,\lambda=1,...,D$.
The eigenvalue distribution $\rho_{\mu}(\lambda_{\mu})$ for each $A_{\mu}$ can still be defined without difficulty, but it only allows one to compute moments of the form
\begin{equation}
\int d\lambda_{\mu}\lambda_{\mu}^{n}\rho_{\mu}(\lambda_{\mu})=\langle\frac{1}{N}\text{Tr}(A_{\mu}^{n})\rangle.
\end{equation}
Thus, for example, if one wishes to compute quantities such as $\langle\frac{1}{N}\text{Tr}(A_{1}A_{2}+A_{2}A_{1})\rangle$, $\rho_{\mu}(\lambda_{\mu})$ is entirely insufficient.
One might then attempt to introduce a corresponding eigenvalue distribution $\rho_{12}(\lambda_{12})$, but this again only yields moments of the form
\begin{equation}
\int d\lambda_{12}\lambda_{12}^{n}\rho_{12}(\lambda_{12})=\langle\frac{1}{N}\text{Tr}((A_{1}A_{2}+A_{2}A_{1})^{n})\rangle.
\end{equation}
As is clear from this, in order to compute arbitrary moments appearing in a two-matrix model, one would need infinitely many different types of eigenvalue distributions.
Although the eigenvalue distribution was introduced to simplify the analysis, this defeats the original purpose.

Another issue is that, historically, eigenvalue distributions have been discussed mainly in the context of Euclidean-type matrix models.
As a result, the theory of eigenvalue distributions for Minkowski-type matrix models is not well established.
In \cite{Maeta:2026oku}, a formal eigenvalue distribution was defined for the one-matrix model and was shown to reproduce perturbative results, but this construction was somewhat ad hoc and does not provide a general solution for Minkowski-type matrix models.
As will be explained in detail in Section \ref{subsubsec:Master_Field_as_an_Equal-Weight_Curve_in_the_Complex_Plane}, even the definition of the eigenvalue distribution itself is a nontrivial problem in Minkowski-type matrix models.
Since one cannot utilize an ill-defined object for numerical computation, this situation makes it difficult to perform reliable numerical analyses for general Minkowski-type matrix models.

\subsection{Large-$N$ Master Field} \label{subsec:Large-N_Master_Field}

\subsubsection{Large-$N$ Factorization and the Master-Field Conjecture} \label{subsubsec:Large-N_Factorization_and_the_Master-Field_Conjecture}

From this viewpoint, general multi-matrix models and Minkowski-type matrix models may appear to be too complicated to handle with conventional methods.
However, this is not necessarily the case.
It is widely believed that, regardless of the number of matrices or the signature of the theory, many matrix models possess a remarkably simple structure in the large-$N$ limit.
Intuitively, the key point is the relation
\begin{equation}
S_{\text{MM}}\sim N^{2},
\end{equation}
that is, apart from some exceptions, the action $S_{\text{MM}}$ of a matrix model typically scales as $N^{2}$.
Writing $S_{\text{MM}}[X]=N^{2}S_{0}[X]$ with $S_{0}\sim{\cal O}(N^{0})$, the partition function can be expressed as
\begin{equation}
Z_{\text{MM}}=\int dX e^{-N^{2}S_{0}[X]}.
\end{equation}
Comparing this with the QFT expression
\begin{equation}
Z_{\text{FT}}=\int{\cal D}\phi e^{-\frac{1}{\hbar}S[\phi]},
\end{equation}
one is naturally led to the correspondence
\begin{equation}
N^{2}\sim\frac{1}{\hbar},
\end{equation}
noting that in general $S[\phi]\sim{\cal O}(\hbar^{0})$.
On the QFT side, the formal limit $\hbar\to0$ corresponds to the semiclassical limit, in which the path integral is dominated by a single configuration and quantum fluctuations disappear.
The same phenomenon occurs in matrix models in the limit $N\to\infty$, and thus the planar large-$N$ limit corresponds to a ``semiclassical'' limit.

Let us analyze this more quantitatively.
For this purpose, the Euclidean one-matrix model $S_{\text{1-MM}}[\phi]$ and its loop equations serve as a simple example.
Here, the loop equations are the Schwinger--Dyson equations for $w_{n}$, and their derivation starts from the trivial identity
\begin{equation}
\int d\phi(\text{tr}(t^{a}\phi^{n-1}))(\text{tr}\phi^{m})e^{-S[\phi]}=\int d\phi'(\text{tr}(t^{a}\phi'{}^{n-1}))(\text{tr}\phi'{}^{m})e^{-S[\phi']}.
\end{equation}
Here $t^{a}$ are generators of $U(N)$ satisfying
\begin{equation} \label{eq:property_of_t^a}
\begin{aligned}
\sum_{a}\text{Tr}(t^{a}A)\text{Tr}(t^{a}B) & =\text{Tr}(AB),\\
\sum_{a}\text{Tr}(t^{a}At^{a}B) & =\text{Tr}A\text{Tr}B,
\end{aligned}
\end{equation}
and taking
\begin{equation}
\phi'=\phi+\epsilon t^{a},
\end{equation}
the terms linear in $\epsilon$ on the right-hand side must vanish, which leads to
\begin{equation}
0  =\sum_{k=0}^{n-2}\langle\text{tr}\phi^{n-k-2}\text{tr}\phi^{k}\text{tr}\phi^{m}\rangle-\langle\text{tr}\phi^{n}\text{tr}\phi^{m}\rangle+g\langle\text{tr}\phi^{n+2}\text{tr}\phi^{m}\rangle+\frac{m}{N^{2}}\langle\text{tr}\phi^{n+m-2}\rangle.
\end{equation}
The first term arises from $\text{tr}(t^{a}\phi'{}^{n-1})\sim\epsilon\text{tr}(t^{a}\phi{}^{n-k-2}t^{a}\phi^{k})$ and corresponds to the splitting of the moment $w_{n}$.
The second and third terms represent the coupling between $\text{tr}(t^{a}\phi^{n-1})$ and the action $S$, and can be regarded as infinitesimal deformations of $w_{n}$.
The fourth term corresponds to the joining of $w_{n}$ and $w_{m}$ and is suppressed by $N^{-2}$ compared to the other terms.
Therefore, in the large-$N$ limit, contributions corresponding to the joining of moments vanish, and such a limit is called the planar limit.
In this case, since the first three terms all contain a common factor $w_{m}$, it suffices to consider
\begin{equation}
\int d\phi\text{tr}(t^{a}\phi^{n-1})e^{-S[\phi]}=\int d\phi'\text{tr}(t^{a}\phi'{}^{n-1})e^{-S[\phi']},
\end{equation}
which leads to
\begin{equation} \label{eq:loopeq_E-type_one-matrix}
0=\sum_{k=0}^{n-2}w_{n-k-2}w_{k}-w_{n}+gw_{n+2}.
\end{equation}
This is the loop equation in the planar limit.
Although we have used the one-matrix model for illustration, the suppression of joining terms by $N^{-2}$ is a universal feature of loop equations in a wide class of matrix models, since it follows from the trace identities involving $t^{a}$ and the scaling $S_{\text{MM}}\sim N^{2}$, and is largely independent of the detailed form of the action. 

From this analysis, an important insight emerges.
As noted above, in the large-$N$ limit, contributions corresponding to the joining of two moments $w_{n}$ and $w_{m}$ become negligible, which implies that correlations between them vanish.
This suggests that
\begin{equation}
\langle\text{tr}{\cal O}_{1}\text{tr}{\cal O}_{2}\rangle\overset{N\to\infty}{=}\langle\text{tr}{\cal O}_{1}\rangle\langle\text{tr}{\cal O}_{2}\rangle
\end{equation}
holds, a phenomenon known as large-$N$ factorization and verified in a wide class of matrix models.
From this relation, it follows that for any matrix $A_{\mu}$ in the theory,
\begin{equation}
\forall n\in\mathbb{N},\quad\langle(\text{tr}A_{\mu})^{n}\rangle-\langle\text{tr}A_{\mu}\rangle^{n}=0
\end{equation}
in the large-$N$ limit.
This means that, when $\text{tr}A_{\mu}$ is regarded as a random variable, its fluctuations completely vanish as $N\to\infty$.
Since the same holds for any single-trace operator $\text{tr}(A_{\mu_{1}}...A_{\mu_{n}})$, this simple observation leads to the expectation that the matrix integral becomes effectively deterministic at the level of single-trace observables in the large-$N$ limit.
In other words, it is expected that there exist matrices $\hat{A}_{\mu}$ satisfying
\begin{equation}
\langle\text{tr}(A_{\mu_{1}}...A_{\mu_{n}})\rangle\overset{N\to\infty}{=}\text{tr}(\hat{A}_{\mu_{1}}...\hat{A}_{\mu_{n}}),
\end{equation}
which are referred to as the \textit{large-$N$ master field} \cite{Witten:1979pi}.
More precisely, for the matrices $A_{\mu}$ in the theory, there exist corresponding matrices $\hat{A}_{\mu}$ in the limit $N\to\infty$ such that the expectation value of any single-trace operator $\text{tr}(A_{\mu_{1}}...A_{\mu_{n}})$ is given simply by the trace $\text{tr}(\hat{A}_{\mu_{1}}...\hat{A}_{\mu_{n}})$.
Since expectation values correspond to statistical averages with respect to $e^{-S}$ or $e^{iS}$, this means that in the limit $N\to\infty$ the theory behaves ``classically''.
This provides a more precise realization of the analogy $N^{2}\sim\frac{1}{\hbar}$.

In this way, large-$N$ matrix models are expected to exhibit a simple structure, originating from the fact that the matrix integral becomes effectively deterministic.
In the case of the one-matrix model, this simplicity is directly captured by the eigenvalue distribution $\rho(\lambda)$.
In multi-matrix models, the eigenvalue distribution is no longer as powerful, but this merely reflects the loss of equivalence between the eigenvalue distribution and the master field; the theory should still possess a simple structure in the limit $N\to\infty$ if the master field indeed exists.

\subsubsection{Master Field as a Generalization of the Eigenvalue Distribution} \label{subsubsec:Master_Field_as_a_Generalization_of_the_Eigenvalue_Distribution}

Before proceeding to a concrete implementation for numerical computations, it is helpful to clarify the relation between the master field and the eigenvalue distribution.
As discussed above, eigenvalue distributions are defined separately for each infinite-dimensional (generally Hermitian or unitary) matrix $X$ in the planar limit, and by integrating with respect to it one can compute $\langle\text{tr}X^{n}\rangle$.
However, it is completely powerless for matrices that cannot be expressed as powers of $X$, and thus is not particularly useful in multi-matrix models.
In contrast, the master field is extremely powerful: once the master fields $\hat{A}_{\mu}$ corresponding to the fundamental matrices $A_{\mu}$ of the theory are known, one can compute \textit{any} expectation value $\langle\text{tr}(A_{\mu_{1}}...A_{\mu_{n}})\rangle$.

From this, one expects that the master fields $\hat{A}_{\mu}$ should contain the information of the eigenvalue distributions of $A_{\mu}$.
In fact, the master field can be regarded as a generalization of the eigenvalue distribution.
Let us verify this in the case of the Euclidean-type one-matrix model.
The master field $\hat{\phi}$ of this model is an infinite-dimensional Hermitian matrix satisfying
\begin{equation}
w_{n}=\text{tr}\hat{\phi}^{n}
\end{equation}
for any $n\in\mathbb{N}$.
Since only the eigenvalues of $\hat{\phi}$ contribute to $w_{n}$, diagonalizing it as
\begin{equation}
U\hat{\phi}U^{\dagger}=\text{diag}(\hat{\lambda}_{1},...,\hat{\lambda}_{N}),
\end{equation}
one finds
\begin{equation} \label{eq:wn_by_lambda}
w_{n}=\frac{1}{N}\sum_{i=1}^{N}\hat{\lambda}_{i}^{n}.
\end{equation}
Comparing this with $w_{n}=\int_{\Gamma}d\lambda\lambda^{n}\rho_{E}(\lambda)$, one sees that $\rho_{E}(\lambda)$ and $\hat{\phi}$ are related by
\begin{equation}
\rho_{E}(\lambda)=\frac{1}{N}\sum_{i=1}^{N}\delta(\lambda-\hat{\lambda}_{i}).
\end{equation}
In other words, the infinitely many eigenvalues $(\hat{\lambda}_{1},...,\hat{\lambda}_{N})$ of the master field $\hat{\phi}$ give rise to the eigenvalue distribution $\rho_{E}(\lambda)$ as a spectral density.
Equivalently, sampling infinitely many eigenvalues according to $\rho_{E}(\lambda)$ reproduces $(\hat{\lambda}_{1},...,\hat{\lambda}_{N})$.

This relation can also be explained in terms of the resolvent $R(z)$.
The resolvent is defined, irrespective of whether the model is Euclidean or Minkowski, by
\begin{equation}
R(z)\equiv\langle\frac{1}{N}\sum_{i=1}^{N}\frac{1}{z-\lambda_{i}}\rangle=\langle\text{tr}(\frac{1}{z-\phi})\rangle.
\end{equation}
Since the eigenvalues $\lambda_{i}$ are real, $\frac{1}{z-\lambda_{i}}$ has poles on the real axis at $z=\lambda_{i}$, and hence for Euclidean matrix models $R(z)$ also has poles on the real axis, which in the large-$N$ limit typically become (finite) branch cuts.
The direct motivation for introducing this function is that it serves as a generating function for the moments $w_{n}$.
Indeed, expanding for $z\to\infty$ using the geometric series, one obtains
\begin{equation}
R(z)=\sum_{n=0}^{\infty}\frac{w_{n}}{z^{n+1}},
\end{equation}
and by the residue theorem
\begin{equation}
w_{n}=\oint_{C}\frac{dz}{2\pi i}z^{n}R(z)
\end{equation}
follows immediately, where $C$ is a sufficiently large counterclockwise contour.
By Cauchy's theorem, the contour $C$ can be continuously deformed as long as it does not cross cuts or poles, and deforming it to wrap around the cuts shows that only the discontinuities across the cuts contribute.
In the Euclidean one-matrix model, this observation leads directly to the eigenvalue distribution $\rho_{E}(\lambda)$.

Returning to the definition of $R(z)$, it is expressed as an expectation value of a trace.
In the planar limit, one may therefore replace it by the master field $\hat{\phi}$ as
\begin{equation}
R(z)=\text{tr}(\frac{1}{z-\hat{\phi}})=\frac{1}{N}\sum_{i=1}^{N}\frac{1}{z-\hat{\lambda}_{i}}.
\end{equation}
Indeed, the residue theorem again gives
\begin{equation} \label{eq:generating_function_of_wn}
\oint_{C}\frac{dz}{2\pi i}z^{n}R(z)=\frac{1}{N}\sum_{i=1}^{N}\hat{\lambda}_{i}^{n}=w_{n}.
\end{equation}
Thus, if the master field $\hat{\phi}$ exists, the poles of $R(z)$ coincide with the eigenvalues $(\hat{\lambda}_{1},...,\hat{\lambda}_{N})$ of the master field.
As is now clear, once the master field $\hat{\phi}$ is known, one can completely reconstruct the eigenvalue distribution $\rho_{E}(\lambda)$ and the resolvent $R(z)$.
This is expected, since $\hat{\phi}$ essentially contains all the information of the system.

\subsubsection{Master Field as an Equal-Weight Curve in the Complex Plane} \label{subsubsec:Master_Field_as_an_Equal-Weight_Curve_in_the_Complex_Plane}

From the preceding calculation, we have seen that the master field $\hat{\phi}$ contains complete information about the eigenvalue distribution $\rho_{E}(\lambda)$.
Can we then define an eigenvalue distribution $\rho_{M}(z)$ also for the Minkowski one-matrix model by using the master field $\hat{\phi}$?
In fact, this turns out not to work as straightforwardly as in the Euclidean case.
If we follow the preceding procedure, the starting point would be to assume, also in the Minkowski matrix model, the existence of a master field $\hat{\phi}$ such that $w_{n}=\text{tr}\hat{\phi}^{n}$ holds.
The difference from the Euclidean case is that, in the Minkowski case, $\hat{\phi}$ is weighted by the complex phase $e^{iS}$, and hence it is no longer necessarily a Hermitian matrix.
Thus, $\hat{\phi}$ becomes a complex matrix.
We assume that it can be diagonalized in the form
\begin{equation}
V\hat{\phi}V^{-1}=\text{diag}(\hat{z}_{1},\ldots,\hat{z}_{N}) .
\end{equation}
Since $\hat{\phi}$ is a complex matrix, the eigenvalues $\hat{z}_{1},\ldots,\hat{z}_{N}$ can in general take complex values; on the other hand, since $\hat{\phi}$ before the weighting was a Hermitian matrix, it is reasonable to assume that it remains diagonalizable in this way after the weighting.

The fact that $\hat{\phi}$ becomes a diagonalizable complex matrix may seem to be a natural consequence, but some care is required.
Recall that, as shown in \eqref{eq:generating_function_of_wn}, the resolvent $R(z)$ is the generating function of the moments $w_{n}$, and that the integration contour $C$ can be continuously deformed.
In the Euclidean one-matrix model, the assumption that $\hat{\phi}$ is Hermitian allowed us to choose $C$ along the real axis, which gave the support of the eigenvalue distribution $\rho_E(\lambda)$.
In the Minkowski one-matrix model, however, there is no such guiding principle for choosing $C$.
Then, in such a situation, is there another ``natural'' way to choose $C$?
This becomes clear if we parametrize $C$ as $C=\{z(t);t\in[0,1]\}$ and compute $w_{n}$.
Since $w_{n}$ is generated by $R(z)$ and can also be written as the average of $\hat{z}_{1}^{n},\ldots,\hat{z}_{N}^{n}$, we obtain
\begin{equation} \label{eq:wn_by_z}
w_{n}=\int_{0}^{1}dt\frac{\dot{z}(t)}{2\pi i}R(z(t))z^{n}(t)=\frac{1}{N}\sum_{i=1}^{N}\hat{z}_{i}^{n} .
\end{equation}
Here, $\dot{z}(t)=\frac{dz(t)}{dt}$.
The rightmost expression is the sum over $i=1,\ldots,N$ divided by $N$, and we notice that this has the same form as a Riemann sum for an integral over the interval $[0,1]$.
Therefore, if we assume that $C$ is chosen so that
\begin{equation} \label{eq:DeffEq_for_z(t)}
\frac{\dot{z}(t)}{2\pi i}R(z(t))=1
\end{equation}
holds, then
\begin{equation}
\frac{1}{N}\sum_{i=1}^{N}\hat{z}_{i}^{n}=\int_{0}^{1}dtz^{n}(t)
\end{equation}
follows.
Since $N$ is infinite here, this is precisely the Riemann-sum prescription.
In other words, a closed curve $C_{\text{const.}}$ satisfying $\frac{\dot{z}(t)}{2\pi i}R(z(t))=1$ is a ``natural'' choice of $C$ when $\hat{\phi}$ is regarded as a complex matrix, and the complex eigenvalues $\hat{z}_{1},\ldots,\hat{z}_{N}$ of $\hat{\phi}$ are distributed on this curve.
Equation \eqref{eq:DeffEq_for_z(t)} is a differential equation for $z(t)$, and its solution is not uniquely determined unless boundary conditions are specified.
In this regard, it is natural to assume that the curve passes through the endpoints $\pm\alpha$ of the resolvent $R(z)$.

It is worth noting that the same logic also applies to the Euclidean one-matrix model.
The fact that the closed curve $C$ in the complex plane can be continuously deformed is exactly the same in the Euclidean case, and if it is chosen as $C_{\text{const.}}$, then $w_{n}=\int_{0}^{1}dtz^{n}(t)$ also holds in the Euclidean one-matrix model.
In this case, $z(t)$ is nothing but the eigenvalues $\hat{z}_{1},\ldots,\hat{z}_{N}$ of the complex matrix $\hat{\phi}$.
In other words, even in the Euclidean one-matrix model, one may also regard $\hat{\phi}$ as a complex matrix, in which case $\hat{z}_{1},\ldots,\hat{z}_{N}$ lie on $C_{\text{const.}}$.
We confirm in Section \ref{subsec:Numerical_Results_one-matrix_model} that this is indeed the case.

This leads to one question: from $w_{n}=\int_{0}^{1}dtz^{n}(t)$, the values $\hat{z}_{1},\ldots,\hat{z}_{N}$ can be regarded as equal-weight points that keep $\frac{\dot{z}(t)}{2\pi i}R(z(t))$ constant, but how should we interpret the eigenvalues $\hat{\lambda}_{1},\ldots,\hat{\lambda}_{N}$ when $\hat{\phi}$ is chosen to be Hermitian?
This can be understood from $w_{n}=\int_{\Gamma}d\lambda\lambda^{n}\rho_{E}(\lambda)$.
Namely, if we define a new integration measure $d\mu$ by \footnote{As a function, $\mu(\lambda)$ can be defined as an antiderivative of $\rho_{E}(\lambda)$ by $\mu(\lambda)=\int_{-\infty}^{\lambda}d\lambda'\rho(\lambda')$.}
\begin{equation}
d\mu\equiv\rho_{E}(\lambda)d\lambda,\quad\lambda=\lambda(\mu),
\end{equation}
then $w_{n}$ can be reduced to the form
\begin{equation}
w_{n}=\int_{0}^{1}d\mu\lambda^{n}(\mu) .
\end{equation}
This has the same form as $w_{n}=\int_{0}^{1}dtz^{n}(t)$, which means that, with respect to such a variable $\mu$, $\hat{\lambda}_{1},\ldots,\hat{\lambda}_{N}$ are precisely equal-weight points.
However, constructing $\mu$ requires the eigenvalue distribution $\rho_{E}(\lambda)$ itself, and in the Minkowski one-matrix model we do not know what the corresponding object is, nor how it is distributed in the complex plane.
Thus, it cannot be used to define an eigenvalue distribution $\rho_{M}(\lambda)$.
Nevertheless, the purpose of our calculation is to estimate the values of $w_{n}$, and for this purpose the complex eigenvalues $\hat{z}_{1},\ldots,\hat{z}_{N}$ are entirely sufficient.
Therefore, in this work we do not pursue the definition of $\rho_{M}(\lambda)$ any further, and instead deal with the complex master field $\hat{\phi}$ and its complex eigenvalues $\hat{z}_{1},\ldots,\hat{z}_{N}$.

The theoretical basis of the analysis so far is that, if the master field $\hat{\phi}$ exists, then $w_{n}$ can be written in the form \eqref{eq:wn_by_lambda} or \eqref{eq:wn_by_z}.
By regarding this as a Riemann-sum prescription, we conclude that $\hat{\lambda}_{1},\ldots,\hat{\lambda}_{N}$ and $\hat{z}_{1},\ldots,\hat{z}_{N}$ are equal-weight points, namely, values on a curve for which the weight (probability density) of $w_{n}$ as a moment is constant.
These statements are confirmed by the numerical results presented in Section \ref{subsec:Numerical_Results_one-matrix_model} and Appendix \ref{app:appendixB}.

Although the preceding discussion has clarified several properties of the master field, the existence of the master field itself remains an assumption at this stage.
Fortunately, the master field is an object that is very well suited to numerical computation, and it is possible to use it to investigate whether the master field itself exists.
The numerical method based on the master field is the regularized master-field approximation (RMA), which we explain below and which is the main subject of this work.

\subsection{Regularized Master-Field Approximation (RMA)} \label{subsec:Regularized_Master-Field_Approximation}

\subsubsection{Approximation by a Finite-Dimensional Regularized Master Field} \label{subsubsec:Approximation_by_a_Finite-Dimensional_Regularized_Master_Field}

The claim that the master field $\hat{A}_{\mu}$ is well suited to numerical computation may sound counterintuitive.
This is because $\hat{A}_{\mu}$ is an infinite-dimensional matrix that can be defined only in the limit $N\to\infty$, and numerical computation cannot handle such an infinity.
How, then, can such a quantity be useful for numerical computation?

Here we adopt a very naive idea.
Although large-$N$ factorization holds exactly only when one takes the limit $N\to\infty$, the $1/N^{2}$ suppression itself remains true even for finite $N$.
Then, even if $N$ is finite, as long as it is sufficiently large, correlations among loops should also be strongly suppressed.
In other words, for such finite but large $N$, one is led to expect that the system should exhibit ``almost deterministic'' behavior.
This may also be inferred from the fact that the weight $e^{-S}$ of the matrix integral develops a delta-function-like peak when $N\gg1$.\footnote{However, it is dangerous to apply such a naive argument based on the matrix-integral weight $e^{-S}$ directly to the Minkowski-type weight $e^{iS}$.
This is because, in such an oscillatory integral, the absolute value of $S$ itself does not carry much significance.}

From here a very simple idea emerges: why not simply regularize the infinite-dimensional matrix $\hat{A}_{\mu}$ at finite dimension?
Since the system behaves ``almost deterministically'' when $N$ is sufficiently large, one may naively expect that, at least in some sense, such a finite-dimensional regularization should be meaningful as an approximate representation.
In this paper, to distinguish this quantity, we denote such a ``large but finite $N$'' by $M$, and we write the master field regularized to $M$ dimensions as $\hat{A}_{\mu}^{(M)}$.

Historically, essentially the same idea was already introduced in \cite{Jevicki:1982jj, Jevicki:1983hb}, and, following the proposal and subsequent development of the matrix bootstrap, detailed numerical studies have recently been carried out using this approach \cite{Koch:2021yeb, Mathaba:2023non}.
These studies performed numerical calculations for the $c=0$ matrix integral, which may be regarded as an analogue of the large-$N$ reduced matrix model, and for the $c=1$ matrix quantum mechanics by introducing ``master variables'' in loop space, corresponding to the matrix elements of $\hat{A}_{\mu}^{(M)}$ in the present formulation.
Our approach, on the other hand, may be interpreted as emphasizing the viewpoint of a finite-dimensional regularization of the master field, while incorporating several modifications to the formulation so that it is also applicable to Minkowski-type theories.
The differences from the previous approaches, as well as the advantages of the present method, will be discussed in detail in Section \ref{subsubsec:Some_Remarks_on_the_RMA}.

The question is by what principle this \textit{regularized master field} $\hat{A}_{\mu}^{(M)}$ should be regularized to finite dimension.
To repeat, the master field $\hat{A}_{\mu}$ is originally a quantity that can be defined only as an infinite-dimensional matrix, and reducing it to finite dimension is a highly nontrivial operation.
Conversely, since this finite-dimensional regularization is literally a regularization, one should in principle be able to remove it by a legitimate procedure, that is, starting from finite $M$, one should be able to recover the original infinite-dimensional matrix $\hat{A}_{\mu}$ uniquely.

Here, following recent studies on bootstrap programs, let us adopt the loop equations as the criterion for regularization (the advantages and caveats of using the loop equations will be discussed in Section \ref{subsubsec:Some_Remarks_on_the_RMA}).
As in the case of the loop equation \eqref{eq:loopeq_E-type_one-matrix} for the 1-matrix model, loop equations are in general an infinite set of identities for moments such as $\langle\text{tr}(A_{\mu_{1}}...A_{\mu_{n}})\rangle$.
Here we write them abstractly as
\begin{equation}
0={\cal L}_{\alpha},
\end{equation}
where $\alpha$ is an abstract label used to distinguish different loop equations.
Since ${\cal L}_{\alpha}$ is a combination of moments such as $\langle\text{tr}(A_{\mu_{1}}...A_{\mu_{n}})\rangle$, and since we have the large-$N$ limit in mind here, we assume that ${\cal L}_{\alpha}$ does not contain the joining terms of moments, namely the terms suppressed by $1/N^{2}$.
For example, if the right hand side of the loop equation \eqref{eq:loopeq_E-type_one-matrix}  of the 1-matrix model is expressed in this notation, then it takes the form
\begin{equation}
{\cal L}_{n}=\sum_{k=0}^{n-2}w_{n-k-2}w_{k}-w_{n}+gw_{n+2},
\end{equation}
where we have used the degree $n$ of the moment as the label $\alpha$.
In this notation, the loop equations can simply be written as
\begin{equation}
0={\cal L}_{n}\quad\text{for}\quad n=1,2,...\,.
\end{equation}
For more general matrix models, the structure is almost the same, and ${\cal L}_{\alpha}$ is given by a combination of moments such as $\langle\text{tr}(A_{\mu_{1}}...A_{\mu_{n}})\rangle$.
If the matrix model under consideration admits a master field $\hat{A}_{\mu}$, then all of these moments are equal to $\text{tr}(\hat{A}_{\mu_{1}}...\hat{A}_{\mu_{n}})$, and hence ${\cal L}_{\alpha}$ can be written in terms of the master field, more precisely, in terms of its components:
\begin{equation}
{\cal L}_{\alpha}={\cal L}_{\alpha}(\hat{A}_{\mu}).
\end{equation}
Let us use this for regularization.
What we need to do is simple: we replace all moments $\text{tr}(\hat{A}_{\mu_{1}}...\hat{A}_{\mu_{n}})$ appearing in the loop equations by \textit{regularized moments}.
Here the regularized moments are defined by
\begin{equation}
\text{tr}_{N}(\hat{A}_{\mu_{1}}...\hat{A}_{\mu_{n}})\overset{\text{regularization}}{\longrightarrow}\text{tr}_{M}(\hat{A}_{\mu_{1}}^{(M)}...\hat{A}_{\mu_{n}}^{(M)}),
\end{equation}
where $\text{tr}_{N}(...)$ and $\text{tr}_{M}(...)$ denote the normalized traces of $N$- and $M$-dimensional matrices, respectively.
Using this, let us write by ${\cal L}_{\alpha}^{(M)}$ the quantity obtained by replacing all moments appearing in ${\cal L}_{\alpha}$ by regularized moments.
For example, in the 1-matrix model one has
\begin{equation}
{\cal L}_{n}^{(M)}=\sum_{k=0}^{n-2}w_{n-k-2}^{(M)}w_{k}^{(M)}-w_{n}^{(M)}+gw_{n+2}^{(M)}.
\end{equation}
Here $w_{n}^{(M)}$ is the regularized moment corresponding to $w_{n}$.
Corresponding to the fact that ${\cal L}_{n}$ was a function of the master field $\hat{\phi}$, more precisely of its matrix components, note that ${\cal L}_{n}^{(M)}$ is a function of the regularized master field $\hat{\phi}^{(M)}$, namely ${\cal L}_{n}^{(M)}={\cal L}_{n}^{(M)}(\hat{\phi}^{(M)})$.
Likewise, in multi-matrix models, ${\cal L}_{\alpha}^{(M)}$ is a function of $\hat{A}_{\mu}^{(M)}$.

As already stated, the regularized master field $\hat{A}_{\mu}^{(M)}$ should be a quantity that approximates $\hat{A}_{\mu}$ in some sense, but only now do we obtain the criterion for that approximation.
Namely, if $\hat{A}_{\mu}^{(M)}$ is close to $\hat{A}_{\mu}$, then one should require that
\begin{equation}
0\approxeq{\cal L}_{\alpha}^{(M)}(\hat{A}_{\mu}^{(M)})
\end{equation}
hold for sufficiently many $\alpha$.
More precisely, the definition of the finite-dimensional regularization of the master field is given by determining the components of $\hat{A}_{\mu}^{(M)}$ so that this condition is satisfied as well as possible.

For a rigorous formulation, it is necessary to clarify the meaning of the approximate relation connected by ``$\approxeq$.''
Here, with numerical implementation in mind, we adopt a least-squares formulation.
Specifically, we define the objective function $F^{(M)}$ by
\begin{equation} \label{eq:Def_of_Objective_Function}
F^{(M)}(\hat{A}_{\mu}^{(M)})\equiv\sum_{\alpha\in{\cal S}^{(M)}}r_{\alpha}|{\cal L}_{\alpha}^{(M)}|^{2}.
\end{equation}
Since ${\cal L}_{\alpha}^{(M)}$ should be a quantity close to zero, or equivalently since $\hat{A}_{\mu}^{(M)}$ should be determined so that this is the case, the objective function $F^{(M)}$ is regarded as a quantity that should be made as close to zero as possible.
Here, $r_{\alpha}$ is a positive real number, introduced mainly as a factor for ensuring convergence in numerical calculations.

Since ${\cal L}_{\alpha}^{(M)}$ is a function of $\hat{A}_{\mu}^{(M)}$, the objective function $F^{(M)}$ defined from it is also a function of $\hat{A}_{\mu}^{(M)}$.
Thus, the number of unknown variables in $F^{(M)}$ is the same as the total number of independent components of $\hat{A}_{\mu}^{(M)}$.
Therefore, if the number of loop equations ${\cal L}_{\alpha}^{(M)}=0$, or more precisely the number of constraints imposed by ${\cal L}_{\alpha}^{(M)}=0$, is smaller than the total number of independent components of $\hat{A}_{\mu}^{(M)}$, then there may exist infinitely many $\hat{A}_{\mu}^{(M)}$ that give $F^{(M)}=0$.
This, however, does not serve the purpose of determining $\hat{A}_{\mu}^{(M)}$.
Therefore, the sum defining $F^{(M)}$ must be taken over sufficiently many $\alpha$ so that it gives a number of constraints greater than or equal to the total number of independent components of $\hat{A}_{\mu}^{(M)}$.

The sum $\sum_{\alpha\in{\cal S}^{(M)}}$ appearing in the definition \eqref{eq:Def_of_Objective_Function} is defined so as to satisfy this condition.
Namely, ${\cal S}^{(M)}$ is a set of loop equations, or more precisely their labels $\alpha$, and its total number of elements is $|{\cal S}^{(M)}|$.
Here, we require that the number of constraints imposed by these $|{\cal S}^{(M)}|$ loop equations ${\cal L}_{\alpha}^{(M)}=0$ be chosen to be greater than or equal to the total number of independent components of $\hat{A}_{\mu}^{(M)}$.
In other words, we choose ${\cal S}^{(M)}$ so that $F^{(M)}=0$ becomes an overdetermined or square system.
With this definition of the objective function $F^{(M)}$, it is natural to expect that $\hat{A}_{\mu}^{(M)}$ minimizing $F^{(M)}$ is determined uniquely.\footnote{From a purely algebraic point of view, even when the number of constraints imposed by ${\cal L}_{\alpha}^{(M)}=0$ balances the number of unknown variables in $F^{(M)}$, $F^{(M)}=0$ need not have a unique solution.
This is because the loop equations ${\cal L}_{\alpha}^{(M)}=0$ are nonlinear equations, and in such a case the corresponding $\hat{A}_{\mu}^{(M)}$ is not uniquely determined.
However, even if such an ambiguity arises, one can always increase the number of constraints and make the system overdetermined, so this does not cause a practical problem.}
This gives a concrete definition of how the infinite-dimensional matrix $\hat{A}_{\mu}$ is regularized into the finite $M$-dimensional matrix $\hat{A}_{\mu}^{(M)}$.

An important point from the viewpoint of regularization is that this definition can be applied, at least formally, for arbitrary $M$.
This gives a concrete prescription for removing the regularization and taking the large-$N$ (or large-$M$) limit.
For example, suppose that we compute $F^{(M)}$ and determine $\hat{A}_{\mu}^{(M)}$, and then determine $\hat{A}_{\mu}^{(M+1)}$ from $F^{(M+1)}$.
In this case, from the viewpoint of the convergence of the numerical results, and hence the consistency of the limiting procedure, $F^{(M+1)}$ should be chosen so as to contain all loop equations that were included in $F^{(M)}$.
Thus, we impose
\begin{equation}
{\cal S}^{(M)}\subseteq{\cal S}^{(M+1)}.
\end{equation}
At the same time, of course, ${\cal S}^{(M+1)}$ should also be chosen so that $F^{(M+1)}=0$ becomes an overdetermined or square system.
The weights $r_{\alpha}$ should also be chosen so as to be consistent, in some sense, with those used in $F^{(M)}$.
How these choices should be made in practice is best decided depending on the model; this will be examined explicitly in Section \ref{subsec:Overall_Setup_one-matrix_model} and \ref{subsec:Overall_Setup_two-matrix_model}.
Once ${\cal S}^{(M+1)}$ and $r_{\alpha}$ are chosen in this way, minimizing $F^{(M+1)}$ generically determines the corresponding $\hat{A}_{\mu}^{(M+1)}$ uniquely.

This procedure can be continued no matter how large $M$ is taken.
In other words, this is precisely the procedure for taking the limit $M\to\infty$.
The question is whether a meaningful limit exists at the end of this procedure.
One possible way to judge this is to examine whether, for any finite natural number $n$,
\begin{equation} \label{eq:Large_M_Limit}
\lim_{M\to\infty}\text{tr}_{M}(\hat{A}_{\mu_{1}}^{(M)}\cdots\hat{A}_{\mu_{n}}^{(M)})
\end{equation}
converges to a definite value for arbitrary $(\mu_{1},\ldots,\mu_{n})$.
Of course, since only finite values of $M$ can be treated numerically, it is impossible to verify convergence in its complete form.
Nevertheless, it is quite realistic to examine the variation of the regularized moments for several different values of $M$ and use this to infer whether the limit exists.
Moreover, matrix models have loop equations, under which most moments are not mutually independent.
Therefore, it is not necessary to check the convergence of all moments, which is a useful point in practice.
In addition to moments, the eigenvalues of $\hat{A}_{\mu}^{(M)}$ also provide a possible diagnostic.
These points will be examined explicitly in Section \ref{subsec:Numerical_Results_one-matrix_model} and \ref{subsec:Numerical_Results_two-matrix_model}.

A virtue of this regularization is that it can be used almost directly for numerical computation.
Namely, $\hat{A}_{\mu}^{(M)}$ minimizing $F^{(M)}$ can be obtained numerically by the least-squares method.
We call this method the regularized master-field approximation (RMA).
The main purpose of the following Section \ref{sec:One-Matrix_Model} and \ref{sec:Two-Matrix_Model} is to confirm that this RMA indeed works for one-matrix and two-matrix models.

\subsubsection{Some Remarks on the RMA} \label{subsubsec:Some_Remarks_on_the_RMA}

Before turning to the actual numerical calculations, we summarize several important points that should be kept in mind when applying the RMA.

\begin{enumerate}
\item[(1)] \textbf{Assumption of the existence of the master field}

First, the present method is formulated by assuming the existence of a master field.
However, the existence of the master field has so far remained a conjecture in general, with the eigenvalue distribution of the Euclidean one-matrix model serving only as a partial example.
Here, we call it ``partial'' because, as explained in Section \ref{subsubsec:Master_Field_as_a_Generalization_of_the_Eigenvalue_Distribution}, the eigenvalue distribution is contained within the concept of the master field.
If the matrix model under consideration does not admit a master field, then the present method loses its justification.
Therefore, when applying this method, one must examine the convergence of the limit in the sense described in \eqref{eq:Large_M_Limit} and check whether the regularized master-field description is plausible.

\item[(2)] \textbf{Necessity of consistency checks}

Another important point is that the present method is intrinsically an approximation.
While the matrix bootstrap method based on positivity gives rigorous bounds on physical quantities, the present method only computes approximate values.
Therefore, the numerical results cannot be accepted unconditionally.
In order to assess whether the obtained results are reliable, it is essential to have consistency checks that can be carried out systematically and consistently.
For example, in \cite{Maeta:2026oku}, consistency checks based on higher-order loop equations and those based on positivity were proposed.
Although we do not pursue this direction in depth in the present work, it remains an important subject for future study.

\item[(3)] \textbf{Finite-dimensional regularization}

A further important point is that the master field $\hat{A}_{\mu}$, which should originally be infinite-dimensional, is regularized at a finite dimension $M$.
Because of this finite-dimensional regularization, solutions that require infinite-dimensional unitary representations are excluded from the outset.
This can become a subtle issue in the two-matrix model defined in \eqref{eq:Action_of_Two-matrix_Models}.
For example, if we set $g=0$ and drop the quartic term, the corresponding equations of motion become
\begin{equation} \label{eq:EOM_of_2-MM}
\begin{cases}
h[A_{2},[A_{2},A_{1}]]+A_{1}=0,\\
h[A_{1},[A_{1},A_{2}]]+A_{2}=0.
\end{cases}
\end{equation}
In fact, these equations of motion admit as a solution the $SO(1,2)$ algebra expressed by the commutation relations
\begin{equation}
\begin{aligned}
[J_{i},J_{j}] & =i\epsilon_{ijk}J^{k},\\
J_{0} & =J^{0},\\
J_{1} & =-J^{1},\\
J_{2} & =-J^{2}.
\end{aligned}
\end{equation}
Namely, one can verify by direct calculation that
\begin{equation}
A_{\mu}=\frac{J_{\mu}}{\sqrt{h}}
\end{equation}
satisfies the equations of motion.
Thus, such a configuration should indeed constitute one saddle point at $g=0$.
However, since $SO(1,2)$ is a noncompact group, an infinite-dimensional representation is required in order for $J_{\mu}$ to be Hermitian, and it is known that no finite-dimensional unitary representation exists.
It follows that such a representation cannot be realized by increasing the finite dimension $M$ one by one, and is not expected to be reached in the $M\to\infty$ limit defined in \eqref{eq:Large_M_Limit}.
In other words, solutions such as the $SO(1,2)$ solution are automatically excluded by the present finite-dimensional regularization.

On the other hand, it should be pointed out that the large-$N$ limit itself can be subtle in certain situations.
This is because, for infinite-dimensional matrices obtained by taking $N\to\infty$, the trace is not necessarily well defined, that is, there can exist matrices that are not of trace class.
In fact, even a diagonal matrix, which is the most basic type of matrix, is in general not trace-class in the $N\to\infty$ limit.
Furthermore, noncompact noncommutative spaces, including the $J_{\mu}$ introduced above and many other solutions proposed in \cite{Chatzistavrakidis:2011su, Manta:2025tcl, Liao:2025yfb}, are in general not trace-class despite their physical interest.
In the infinite-dimensional matrix integral $\int dA\,{\cal O}(A)e^{-N\text{Tr}(\cdots)}$, this raises a subtle issue as to whether such non-trace-class matrices should be excluded, or whether they should be taken into account, at least formally.

From a general standpoint, it is perhaps natural, prioritizing physical relevance, to include such non-trace-class matrices in the matrix integral.
However, this in turn leads to another problem.
The reason why algebras of the type mentioned above are important in the context of matrix models is that they can serve as solutions of the equations of motion, for example, \eqref{eq:EOM_of_2-MM}.
On the other hand, in deriving the equations of motion, one typically needs to use the cyclic invariance of the trace.
For matrices that are not of trace class, however, the trace itself is not well defined, and hence the cyclic invariance is not even meaningful.
Therefore, it appears logically inconsistent to assume non-trace-class algebras as solutions of the equations of motion.
Indeed, when such non-trace-class solutions are substituted back into the action $S$, the resulting expression typically diverges or becomes ill-defined.

In this regard, the large-$N$ (or large-$M$) limit defined in the present work is as follows.
The finite-dimensionally regularized matrices $\hat{A}_{\mu}^{(M)}$ can be defined for any positive integer $M$, and if a limit exists in some sense as $M$ is increased, that limit is taken to define the large-$N$ limit.
Regardless of the physical validity of this definition, it removes most of the subtleties associated with the usual large-$N$ limit, at least from a mathematical point of view.
In what follows, the large-$N$ limit considered in this paper refers to this particular limiting procedure.
Conversely, when one wishes to treat, in particular numerically, algebras such as the Moyal-Weyl noncommutative-plane solution or the $SO(1,2)$ and $SO(1,3)$ algebras, it appears necessary to define an appropriate large-$N$ limit separately.

\item[(4)] \textbf{Modifications to the previous approach}

The idea of using a regularized master field for approximate calculations goes back to \cite{Jevicki:1982jj, Jevicki:1983hb}.
There, variables called ``master variables'' were introduced, which essentially correspond to the matrix elements of $\hat{A}_{\mu}^{(M)}$ in the present formulation.
More recently, numerical calculations were carried out in \cite{Koch:2021yeb, Mathaba:2023non} for the $c=0$ matrix integral, which is an analogue of the large-$N$ reduced matrix model, and for a toy model of the $c=1$ matrix quantum mechanics.
More concretely, the effective action or effective Hamiltonian of the theory was derived, and matrix configurations extremizing it were obtained numerically.\footnote{
Here the term ``effective action'' does not refer to the usual 1PI effective action defined by the Legendre transform of the free energy.
Rather, it refers to the quantity obtained by adding to the classical action the contribution from the Jacobian that appears when the matrix variables are gauge fixed or transformed into gauge-invariant loop variables.
}
In other words, their approach amounts to solving numerically the equations of motion derived from the effective action or effective Hamiltonian.

Let us now turn to our method.
A major difference from the previous approach is that we use the loop equations, rather than the equations of motion, for the optimization.
To explain the reason for this, we first need to point out that in a large-$N$ reduced matrix model without time $t$, there is no a priori notion of energy.
Even in such a situation, one can still write down equations of motion, but their solutions give extrema of the value of the effective action itself, rather than of the energy.
The subtlety arises when there are multiple such local solutions.
As long as the matrix model under consideration is Euclidean-type, this does not cause a serious problem.
This is because, if the weight of the matrix integral is given by $e^{-S_{\text{eff}}}$, the configuration minimizing $S_{\text{eff}}$ should dominate in the large-$N$ limit.

On the other hand, this becomes a serious issue when one considers a Minkowski-type large-$N$ reduced matrix model.
This is because, if the weight of the matrix integral is given by $e^{iS_{\text{eff}}}$, the magnitude of $S_{\text{eff}}$ itself should have no direct significance.
From the viewpoint of numerical computation, it is still possible, even for a large-$N$ reduced matrix model, to optimize $\hat{A}_{\mu}^{(M)}$, or the master variables, by using the equations of motion.
The real problem is that, when several configurations are obtained as a result of the optimization, there is no clear selection principle for choosing the correct one among them.
In the case of matrix quantum mechanics, by contrast, one can define the corresponding effective Hamiltonian.
Hence, at least in principle, there exists a firm criterion for identifying the correct saddle, namely that the true vacuum is given by the configuration minimizing the effective energy among all saddle points.
This is equivalent to taking the lowest-energy state of the quantum theory as the vacuum, and in this sense the criterion has a physical basis.\footnote{
However, even in matrix quantum mechanics, if the model is Minkowski-type, the corresponding master field becomes complex.
Therefore, the loop-space positivity constraint $\Omega(C,C')\ge0$ pointed out in \cite{Koch:2021yeb} is not automatically satisfied.
}
In the case of a large-$N$ reduced matrix model, however, the most we can identify is the configuration minimizing the effective action, and there is no guarantee that this configuration corresponds to the true saddle of the theory.

Thus, the difficulty of the Minkowski-type large-$N$ reduced matrix model is that a solution of the equations of motion does not necessarily correspond to the dominant saddle, and that there is no criterion, namely energy, for deciding whether it does.
At the same time, tracing the issue back to its origin, one may also view the problem as stemming from the fact that the equations of motion can admit multiple solutions in the first place.
This is precisely why we use the loop equations rather than the equations of motion.
Unlike the equations of motion, one can generate arbitrarily many loop equations, and it is always possible to make $F^{(M)}=0$ an overdetermined system.
For such an overdetermined system, there is generally no solution of $F^{(M)}=0$, namely no matrix configuration satisfying all of the given loop equations.
Instead, one considers a configuration that minimizes $F^{(M)}$ as much as possible, or equivalently, one that satisfies the given loop equations as well as possible.
For an overdetermined system, this configuration is uniquely determined except in special exceptional situations.
In this way, the problem of multiple solutions is theoretically resolved.
In practice, however, it is not easy to find numerically the matrix configuration that minimizes $F^{(M)}$.
How to obtain the correct matrix configuration within the present formulation, including the issue of multiple solutions, will be discussed in detail in the analysis of Section \ref{subsubsec:Minkowski-type_One-Matrix_Model} and \ref{subsubsec:Minkowski-type_two-Matrix_Model}.
Other properties of the loop equations are explained in the following.

\item[(5)] \textbf{Several properties of the loop equations}

The loop equations, or more generally the Schwinger-Dyson equations, provide powerful constraints as nonperturbative equations; however, as has been repeatedly emphasized in previous studies, they cannot by themselves determine the values of physical quantities or their functional forms, at least as long as only finitely many equations are considered.
This is because the loop equations allow infinitely many unphysical solutions.
For example, in the one-matrix model, the values of $w_{1}$ and $w_{2}$ cannot be determined from the loop equations, and hence the values of $w_{n}$ remain undetermined.
It is an artifact caused by forgetting the origin of the loop equations and regarding them as a purely algebraic system of equations.

Existing approaches can all be understood as attempts to remove such unphysical solutions in one way or another, in order to make maximal use of the constraining power of the loop equations.
This removal has been achieved by positivity, by the cut structure of the resolvent, by the eigenvalue distribution, or by the asymptotic behavior of $w_{n}$.
In the present work, by contrast, we expect that the assumption that the expectation value $\langle\text{tr}(A_{\mu_{1}}\cdots A_{\mu_{n}})\rangle$ can be written simply as the trace $\text{tr}(\hat{A}_{\mu_{1}}\cdots\hat{A}_{\mu_{n}})$, namely, the assumption that the master field $\hat{A}_{\mu}$ exists, plays the role of excluding such unphysical solutions.
This is because the existence, or conjectured existence, of $\hat{A}_{\mu}$ originates from the fact that the theory is defined by a matrix integral, and therefore constitutes information brought in from outside the loop equations regarded as algebraic equations.

Once unphysical solutions have been excluded, the loop equations can become an extremely powerful tool for constraining the values of lower-order moments.
For example, Ref. \cite{Kazakov:2021lel} derived the allowed region for lower-order moments in a two-matrix model using rigorous inequality bounds, where the values were constrained with a precision reaching the seventh decimal place.
Given that the variables appearing in the loop equations, namely the moments $\langle\text{tr}(A_{\mu_{1}}\cdots A_{\mu_{n}})\rangle$, constitute an infinite set, whereas the bootstrap approach considers only finitely many loop equations, it is by no means obvious that such a high level of precision can be achieved.
This is presumably a consequence of the (weak) hierarchical structure of the loop equations.
While the loop equations relate moments of different orders, the coupling between low-order and higher-order moments is relatively weak, whereas moments of similar order are more strongly connected to one another.
Because of this structure, low-order loop equations are expected to play the dominant role in constraining lower-order moments, while the contribution of very high-order loop equations becomes comparatively less important.

From a practical point of view, this hierarchical structure of the loop equations is highly advantageous.
A computer can handle only finitely many variables, and therefore it is, in principle, impossible to determine or approximate infinitely many moments simultaneously with high precision.
By exploiting the hierarchical structure of the loop equations, however, one may reasonably expect that at least the lower-order moments can be computed accurately, albeit at the expense of the accuracy of higher-order moments.
For example, in the one-matrix model, it has been shown that the eigenvalue distribution can be approximated with very high accuracy by using only the low-order loop equations \cite{Maeta:2026oku}.
This, in turn, provides an accurate approximation to the lower-order moments, whereas the accuracy deteriorates for higher-order moments because they are much more sensitive to errors in the approximate eigenvalue distribution.
Moreover, within the matrix-bootstrap approach, it has been rigorously shown that combining the low-order loop equations with the positivity constraints leads to strong constraints on the values of the lower-order moments \cite{Kazakov:2021lel}.
Although RMA does not make use of positivity, relying instead on the existence of the master field as the basis of the approximation, the hierarchical structure of the loop equations themselves remains unchanged.
Furthermore, since the positivity constraints are not expected to impose overwhelmingly strong restrictions, one may likewise expect RMA to provide a good approximation, at least for the lower-order moments.

\item[(6)] \textbf{Roles of $r_{\alpha}$ and ${\cal S}^{(M)}$}

Finally, let us comment on the roles of $r_{\alpha}$ and ${\cal S}^{(M)}$ in \eqref{eq:Def_of_Objective_Function}.
The regularized master field $\hat{A}_{\mu}^{(M)}$ is determined as a matrix that minimizes the objective function $F^{(M)}$.
Since the definition of $F^{(M)}$ contains the weights $r_{\alpha}$ and the set ${\cal S}^{(M)}$, the resulting $\hat{A}_{\mu}^{(M)}$ depends on them.
This is a kind of artifact associated with the regularization scheme.
If the theory has a master field $\hat{A}_{\mu}$, then the dependence on $r_{\alpha}$ and ${\cal S}^{(M)}$ should disappear in the $M\to\infty$ limit.
On the other hand, the rate of convergence as $M$ is increased depends strongly on the choices of $r_{\alpha}$ and ${\cal S}^{(M)}$.
Since $F^{(M)}$ and $\hat{A}_{\mu}^{(M)}$ are also used for numerical computation, choosing $r_{\alpha}$ and ${\cal S}^{(M)}$ too arbitrarily may undermine the practical usefulness of the RMA as an approximation method.
Therefore, in Section \ref{subsec:Overall_Setup_one-matrix_model} and \ref{subsec:Overall_Setup_two-matrix_model}, we propose suitable choices of $r_{\alpha}$ and ${\cal S}^{(M)}$, and verify that the approximation works well under these choices.
\end{enumerate}

\section{One-Matrix Model} \label{sec:One-Matrix_Model}

\subsection{Overall Setup} \label{subsec:Overall_Setup_one-matrix_model}

In Section \ref{sec:Method}, we have explained the concept of the RMA and its concrete implementation procedure.
Based on this, we now apply the method to simple examples, namely Hermitian one- and two-matrix models, in order to demonstrate its usefulness.
In this Section \ref{sec:One-Matrix_Model}, we consider the Euclidean and Minkowski one-matrix models.
We begin by recalling the definitions of the action and the partition functions:
\begin{equation}
\begin{aligned}
S_{\text{1-MM}}[\phi] & =N\text{Tr}\left(\frac{1}{2}\phi^{2}-\frac{g}{4}\phi^{4}\right),\\
Z_{E} & \equiv\int d\phi\, e^{-S_{\text{1-MM}}[\phi]},\\
Z_{M} & \equiv\int d\phi\, e^{iS_{\text{1-MM}}[\phi]}.
\end{aligned}
\end{equation}
The loop equations derived from these definitions take the following forms in the Euclidean and Minkowski cases, respectively:
\begin{equation}
\begin{aligned}
0 & =\sum_{k=0}^{n-2}w_{n-k-2}w_{k}-w_{n}+gw_{n+2}={\cal L}_{E,n},\\
0 & =\sum_{k=0}^{n-2}w_{n-k-2}w_{k}+iw_{n}-igw_{n+2}={\cal L}_{M,n}.
\end{aligned}
\end{equation}
Here, the notation ${\cal L}_{E,n}$ and ${\cal L}_{M,n}$ is the same as that introduced in Section \ref{subsubsec:Approximation_by_a_Finite-Dimensional_Regularized_Master_Field}, so that the loop equations can simply be written as ${\cal L}_{E,n}=0$ and ${\cal L}_{M,n}=0$.
Since we are taking the limit $N\to\infty$, we neglect the joining terms suppressed by $1/N^{2}$.

The starting point of the RMA is the existence of the master field.
Accordingly, we assume here that there exists an infinite-dimensional matrix $\hat{\phi}$ such that
\begin{equation}
w_{n}=\text{tr}\hat{\phi}^{n}
\end{equation}
holds for arbitrary $n\in\mathbb{N}$.
It should be noted, however, that this is merely an assumption at this stage, and should be checked or proven by numerical or analytical calculations.

A special feature of the one-matrix model is that, since there is only one matrix, only the diagonal components of $\hat{\phi}$ contribute to the moments $w_{n}$.
We therefore diagonalize $\hat{\phi}$ from the outset.
In Section \ref{sec:One-Matrix_Model}, for both the Euclidean and Minkowski cases, we assume the ansatz
\begin{equation}
\hat{\phi}=\text{diag}(\hat{z}_{1},\ldots,\hat{z}_{N}) .
\end{equation}
Of course, in the Euclidean case it is also possible to perform the RMA using the real-eigenvalue ansatz $(\hat{\lambda}_{1},\ldots,\hat{\lambda}_{N})$.
However, since the issue of false convergence associated with different initial conditions becomes important in the analysis of the Minkowski two-matrix model in Section \ref{subsubsec:Minkowski-type_two-Matrix_Model}, and since this issue is more naturally discussed using the complex-eigenvalue ansatz $(\hat{z}_{1},\ldots,\hat{z}_{N})$, we adopt the present choice in order to maintain a coherent flow of discussion.
The numerical results obtained using the real-eigenvalue ansatz $(\hat{\lambda}_{1},\ldots,\hat{\lambda}_{N})$ are presented in Appendix \ref{app:appendixB}.
We also note that, when using $(\hat{\lambda}_{1},\ldots,\hat{\lambda}_{N})$, one must choose the weights $r_{n}$ in the objective function in a somewhat ad hoc manner.

The next step is to regularize the master field $\hat{\phi}$ to a finite $M$-dimensional matrix and introduce the corresponding regularized moments:
\begin{equation}
\begin{aligned}
\hat{\phi} & \overset{\text{regularization}}{\longrightarrow}\hat{\phi}^{(M)},\\
w_{n}=\text{tr}_{N}\hat{\phi}^{n} & \overset{\text{regularization}}{\longrightarrow}\text{tr}_{M}(\hat{\phi}^{(M)})^{n}=w_{n}^{(M)}.
\end{aligned}
\end{equation}
Here, we also diagonalize $\hat{\phi}^{(M)}$ in accordance with $\hat{\phi}$, and then the component representation of $w_{n}^{(M)}$ becomes
\begin{equation}
\begin{aligned}
\hat{\phi}^{(M)} & =\text{diag}(\hat{z}_{1}^{(M)},\ldots,\hat{z}_{M}^{(M)}),\\
w_{n}^{(M)} & =\frac{1}{M}\sum_{i=1}^{M}(\hat{z}_{i}^{(M)})^{n}.
\end{aligned}
\end{equation}
The quantities ${\cal L}_{n}^{(M)}$ are obtained by replacing the moments $w_{n}$ appearing in ${\cal L}_{n}$, namely in the right-hand side of the loop equations, with these regularized moments $w_{n}^{(M)}$:
\begin{equation}
\begin{aligned}
{\cal L}_{E,n}^{(M)} & =\sum_{k=0}^{n-2}w_{n-k-2}^{(M)}w_{k}^{(M)}-w_{n}^{(M)}+gw_{n+2}^{(M)},\\
{\cal L}_{M,n}^{(M)} & =\sum_{k=0}^{n-2}w_{n-k-2}^{(M)}w_{k}^{(M)}+iw_{n}^{(M)}-igw_{n+2}^{(M)}.
\end{aligned}
\end{equation}
Using these quantities, we define the objective function by
\begin{equation} \label{eq:Objective_Function_for_1-MM}
F^{(M)}\equiv\sum_{n=1}^{\Lambda}r^{n}|{\cal L}_{n}^{(M)}|^{2}.
\end{equation}
Here, we simply take the sum over $n=1,2,\ldots$ in order, which corresponds to ${\cal S}^{(M)}$ in \eqref{eq:Def_of_Objective_Function}.
This choice is the most straightforward one, but the main reason for adopting it is the asymptotic behavior of the regularized moments $w_{n}^{(M)}$.
Indeed, since the present $w_{n}^{(M)}$ is given as the average of the $n$th powers of the eigenvalues $\hat{z}_{i}^{(M)}$, it is expected to grow exponentially as $w_{n}^{(M)}\sim a^{n}$ for $n\gg1$.
Accordingly, ${\cal L}_{n}^{(M)}$ should also take such extremely large values, which is not desirable for carrying out the RMA stably.
This is because, in that case, only loop equations ${\cal L}_{n}^{(M)}$ with large $n$ would effectively contribute to $F^{(M)}$, thereby undermining the effectiveness of the RMA as an approximation method.

For this reason, in the present case it is desirable to keep $n$ as small as possible, and this is achieved simply by summing over $n=1,\ldots,\Lambda$.
Here, $\Lambda$ denotes the highest order of the loop equations taken into account, and the maximum order $\Lambda$ of the loop equations is chosen so as to satisfy $\Lambda\ge M$, in accordance with the requirement that $F^{(M)}=0$ be an overdetermined or square system.
As for the weights $r_{n}$, we introduce a positive real number $r$ satisfying $0<r<1$ and set $r_{n}=r^{n}$.
This choice is made, as explained above, because the behavior $w_{n}^{(M)}\sim a^{n}$ is expected for $n\gg1$.

\subsection{Numerical Results} \label{subsec:Numerical_Results_one-matrix_model}

Under the setup introduced in Section \ref{subsec:Overall_Setup_one-matrix_model}, we performed numerical calculations using the RMA.
In this work, we performed the least-squares minimization using the \texttt{least\_squares} function provided by SciPy~\cite{virtanen2020scipy}. We employed the \texttt{trf} algorithm, and set \texttt{ftol}, \texttt{xtol}, and \texttt{gtol} all to $10^{-14}$, with \texttt{max\_nfev} set to $20000$.
The value of $g$ is fixed to $g=-1$, and four matrix sizes, $M=10,20,30,40$, are examined.
For the objective function $F^{(M)}$, we always choose $\Lambda=M$ and set $r=0.8$ throughout this work.
The optimization is initialized from random initial conditions obtained by independently sampling the real and imaginary parts of $\hat{z}_{1}^{(M)},\ldots,\hat{z}_{M}^{(M)}$ from Gaussian distributions with zero mean and standard deviation $0.01$.
In practice, however, the results of the RMA method depend on the initial condition, which can sometimes lead to undesirable spurious solutions.
In Section \ref{subsubsec:Euclidean-type_One-Matrix_Model}, we briefly illustrate the effects of different initial conditions using the Euclidean one-matrix model, while a more detailed discussion will be given later in Section \ref{subsubsec:Minkowski-type_two-Matrix_Model} in the context of the two-matrix model computations.
The results for the Euclidean and Minkowski cases are presented in Section \ref{subsubsec:Euclidean-type_One-Matrix_Model} and \ref{subsubsec:Minkowski-type_One-Matrix_Model}, respectively.

\subsubsection{Euclidean-type One-Matrix Model} \label{subsubsec:Euclidean-type_One-Matrix_Model}

\begin{figure}[t] 
\centering
\includegraphics[width=0.7\textwidth]{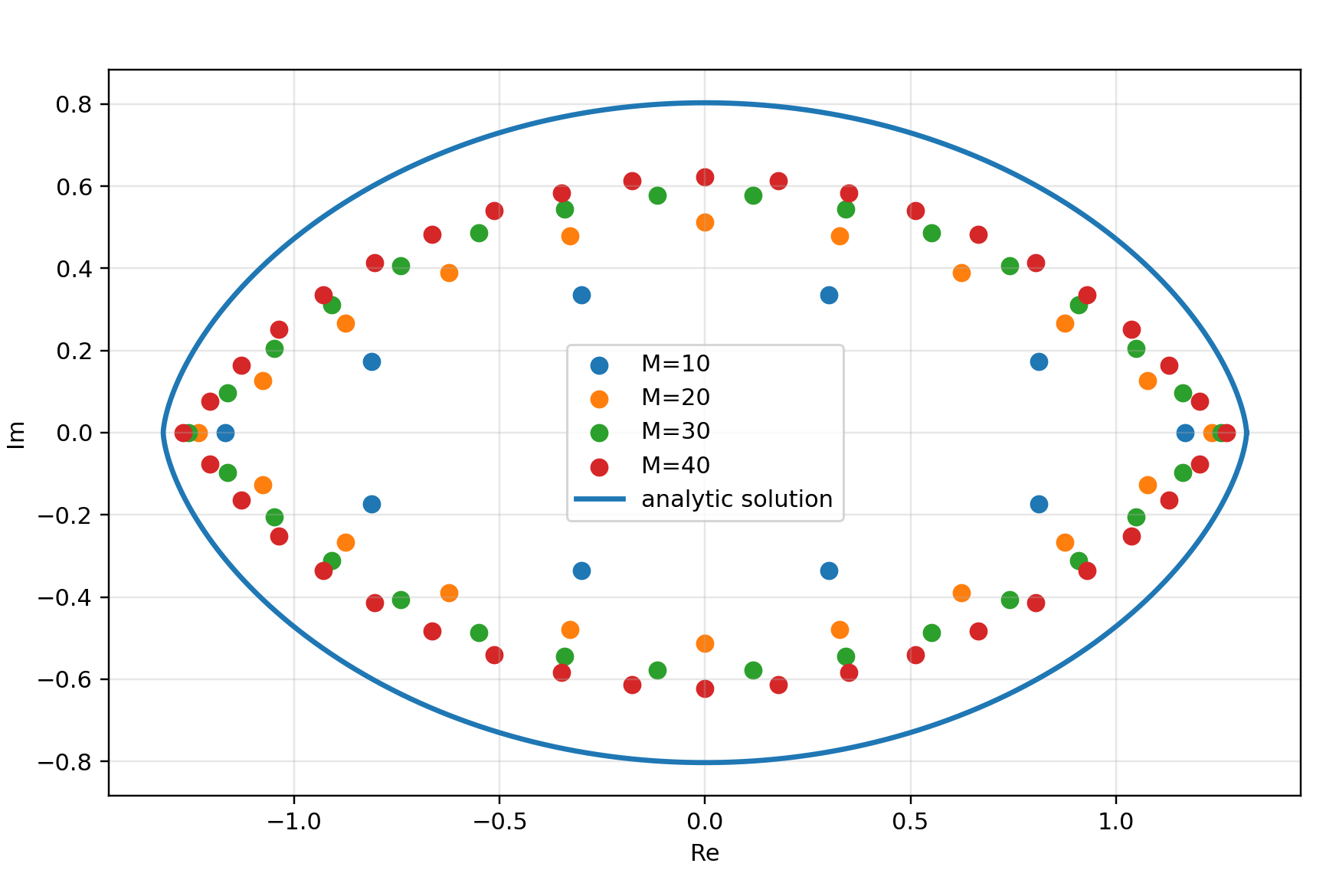}
\caption{
The eigenvalues $\hat{z}_{1}^{(M)},\ldots,\hat{z}_{M}^{(M)}$ obtained by the RMA for $g=-1$ with $M=10,20,30,40$, plotted in the complex plane.
The cases $M=10,20,30,40$ correspond to blue, orange, green, and red points, respectively.
The solid blue curve represents a closed contour satisfying $\frac{\dot{z}(t)}{2\pi i}R_{E}(z(t))=1$ and passing through the endpoints $\pm\alpha$ of $R_{E}(z)$.
}
\label{fig:E-type_1MM_Eigenvalues_complex}
\end{figure}

\begin{table}[t]
\centering
\begin{tabular}{cccccc}
\hline
\(M\)
&
\(w_2^{(M)}\)
&
\(\left|w_2^{\rm exact}-w_2^{(M)}\right|\)
&
\(z_{\max}^{(M)}\)
&
\(\left|\alpha-z_{\max}^{(M)}\right|\)
&
\(F_E^{(M)}\)
\\
\hline
10
&
0.5165177298
&
\(3.665\times 10^{-4}\)
&
1.168
&
0.150
&
\(1.261\times 10^{-31}\)
\\
20
&
0.5161486236
&
\(2.609\times 10^{-6}\)
&
1.233
&
0.085
&
\(7.633\times 10^{-31}\)
\\
30
&
0.5161512617
&
\(2.882\times 10^{-8}\)
&
1.257
&
0.061
&
\(4.458\times 10^{-30}\)
\\
40
&
0.5161512326
&
\(2.963\times 10^{-10}\)
&
1.270
&
0.048
&
\(3.184\times 10^{-29}\)
\\
\hline
\end{tabular}
\caption{
Detailed numerical results of the RMA for $g=-1$ with $M=10,20,30,40$.
Here, $z_{\text{max}}^{(M)}$ denotes the element of $\hat{z}_{1}^{(M)},\ldots,\hat{z}_{M}^{(M)}$ with the largest real part, while $\alpha$ denotes one of the endpoints of the cut of the resolvent $R_{E}(z)$ that takes a positive real value.
}
\label{tab:one_matrix_euclidean_rma}
\end{table}

For fixed $g=-1$, the approximate eigenvalues $\hat{z}_{1}^{(M)},\ldots,\hat{z}_{M}^{(M)}$ computed for $M=10,20,30,40$ are shown in Figure \ref{fig:E-type_1MM_Eigenvalues_complex}.
These eigenvalues lie in the complex plane, and the points are colored according to the corresponding value of $M$.
The solid blue curve represents the closed curve $C_{\text{const.}}$ that satisfies $\frac{\dot{z}(t)}{2\pi i}R_{E}(z(t))=1$ and passes through the endpoints $\pm\alpha$ of $R_{E}(z)$.
In Section \ref{subsubsec:Master_Field_as_a_Generalization_of_the_Eigenvalue_Distribution}, we argued that the eigenvalues $\hat{z}_{1},\ldots,\hat{z}_{N}$ of the master field should coincide with this curve, and indeed $\hat{z}_{1}^{(M)},\ldots,\hat{z}_{M}^{(M)}$ approach $C_{\text{const.}}$ as $M$ increases.
The resolvent $R_{E}(z)$ of the Euclidean one-matrix model is obtained analytically as
\begin{equation}
R_{E}(z)=\frac{1}{2}\left[z-gz^{3}-\sqrt{(z-gz^{3})^{2}+4\{g(w_{2}^{2}+zw_{1}+z^{2})-1\}}\right],
\end{equation}
and, assuming a one-cut structure, it is given by
\begin{equation}
\begin{aligned}
R_{E}(z) & =\frac{1}{2}\left[z-gz^{3}-\left(1-\frac{\alpha^{2}g}{2}-gz^{2}\right)\sqrt{(z-\alpha)(z+\alpha)}\right],\\
w_{2} & =\frac{\alpha^{4}-\alpha^{6}g}{16},\\
\alpha & =\sqrt{\frac{2(1-\sqrt{1-12g})}{3g}}.
\end{aligned}
\end{equation}
Using these expressions, one can compute $C_{\text{const.}}$ and various physical quantities exactly.
For a more detailed analysis of the one-matrix model, see \cite{DiFrancesco:1993cyw}.

As shown in Figure \ref{fig:E-type_1MM_Eigenvalues_complex}, even at $M=40$ there remains a non-negligible discrepancy between $\hat{z}_{1}^{(M)},\ldots,\hat{z}_{M}^{(M)}$ and $C_{\text{const.}}$, suggesting that it is difficult to approximate $C_{\text{const.}}$ efficiently by the RMA.
However, this does not undermine the value of the RMA as an approximation method; what we really want to know is not the closed curve $C_{\text{const.}}$ itself, but the values of the physical observables $w_{n}$.
Detailed numerical results, including these quantities, are summarized in Table \ref{tab:one_matrix_euclidean_rma}.
It is noteworthy that, despite the deviation from $C_{\text{const.}}$, even the smallest case $M=10$ reproduces the value of $w_{2}$ well.
The endpoint \(\alpha\) is also approximated by \(z_{\text{max}}^{(M)}\), defined as the eigenvalue with the largest real part, with an error of around \(10\%\) already at \(M=10\).
Overall, the RMA can be regarded as a highly efficient computational method for quantities that are physically or practically important.

\begin{table}[t]
\centering
\begin{tabular}{cccc}
\hline
sample & $w_{2}^{(20)}$ & $F^{(20)} \sim 10^{-30}$ ? & Exact? \\
\hline
seed 0 & $0.4805 + 0.1375i$ & Yes & No \\
seed 1 & $-1.0031 - 0.0699i$ & No & No \\
seed 2 & $-0.5291 - 0.5499i$ & No & No \\
seed 3 & $-2.7763 + 2.4900i$ & No & No \\
seed 4 & $-0.8518 + 0.3145i$ & No & No \\
seed 5 & $0.5162$ & Yes & Yes \\
seed 6 & $0.5162$ & Yes & Yes \\
seed 7 & $-0.7440 - 0.1631i$ & No & No \\
seed 8 & $0.1057 + 0.1874i$ & Yes & No \\
seed 9 & $-1.2720 - 0.1026i$ & No & No \\
seed 10 & $-0.5585 - 0.5577i$ & No & No \\
seed 11 & $-0.4841 + 0.0135i$ & Yes & No \\
seed 12 & $0.5162$ & Yes & Yes \\
seed 13 & $-0.4635 - 0.6845i$ & No & No \\
seed 14 & $-0.9523 - 0.2948i$ & No & No \\
\hline
\end{tabular}
\caption{
For $g=-1$, we construct an underdetermined system with $M=20$ and $\Lambda=18$, and perform the RMA procedure 15 times using randomized initial conditions.
More specifically, the real and imaginary parts of $\hat{z}_{1}^{(M)},...,\hat{z}_{M}^{(M)}$ are independently sampled from Gaussian distributions with standard deviation $\sigma=0.8$, corresponding to seed 0--14, respectively.
The columns ``$F^{(20)}\sim10^{-30}$ ?'' and ``Exact?'' indicate whether the objective function decreases to $F^{(20)}\sim10^{-30}$ and whether $w_{2}^{(18)}$ reproduces the exact value, respectively.
}
\label{tab:e1mm_m20_l18_seed_dependence}
\end{table}

The results presented above were obtained by initializing all of $\hat{z}_{1}^{(M)},\ldots,\hat{z}_{M}^{(M)}$ with small random fluctuations around zero.
On the other hand, if large fluctuations are introduced into the initial values of $\hat{z}_{1}^{(M)},...,\hat{z}_{M}^{(M)}$, the optimization does not necessarily converge to the optimal solution.
More specifically, the system can become trapped in a local solution for which the objective function $F^{(M)}$ does not decrease sufficiently.
Naturally, such matrix configurations do not correctly reproduce the analytic solution for $w_{n}$.
As an example, we consider here an underdetermined system with $M=20$ and $\Lambda=18$, and perform the RMA procedure with randomized initial conditions.
More specifically, the real and imaginary parts of $\hat{z}_{1}^{(M)},...,\hat{z}_{M}^{(M)}$ are independently sampled from Gaussian distributions with standard deviation $\sigma=0.8$.
Table \ref{tab:e1mm_m20_l18_seed_dependence} summarizes the results of 15 independent RMA runs performed with these random initial conditions.

Among the 15 different initial conditions, only three cases, namely Seed 5, 6, and 12, converged to the correct solution.
In these cases, the objective function $F^{(M)}$ decreased down to $10^{-30}$, meaning that the resulting configurations $\hat{z}_{1}^{(M)},...,\hat{z}_{M}^{(M)}$ provide almost exact solutions of the 18 loop equations.
At present, we are dealing with an underdetermined system with $M=20$ and $\Lambda=18$, so there is no obvious guarantee that the approximation should succeed.
Nevertheless, the correct approximate solution can in fact be reproduced with high precision.
We also note that when smaller fluctuations with $\sigma\lessapprox0.5$ were used, the optimization always converged to the correct solution.

On the other hand, the remaining 12 initial conditions all converged to incorrect configurations, and the resulting values of $w_{2}^{(M)}$ were far from the exact solution.
Interestingly, however, among them, Seed 0, 8, and 11 still achieved $F^{(M)}\sim10^{-30}$.
This occurs partly because we are considering an underdetermined system, and also because the loop equations form a nonlinear system that may admit multiple solutions even in a square system.
In order for the RMA method to become a reliable computational framework, some mechanism must therefore exist to eliminate such spurious solutions.

\begin{table}[t]
\centering
\begin{tabular}{ccccc}
\hline
$\Lambda$ & seed 0 & seed 5 & seed 8 & seed 11 \\
\hline
18 & $0.4805 + 0.1375i$ & $0.5162$ & $0.1057 + 0.1874i$ & $-0.4841 + 0.0135i$ \\
19 & $0.5375 + 0.0688i$ & $0.5162$ & $0.1060 + 0.1879i$ & $-0.5754 + 0.1137i$ \\
20 & $0.3571 + 0.1377i$ & $0.5161$ & $0.1101 + 0.2065i$ & $-0.5468 + 0.1811i$ \\
21 & $0.3599 + 0.1451i$ & $0.5161$ & $0.1192 + 0.2207i$ & $-0.5842 + 0.2316i$ \\
\hline
\end{tabular}
\caption{
For $g=-1$ and $M=20$, we perform the RMA procedure for Seed 0, 5, 8, and 11 with $\Lambda=19,20,21$.
The resulting values of $w_{2}^{(20)}$ are summarized in a table.
When optimizing at $\Lambda+1$, the computation is started using the optimized configurations $\hat{z}_{1}^{(M)},...,\hat{z}_{M}^{(M)}$ obtained at $\Lambda$ as the initial conditions.
The results for $\Lambda=18$ are taken directly from those computed in Table \ref{tab:e1mm_m20_l18_seed_dependence}.
}
\label{tab:e1mm_seed_comparison_lambda18_21}
\end{table}

To investigate this issue, we next take the matrix configurations $\hat{z}_{1}^{(M)},...,\hat{z}_{M}^{(M)}$ obtained from Seed 0, 8, and 11, together with Seed 5 as a reference corresponding to the exact solution, and use them as initial conditions for new RMA computations with $\Lambda=19$.
The resulting configurations are then reused as initial conditions for the case $\Lambda=20$.
This procedure is repeated up to $\Lambda=21$.
The results are summarized in Table \ref{tab:e1mm_seed_comparison_lambda18_21}.
Although not explicitly displayed there, all configurations $\hat{z}_{1}^{(M)},...,\hat{z}_{M}^{(M)}$ achieve $F^{(M)}\sim10^{-30}$ at $\Lambda=20$.
On the other hand, for the overdetermined case $\Lambda=21$, none of them achieve $F^{(M)}\sim10^{-30}$.
Therefore, the incorrectly converged samples Seed 0, 8, and 11 cannot be excluded solely on the basis of the convergence of $F^{(M)}$. \footnote{Among the 15 seeds, however, 9 failed to achieve $F^{(M)}\sim10^{-30}$ even in the underdetermined regime, and all of them resulted in unsuccessful approximations.
This suggests that whether or not $F^{(M)}\sim10^{-30}$ is achieved still may serve as a useful diagnostic criterion.}

Fortunately, another criterion can be used here to judge the validity of the approximation: one may examine the stability of the value of $w_{2}^{(M)}$.
For Seed 5, which corresponds to the exact solution, the value of $w_{2}^{(M)}$ remains nearly unchanged as $\Lambda$ is varied.
This is natural, given that the approximation was already highly accurate at $\Lambda=18$.
From another viewpoint, such configurations already lie close to the ``physical'' solution of the loop equations even at the stage of an underdetermined system.
As a result, they are largely insensitive to the details of the regularization scheme, or may be regarded as being ``physically protected.''

By contrast, when one starts from incorrect configurations $\hat{z}_{1}^{(M)},...,\hat{z}_{M}^{(M)}$, the value of $w_{2}^{(M)}$ can change significantly even when $\Lambda$ is increased by only one, namely when only a single additional loop equation is included.
This suggests that such configurations do not correspond to physical solutions, but instead happen to satisfy the loop equations accidentally, so that even a slight modification of the regularization scheme drastically changes the resulting values.

In summary, the RMA method suffers from the problem that the converged approximate values may depend on the initial conditions.
The first criterion that should be examined in determining whether a given approximation is reliable is whether $F^{(M)}\sim10^{-30}$ is achieved.
In particular, in underdetermined systems where some configurations satisfy $F^{(M)}\sim10^{-30}$ while others satisfy $F^{(M)}\gg10^{-30}$, the validity of the latter class of approximations becomes highly questionable.
If this criterion is still insufficient to distinguish the solutions, namely when two or more configurations achieve $F^{(M)}\sim10^{-30}$ even in a square system, it is useful to increase the number of loop equations and examine the stability of the resulting moments.
In Section \ref{sec:Two-Matrix_Model}, we make use of these observations in the approximate computations of the two-matrix model.

\subsubsection{Minkowski-type One-Matrix Model} \label{subsubsec:Minkowski-type_One-Matrix_Model}

\begin{figure}[t] 
\centering
\includegraphics[width=0.7\textwidth]{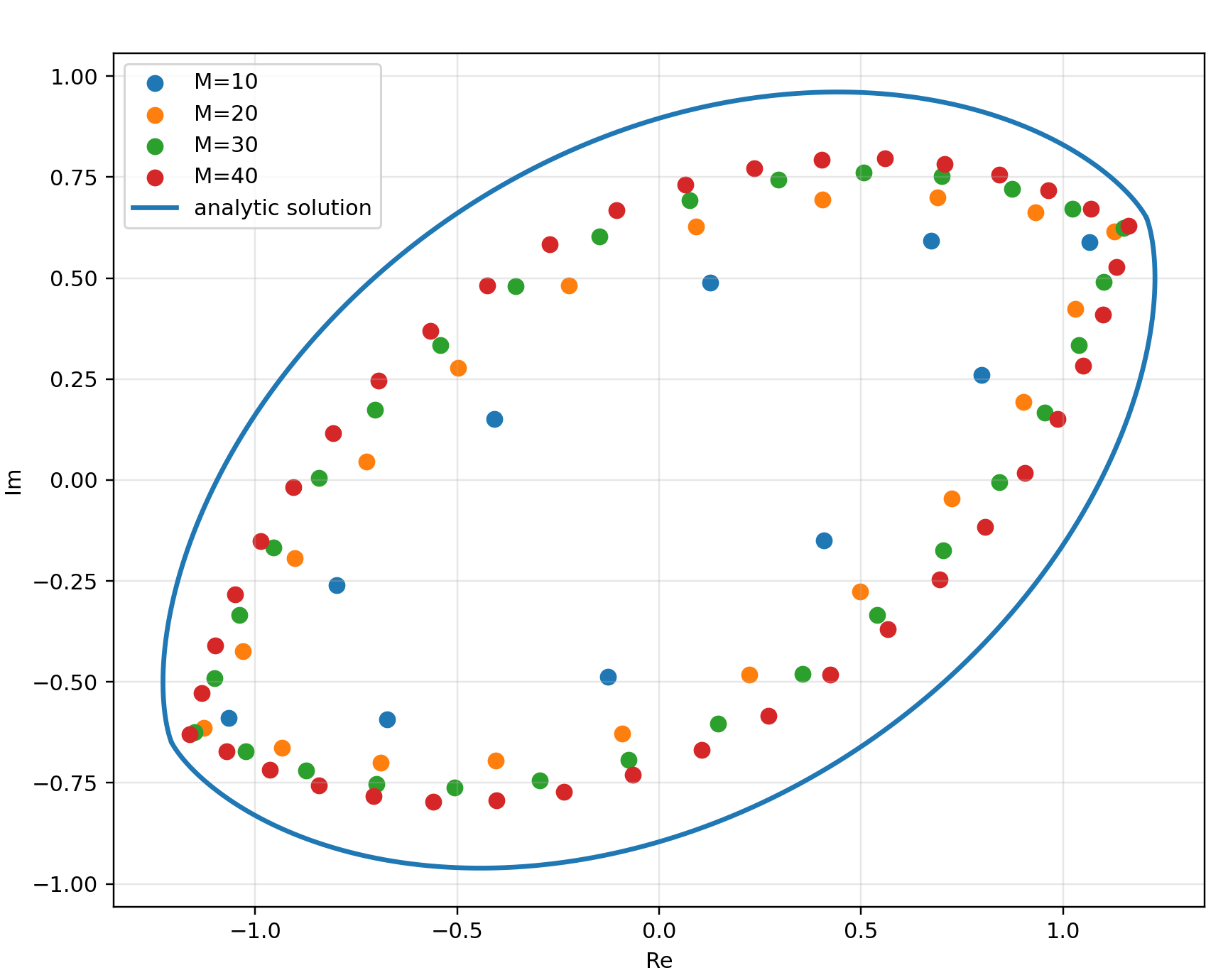}
\caption{
The approximate eigenvalues $\hat{z}_{1}^{(M)},\ldots,\hat{z}_{M}^{(M)}$ obtained by the RMA for $g=-1$ with $M=10,20,30,40$, plotted in the complex plane.
The cases $M=10,20,30,40$ correspond to blue, orange, green, and red points, respectively.
The solid blue curve represents a closed contour satisfying $\frac{\dot{z}(t)}{2\pi i}R_{M}(z(t))=1$ and passing through the endpoints $\pm\alpha$ of $R_{M}(z)$.
}
\label{fig:M-type_1MM_Eigenvalues}
\end{figure}

For the Minkowski one-matrix model, the approximate eigenvalues $\hat{z}_{1}^{(M)},\ldots,\hat{z}_{M}^{(M)}$ computed at fixed $g=-1$ for $M=10,20,30,40$ are shown in Figure \ref{fig:M-type_1MM_Eigenvalues}.
Detailed numerical values are summarized in Table \ref{tab:one_matrix_minkowski_rma}.
The resolvent $R_{M}(z)$ of this model is obtained as
\begin{equation}
R_{M}(z)=\frac{1}{2}\left[-i(z-gz^{3})+\sqrt{-(z-gz^{3})^{2}-4i\{g(w_{2}^{2}+z^{2})-1\}}\right],
\end{equation}
and, assuming a one-cut structure, it is given by
\begin{equation}
\begin{aligned}
R_{M}(z) & =\frac{1}{2}\left[-i(z-gz^{3})-i\left(gz^{2}+\frac{\alpha^{2}g}{2}-1\right)\sqrt{(z-\alpha)(z+\alpha)}\right],\\
w_{2}^{\text{formal}} & =\frac{-i\alpha^{4}+i\alpha^{6}g}{16},\\
\alpha & =\sqrt{\frac{2(1\pm i\sqrt{-1+12ig})}{3g}}.
\end{aligned}
\end{equation}
For the sign $\pm$ inside the square root defining the endpoint $\alpha$, we choose the plus sign for positive $g$ and the minus sign for negative $g$.
Here, $w_{2}^{\text{formal}}$ denotes the ``formal solution'' for $w_{2}$ obtained under the one-cut assumption.
Unlike in the Euclidean case, one should note that the one-cut assumption is not naively justified in the Minkowski one-matrix model.
This is because the poles of the resolvent in the Euclidean model generally lie on the real axis, and the line segment connecting them is interpreted as the support of the eigenvalue distribution, whereas this interpretation does not apply in the Minkowski model.

\begin{table}[t] 
\centering
\setlength{\tabcolsep}{4pt}

\begin{tabular}{cccccccc}
\hline
\(M\)
&
\(\operatorname{Re} w_2^{(M)}\)
&
\(\operatorname{Im} w_2^{(M)}\)
&
\(\left|w_2^{\rm formal}-w_2^{(M)}\right|\)
&
\(\operatorname{Re}\hat z_{\max}^{(M)}\)
&
\(\operatorname{Im}\hat z_{\max}^{(M)}\)
&
\(\left|\alpha-\hat z_{\max}^{(M)}\right|\)
&
\(F_M^{(M)}\)
\\
\hline
10
&
0.276860
&
0.493725
&
\(8.578\times 10^{-4}\)
&
1.066
&
0.589
&
0.153
&
\(1.171\times 10^{-31}\)
\\
20
&
0.277645
&
0.493383
&
\(1.282\times 10^{-5}\)
&
1.127
&
0.615
&
0.087
&
\(1.067\times 10^{-30}\)
\\
30
&
0.277642
&
0.493371
&
\(2.981\times 10^{-7}\)
&
1.150
&
0.624
&
0.062
&
\(1.572\times 10^{-29}\)
\\
40
&
0.277641
&
0.493371
&
\(8.286\times 10^{-9}\)
&
1.162
&
0.630
&
0.049
&
\(2.933\times 10^{-28}\)
\\
\hline
\end{tabular}
\caption{
Detailed numerical results of the RMA for $g=-1$ with $M=10,20,30,40$.
Here, $z_{\text{max}}^{(M)}$ denotes the element of $\hat{z}_{1}^{(M)},\ldots,\hat{z}_{M}^{(M)}$ with the largest real part, while $\alpha$ denotes the endpoint of the cut of the resolvent $R_{M}(z)$ whose real part is positive.
}
\label{tab:one_matrix_minkowski_rma}
\end{table}

Fortunately, in the RMA approach, it is not necessary to impose such an assumption on the cut.
Therefore, by comparing $w_{2}^{(M)}$ computed via RMA with $w_{2}^{\text{formal}}$, one can examine the validity of the one-cut assumption in the Minkowski-type model.
The detailed results are summarized in Table \ref{tab:one_matrix_minkowski_rma}.
Already at $M=10$, the absolute error is at the level of about $0.1\%$, and the accuracy further improves as $M$ increases.
For $M=40$, the difference from $w_{2}^{\text{formal}}$ is almost negligible.
More explicitly, the numerical values are given by
\begin{equation}
\begin{aligned}
w_{2}^{(40)} & =0.2776418499136...+0.493371605...i,\\
w_{2}^{\text{formal}} & =0.2776418499117...+0.493371613...i.
\end{aligned}
\end{equation}
The agreement is remarkably good, with the real and imaginary parts coinciding up to the 11th and 7th decimal places, respectively.
The fact that $w_{2}^{(M)}$ reproduces $w_{2}^{\text{formal}}$ with such high precision strongly suggests the following two points: that the present approximation method based on the master field works effectively; and that the one-cut assumption in the Minkowski-type one-matrix model is reasonable.
In particular, the former provides strong evidence for the validity of the regularized master-field description in the Minkowski-type one-matrix model.

Finally, no qualitative difference from the Euclidean case was observed regarding the dependence on the initial conditions or the behavior in underdetermined systems.

\section{Two-Matrix Model} \label{sec:Two-Matrix_Model}

In Section \ref{sec:One-Matrix_Model}, we have verified that the RMA works in the simplest case of the one-matrix model.
This simultaneously provides strong evidence that there exists the limit $M\to\infty$ in the sense of \eqref{eq:Large_M_Limit}.
However, in the one-matrix model, the master field and the eigenvalue distribution carry almost equivalent information, and in this sense the result may not be particularly surprising.

The situation changes drastically in multi-matrix models.
Because of the noncommutativity among matrices, the master field, if it exists, contains far richer information than a mere eigenvalue distribution.
Whether such an object actually exists is by no means obvious, even after completing the analysis of the one-matrix model.
In this Section \ref{sec:Two-Matrix_Model}, we therefore apply the RMA to the two-matrix model, in order to investigate the existence of the master field and to approximate various physical quantities.

\subsection{Overall Setup} \label{subsec:Overall_Setup_two-matrix_model}

The action of the theory is given, following \cite{Kazakov:2021lel}, by
\begin{equation} \label{eq:Action_of_Two-matrix_Models}
\begin{aligned}
S_{E} & =N\text{Tr}\left\{-\frac{h}{4}\delta^{\mu\rho}\delta^{\nu\lambda}[A_{\mu},A_{\nu}][A_{\rho},A_{\lambda}]+\frac{1}{2}\delta^{\mu\nu}A_{\mu}A_{\nu}+\frac{g}{4}\delta^{\mu\nu}A_{\mu}^{2}A_{\nu}^{2}\right\},\\
S_{M} & =N\text{Tr}\left\{-\frac{h}{4}\eta^{\mu\rho}\eta^{\nu\lambda}[A_{\mu},A_{\nu}][A_{\rho},A_{\lambda}]-\frac{1}{2}\eta^{\mu\nu}A_{\mu}A_{\nu}-\frac{g}{4}\eta^{\mu\nu}A_{\mu}^{2}A_{\nu}^{2}\right\}.
\end{aligned}
\end{equation}
Here, $\eta^{\mu\nu}$ is the $1+1$-dimensional Minkowski metric, which we take to be $\eta=\text{diag}(+1,-1)$.
The Greek indices $\mu,\nu,\rho,\lambda$ are understood to take values $1,2$ in both the Euclidean and Minkowski cases.
Note that for $g\ne0$, the $SO(2)$ and $SO(1,1)$ symmetries, corresponding respectively to matrix rotations and Lorentz transformations, are explicitly broken in both models.\footnote{If one removes the terms consisting of $A_{\mu}A_{\nu}$ and $A_{\mu}^{2}A_{\nu}^{2}$ from this action, and furthermore sets the target-space dimension to $D=10$, one obtains the bosonic part of the IKKT matrix model.
For the IKKT matrix model, the names ``Euclidean'' and ``Lorentzian'' are standard, and the latter comes from the fact that the action is invariant under $1+9$-dimensional Lorentz transformations in target space.
By contrast, the action $S_{M}$ of the present 2-matrix model does not in general possess Lorentz symmetry, and to avoid confusion we therefore use the name ``Minkowski''.}

Let us next derive the loop equations.
The loop equations are a kind of Schwinger-Dyson equation, and in the 1-matrix model they were equations for $w_{n}=\langle\text{tr}\phi^{n}\rangle$.
Since this $w_{n}$ is nothing but the $n$th moment of the eigenvalue distribution $\rho_{E}(\lambda)$ when the latter is regarded as a probability distribution, it is naturally called a ``moment''.
In the 2-matrix model, or more generally in multi-matrix models, the straightforward generalization is that expectation values of the form $\langle\text{tr}(A_{\mu_{1}}...A_{\mu_{n}})\rangle$ appear in the loop equations.
For convenience, let us introduce the notation
\begin{equation}
\begin{aligned}
w\{\mu_{1}...\mu_{n}\} & \equiv\langle\text{tr}(A_{\mu_{1}}...A_{\mu_{n}})\rangle=\langle\frac{1}{N}\text{Tr}(A_{\mu_{1}}...A_{\mu_{n}})\rangle.
\end{aligned}
\end{equation}
In this paper, we shall also refer to $w\{\mu_{1}...\mu_{n}\}$ as a moment.
\footnote{In some references, $\text{tr}(A_{\mu_{1}}...A_{\mu_{n}})$ is also called a Wilson loop.}
The loop equations of the 2-matrix model, or of a general multi-matrix model, are precisely the Schwinger-Dyson equations for such moments $w\{\mu_{1}...\mu_{n}\}$.
Their derivation is basically the same as in the 1-matrix model, and can be obtained by taking the variation of $A_{\mu}$ to be
\begin{equation} \label{eq:Variation_of_A}
A'_{\rho}=\begin{cases}
A_{\mu}+\epsilon t^{a} & (\rho=\mu)\\
A_{\rho} & (\rho\ne\mu)
\end{cases}
\end{equation}
where $t^{a}$ is, as in \eqref{eq:property_of_t^a}, a generator of $U(N)$.
As a result, the loop equations take the following form:
\begin{equation}
\begin{aligned}
\text{Euclidean type}:0 & =\sum_{k=1}^{n}\delta_{\mu_{k},\mu}w\{\mu_{1}...\mu_{k-1}\}w\{\mu_{k+1}...\mu_{n}\}\\
 & \qquad-h\sum_{\nu=1}^{2}w\{\mu_{1}...\mu_{n}[\nu,[\nu,\mu]]\}\\
 & \qquad-w\{\mu_{1}...\mu_{n}\mu\}\\
 & \qquad-gw\{\mu_{1}...\mu_{n}\mu\mu\mu\},\\
\text{Minkowski type}:0 & =\sum_{k=1}^{n}\delta_{\mu_{k},\mu}w\{\mu_{1}...\mu_{k-1}\}w\{\mu_{k+1}...\mu_{n}\}\\
 & \qquad-ih\sum_{\nu=1}^{2}w\{\mu_{1}...\mu_{n}[\nu,[\nu,\mu]]\}\\
 & \qquad-i\eta^{\mu\nu}w\{\mu_{1}...\mu_{n}\nu\}\\
 & \qquad-i\eta^{\mu\nu}gw\{\mu_{1}...\mu_{n}\nu\nu\nu\}.
\end{aligned}
\end{equation}
For the derivation, one may start from $\int dA\text{tr}(t^{a}A_{\mu_{1}}...A_{\mu_{n}})e^{tS}$ and take the variation.
Here, $t$ takes $i$ or $-1$.
The commutator in the middle means
\begin{equation}
\begin{aligned}
w\{\mu_{1}...\mu_{n}[\nu,[\nu,\mu]]\} & =w\{\mu_{1}...\mu_{n}\nu[\nu,\mu]\}-w\{\mu_{1}...\mu_{n}[\nu,\mu]\nu\}\\
 & =w\{\mu_{1}...\mu_{n}\nu\nu\mu\}+w\{\mu_{1}...\mu_{n}\mu\nu\nu\}-2w\{\mu_{1}...\mu_{n}\nu\mu\nu\}.
\end{aligned}
\end{equation}
Here, again for notational simplicity, we denote the right-hand side of the loop equations by ${\cal L}_{n;\alpha_{n}}$.
In this notation, the loop equation is written as ${\cal L}_{n;\alpha}=0$, which we refer to in this work as the loop equation of order $n$.
Here $n$ denotes the degree of the moment $w\{\mu_{1}...\mu_{n}\}$ under consideration, while $\alpha_{n}$ is an abstract label used to distinguish loop equations of the same degree.
More specifically, the label $\alpha_{n}$ distinguishes not only the type of moment $w\{\mu_{1}...\mu_{n}\}$ under consideration, but also the direction in which the variation \eqref{eq:Variation_of_A} is applied.
In deriving the loop equations, one takes $\text{tr}(t^{a}A_{\mu_{1}}...A_{\mu_{n}})$ as the integrand and applies the variation \eqref{eq:Variation_of_A} to it.
Therefore, if one ignores cyclic invariance and possible linear dependencies among the loop equations, there are $2^{n+1}$ loop equations at order $n$.

Having obtained the loop equations, we proceed to regularize the master field using them.
The implementation is essentially the same as in the one-matrix model.
As a starting point, we assume that a master field exists in the large-$N$ limit, namely that there exists an infinite-dimensional matrix $\hat{A}_{\mu}$ such that, for any moment $w\{\mu_{1}\cdots\mu_{n}\}$,
\begin{equation}
w\{\mu_{1}\cdots\mu_{n}\}=\text{tr}(\hat{A}_{\mu_{1}}\cdots\hat{A}_{\mu_{n}})
\end{equation}
holds.
In what follows, we assume that such matrices $\hat{A}_{\mu}$ exist both in the Euclidean and Minkowski cases.

Unlike in the one-matrix model, however, we choose $\hat{A}_{\mu}$ to be Hermitian matrices in the Euclidean case.
There are two reasons for this choice.
First, in contrast to the one-matrix model, the complexification of the master field is not straightforwardly justified in the two-matrix model.
In the one-matrix model, the eigenvalues $\hat{z}_{1},\ldots,\hat{z}_{N}$ of the complexified master field admit a clear interpretation: they are distributed along a closed curve $C_{\text{const.}}$ satisfying $\frac{\dot{z}(t)}{2\pi i}R(z(t))=1$.
In the two-matrix model, however, there are two matrices, and correspondingly infinitely many resolvents, so such a simple interpretation is no longer available.
Second, even if the master field is kept Hermitian, the optimization works as well as in the Minkowski case.
More precisely, the objective function $F^{(M)}$ attains values as small as those obtained in the Minkowski case.
Since the optimization performs well, there is no need to introduce additional assumptions that would unnecessarily complicate the discussion.

The next step is to regularize them at finite dimension.
As notation, let us introduce the regularized moment $w^{(M)}\{\mu_{1}...\mu_{n}\}$ by
\begin{equation}
w^{(M)}\{\mu_{1}...\mu_{n}\}\equiv\text{tr}_{M}(\hat{A}_{\mu_{1}}^{(M)}...\hat{A}_{\mu_{n}}^{(M)})=\frac{1}{M}\text{Tr}(\hat{A}_{\mu_{1}}^{(M)}...\hat{A}_{\mu_{n}}^{(M)}).
\end{equation}
This $w^{(M)}\{\mu_{1}...\mu_{n}\}$ should satisfy the loop equations to high accuracy, and we require that
\begin{equation}
\begin{aligned}
{\cal L}_{n;\alpha_{n}}^{(M)} & \approxeq0
\end{aligned}
\end{equation}
hold for sufficiently many $\alpha_n$.
Of course, ${\cal L}_{n;\alpha_{n}}^{(M)}$ is obtained from ${\cal L}_{n;\alpha_{n}}$ by replacing $w\{\mu_{1}...\mu_{n}\}$ with $w^{(M)}\{\mu_{1}...\mu_{n}\}$.
More specifically, we first define the objective function $F^{(M)}$ by
\begin{equation} \label{eq:F_of_E-type_2MM}
F^{(M)}\equiv\sum_{n=0}^{\Lambda}\sum_{\alpha_{n}}r_{n}|{\cal L}_{n;\alpha_{n}}^{(M)}|^{2}.
\end{equation}
The configuration that gives the minimum of this $F^{(M)}$ is precisely the definition of $\hat{A}_{\mu}^{(M)}$.
As for the details of $F^{(M)}$, in the present work we adopt the prescription of including all loop equations in order from the lower-degree ones upward.
The highest degree $n=\Lambda$ is chosen so that $F^{(M)}=0$ becomes either a square system or an overdetermined system.
Since one of the two matrices $\hat{A}_{\mu}^{(M)}$ can always be diagonalized by an appropriate gauge choice, the number of unknown variables can already be reduced at that stage to $M(M+1)$ in the Euclidean case and to $2M(M+1)$ in the Minkowski case.
In fact, some gauge freedom still remains even after this gauge fixing, and this point will be discussed in detail in Section \ref{subsubsec:Minkowski-type_two-Matrix_Model}.

As for the weights $r_{n}$, in the present numerical calculation we choose
\begin{equation} \label{eq:rn_for_2-MM}
r_{n}=(\text{total number of loop equations of degree }n)^{-1}.
\end{equation}
The reason is as follows: as mentioned above, the total number of loop equations of degree $n$ is roughly of order $2^{n}$, and therefore, if one were to take $r_{n}=1$, the contributions of the low-degree loop equations would effectively disappear.
However, the essential point of the loop equations is that high-degree moments can be expressed in terms of low-degree moments.
What this means is that if the deviations in the low-degree loop equations ${\cal L}_{n;\alpha_{n}}^{(M)}\approxeq0$ are large, then even if the high-degree loop equations are satisfied with high precision, the result cannot be trusted at all.
In that sense, the importance of each low-degree loop equation is far greater than that of a high-degree one.
From this consideration, in order to determine the regularized master field $\hat{A}_{\mu}^{(M)}$ accurately, that is, to accelerate the convergence of the limit in the sense of \eqref{eq:Large_M_Limit}, it is better to choose weights $r_{n}$ such that lower-degree loop equations are given greater emphasis.
The choice adopted in \eqref{eq:rn_for_2-MM} provides one concrete realization of this idea.

We have now introduced the basic form of the objective function for the two-matrix model.
However, with the present formulation, the number of loop equations is restricted to $\sum_{n=0}^{\Lambda}2^{n+1}=2^{\Lambda+2}-2$, so that the condition $F^{(M)}=0$ is in most cases either a highly overdetermined or a highly underdetermined system.
To anticipate the numerical results, the definition \eqref{eq:F_of_E-type_2MM} works sufficiently well for the Euclidean two-matrix model, whereas the Minkowski-type model exhibits unstable behavior.
More specifically, the dependence on the initial conditions becomes much stronger, making it impossible to immediately trust the obtained results.
For this reason, in the Minkowski-type two-matrix model we introduce a new objective function $F^{(M)}$, based on \eqref{eq:F_of_E-type_2MM} but supplemented with further improvements.

\subsection{Numerical Results} \label{subsec:Numerical_Results_two-matrix_model}

In the present setup, the numerical parameters that must be specified by hand are $g$ and $h$ in the action $S$, and $M$ and $\Lambda$ in the objective function $F^{(M)}$.
Here we mainly study the case $g=h=1$.
For the Minkowski-type model, we also compute the cases with $|g|,|h|\ll1$ in order to compare the results with perturbation theory.
As in the one-matrix model, we use SciPy for the numerical computations.
The settings of ftol, xtol, gtol, and max\_nfev are also unchanged from the one-matrix case.
Unless stated otherwise, the initial configuration is generated by assigning each matrix element of the regularized master field $\hat{A}_{\mu}^{(M)}$ a random value whose real and imaginary parts are independently drawn from Gaussian distributions with mean zero and standard deviation $0.01$.
For the Euclidean two-matrix model, however, $\hat{A}_{\mu}^{(M)}$ is required to be Hermitian, and the initial configuration is generated so as to satisfy the Hermiticity condition.
The results for the Euclidean and Minkowski cases are presented in Section \ref{subsubsec:Euclidean-type_two-Matrix_Model} and \ref{subsubsec:Minkowski-type_two-Matrix_Model}, respectively.

\subsubsection{Euclidean-type Two-Matrix Model} \label{subsubsec:Euclidean-type_two-Matrix_Model}

\begin{figure}[t] 
\centering
\includegraphics[width=0.8\textwidth]{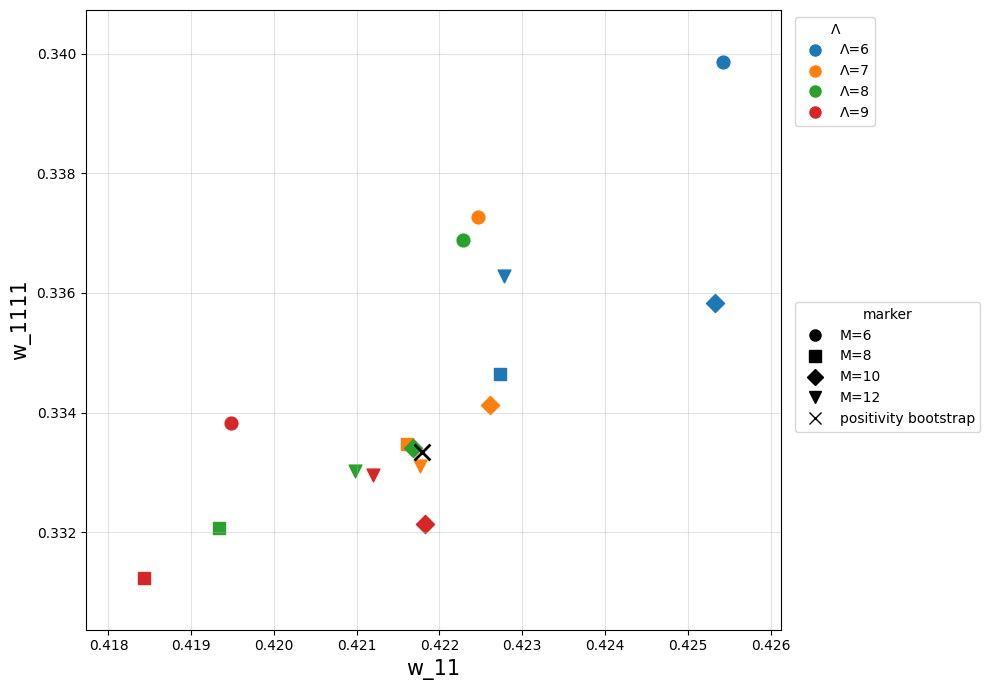}
\caption{
Numerical results for the Euclidean action $S_{E}$ with $g=h=1$, obtained for a total of 16 parameter choices with $M=6,8,10,12$ and $\Lambda=6,7,8,9$.
The horizontal axis shows $w^{(M)}\{11\}$ and the vertical axis shows $w^{(M)}\{1111\}$.
The values of $M$ and $\Lambda$ are distinguished by marker shapes and colors, respectively.
The cross mark corresponds to the value obtained in \cite{Kazakov:2021lel} by the bootstrap method based on positivity constraints.
More precisely, the result in \cite{Kazakov:2021lel} was obtained not as a single value but as an allowed region; however, since it was determined with an accuracy better than the third decimal place, it is represented here simply as a point.
}
\label{fig:E-type_w11_w1111_plot_M=even_Lambda=6-9}
\end{figure}

For the matrix size $M$, we consider the four cases $M=6,8,10,12$, and choose the corresponding values of $\Lambda$ as $\Lambda=6,7,8,9$.
For the smallest value $\Lambda=6$, one obtains $2^{8}-2=254$ loop equations if cyclic invariance and possible dependencies among the loop equations are ignored.
However, once the linear dependencies among the loop equations are taken into account, the effective number of constraints is actually smaller than the number of unknown variables in the case $M=12$.
In other words, in this case the system $F^{(M)}=0$ is underdetermined.
We will briefly comment on this point at the end of this section.

The numerical results are shown in Figure \ref{fig:E-type_w11_w1111_plot_M=even_Lambda=6-9}.
The cross mark in the figure corresponds to the value obtained in \cite{Kazakov:2021lel} by the bootstrap method based on the positivity constraint, \footnote{More precisely, this was obtained not as a single value but as an allowed region.
However, since it was determined with an accuracy better than the third decimal place, we represent it here simply as a point. In addition, in \cite{Kazakov:2021lel}, cyclic invariance of the trace was assumed, and therefore solutions such as $SO(1,2)$ were excluded from the outset, just as in the RMA method.} and values closer to this point provide better approximations.
Corresponding to the fact that $F^{(M)}=0$ is a strongly overdetermined system for almost all combinations of $M$ and $\Lambda$, the value of $F^{(M)}$ decreases only down to about $10^{-4}\sim10^{-6}$.
As an exception, however, the case $\Lambda=6$ and $M=12$ uniquely achieves $F^{(M)}\sim10^{-30}$, as will be discussed later.

The overall trend is that the approximation becomes more accurate for larger $M$, whereas increasing $\Lambda$ does not necessarily seem to improve the accuracy.
This is presumably because increasing the matrix size directly corresponds to approaching the large-$N$ limit, while increasing $\Lambda$ merely makes the system more strongly overdetermined and therefore does not affect the approximation accuracy as much.
More precisely, although there is a tendency for larger $M$ to give better accuracy, one should also note that the results do not approach the allowed region monotonically as $M$ is increased.

Nevertheless, even in the least accurate case, $M=6$ and $\Lambda=6$, the errors in $w^{(M)}\{11\}$ and $w^{(M)}\{1111\}$ are only about one or two percent in absolute value.
Considering the low computational cost, this accuracy is very good, and indicates that the RMA is an efficient method for approximating moment values.
As for the convergence as $M\to\infty$, the results do not suggest that such a limit fails to exist in the two-matrix model.
Rather, it seems more likely that the approximation by the RMA is already so efficient that the accuracy saturates beyond a certain value of $M$.
In any case, it is clear that the RMA gives a very good approximation for the Euclidean two-matrix model, which strongly suggests the validity of the regularized master-field description.

Let us comment on the differences from the Euclidean one-matrix model.
In the one-matrix case, approximate computations based on a Hermitian regularized master field $\hat{\phi}^{(M)}$ did not work particularly well, and several artificial assumptions were required in order to improve the numerical accuracy (see Appendix \ref{app:appendixB}).
By contrast, in the Euclidean two-matrix model, efficient approximate computations can be achieved while keeping the regularized master fields $\hat{A}_{\mu}^{(M)}$ Hermitian.
Moreover, the dependence on the initial conditions is much weaker in the Euclidean two-matrix model.
For example, as in the one-matrix case, suppose that the initial conditions for $\hat{A}_{\mu}^{(M)}$ are randomly sampled from Gaussian distributions with mean zero and variance one.
Then, for $M=\Lambda=6$, the resulting values of $w^{(6)}\{11\}$ and the objective function $F^{(6)}$ are
\begin{equation}
\begin{aligned}
\text{sample 1}:w^{(6)}\{11\} & =0.4180399...,\quad F^{(6)}=3.793753...\times10^{-4},\\
\text{sample 2}:w^{(6)}\{11\} & =0.4171241...,\quad F^{(6)}=1.464419...\times10^{-4},\\
\text{sample 3}:w^{(6)}\{11\} & =0.4171276...,\quad F^{(6)}=1.464419...\times10^{-4}.
\end{aligned}
\end{equation}
The case $M=\Lambda=6$ corresponds to a strongly overdetermined system, and accordingly the optimization only converges down to about $F^{(6)}\sim10^{-4}$.
Nevertheless, all values of $w^{(6)}\{11\}$ converge to almost the same result.
Comparing this with the results for the Minkowski-type two-matrix model in Section \ref{subsubsec:Minkowski-type_two-Matrix_Model}, we conjecture that this insensitivity to the initial conditions and the relatively high numerical accuracy may originate from the use of Hermitian matrices for $\hat{A}_{\mu}^{(M)}$.
A more detailed investigation will be necessary to clarify this point.

Finally, let us briefly discuss the case $\Lambda=6$ and $M=12$.
As will be explained in the next Section \ref{subsubsec:Minkowski-type_two-Matrix_Model}, in the Euclidean two-matrix model, the completely gauge-fixed matrices $\hat{A}_{1}^{(M)}$ and $\hat{A}_{2}^{(M)}$ contain a total of $M^{2}+1$ real variables (or complex variables in the Minkowski case).
Therefore, for $M=12$, the number of unknown variables is 145.
On the other hand, regarding the loop equations, $\Lambda=6$ generates 254 equations once duplicated equations are excluded.
However, when the rank of the Jacobian matrix is numerically evaluated using the optimized configurations $\hat{A}_{1}^{(M)}$ and $\hat{A}_{2}^{(M)}$, the result is 133.
In other words, at least 12 additional independent equations would be necessary in order to determine the solution of $F^{(12)}=0$ uniquely.
Nevertheless, the numerical results still reproduce the positivity-based allowed region with high accuracy.
As in the one-matrix model, this may therefore be interpreted as an example in which approximate computations can still work even for an underdetermined system.

\subsubsection{Minkowski-type Two-Matrix Model} \label{subsubsec:Minkowski-type_two-Matrix_Model}

Numerical computations in the Minkowski-type two-matrix model are more difficult than in the Euclidean case because the numerical results tend to depend strongly on the initial conditions.
Although the precise reason for this behavior is unclear, it is plausible that the Euclidean two-matrix model is in fact exceptional, since a sensitivity to the initial conditions was already observed in the one-matrix model, both in the Euclidean and Minkowski cases.
Moreover, for the Minkowski-type two-matrix model, neither exact analytic solutions nor positivity-based bounds are currently known, so there are no obvious reference values against which the RMA results can be compared directly.

We therefore adopt a strategy in which several different checks are combined in order to assess the reliability of the results.
In the following, we proceed according to the steps below:
\begin{enumerate}
\item compare the RMA results with perturbative calculations in the regime $g=0$, $|h|\ll1$;

\item gradually increase the values of $g$ and $h$ up to $g=h=1.0$;

\item examine the stability of the approximate values under changes in the number of loop equations;

\item examine the convergence of the approximate values under changes in the matrix size $M$;

\item compute the eigenvalues of $\hat{A}_{\mu}$ for different values of $M$.
\end{enumerate}

First, as a reference point for comparison with the RMA results, we compute the perturbative expansion of $w\{11\}$ up to first order in $h$ for $g=0$ and $|h|\ll1$.
The result is
\begin{equation} \label{eq:w11_perturbative}
w\{11\}=-i-2h+\cdots.
\end{equation}
Note that in perturbation theory, odd powers of $h$ contribute to $\text{Re}w\{11\}$, whereas even powers contribute to $\text{Im}w\{11\}$ (see Appendix \ref{app:appendixA} for details of the calculation).
If the RMA method reproduces this behavior for $g=0$ and $|h|\ll1$, it is reasonable to expect that the optimization is functioning correctly.

The optimized $\hat{A}_{\mu}^{(M)}$ obtained in this way are then reused as initial conditions for the RMA computations in the region $|g|,|h|\ll1$.
The rationale is that extrapolating from successful solutions at $g=0$ into the small-$|g|$ region should provide more reliable numerical computations than choosing completely random initial conditions.
Once this procedure succeeds, one can gradually increase the values of $g$ and $h$.

Of course, perturbative results cannot be used as guidance once $g$ and $h$ become large.
Instead, we make use of the lessons learned from the numerical analysis of the one-matrix model: one should examine whether the objective function $F^{(M)}$ decreases down to the level of machine precision, namely $10^{-30}$.
In the one-matrix model, even when $F^{(M)}=0$ formed an underdetermined system, the success or failure of the optimization could still be judged to some extent by whether or not $F^{(M)}\sim10^{-30}$ was achieved.
To make such a criterion meaningful, however, one must control the number of constraints appearing in $F^{(M)}=0$ more carefully.
For this reason, we modify the definition of $F^{(M)}$ to
\begin{equation}
F^{(M)}\equiv\sum_{\alpha_{n}}^{X}r_{n}|{\cal L}_{n;\alpha_{n}}^{(M)}|^{2}.
\end{equation}
Here $\sum_{\alpha_{n}}^{X}$ means that only the first $X$ loop equations, ordered from lower to higher degree, are included in the sum.
In this case, the loop equations of the highest order $\Lambda$ are incorporated only partially.
For convenience, we shall treat these highest-order $\Lambda$ loop equations as having the same order as the loop equations with degree $n=\Lambda-1$, and assign them the corresponding weight $r_{\Lambda-1}$.
Using this modified objective function, it is expected that the success or failure of the optimization can at least partially be judged by examining whether $F^{(M)}\sim10^{-30}$ is achieved.
Note, however, that the $X$ loop equations are not necessarily all independent.
The number of independent equations in the system ${\cal L}_{n;\alpha_{n}}^{(M)}=0$ can be determined from the number of nonzero singular values of the corresponding Jacobian matrix.
More specifically, one may examine the singular values of the Jacobian evaluated around the RMA solution $\hat{A}_{\mu}^{(M)}$.

At the same time, gauge fixing of $\hat{A}_{\mu}^{(M)}$ is also important in counting the degrees of freedom.
At present, $\hat{A}_{1}^{(M)}$ has already been fixed to be diagonal.
However, there still remains residual gauge freedom.
Indeed, under the similarity transformation $\hat{A}_{\mu}^{(M)}\mapsto D\hat{A}_{\mu}^{(M)}D^{-1}$ with an arbitrary complex diagonal matrix $D=\text{diag}(d_{1},...,d_{M})$, the matrix $\hat{A}_{1}^{(M)}$ and the moments $w\{\mu_{1}...\mu_{n}\}$ remain invariant, whereas $\hat{A}_{2}^{(M)}$ changes.
This precisely corresponds to the residual gauge redundancy.
At first sight, this appears to provide $M$ complex degrees of freedom.
However, when $d_{1}=\cdots=d_{M}$, one has $\hat{A}_{2}^{(M)}=D\hat{A}_{2}^{(M)}D^{-1}$, so the correct number of independent complex gauge parameters is actually $M-1$.
As a result, the number of unknown variables appearing in $F^{(M)}$ becomes $M^{2}+1$.
One possible gauge fixing that realizes this reduction is, generically, the form
\begin{equation}
\hat{A}_{2}^{(M)}=\begin{pmatrix}* & 1\\
 & * & \ddots & \\
 &  & \ddots & 1\\
 &  &  & *
\end{pmatrix}.
\end{equation}
However, this gauge fixing can cause difficulties in numerical computations.
Under the diagonal similarity transformation $\hat{A}_{2}^{(M)}\mapsto D\hat{A}_{2}^{(M)}D^{-1}$, the matrix element $(\hat{A}_{2}^{(M)})_{i,i+1}$ transforms as
\begin{equation}
(\hat{A}_{2}^{(M)})_{i,i+1}\mapsto\frac{d_{i}}{d_{i+1}}(\hat{A}_{2}^{(M)})_{i,i+1}.
\end{equation}
Therefore, to impose the gauge condition, one chooses $d_{i}$ and $d_{i+1}$ such that $d_{i}d_{i+1}^{-1}(\hat{A}_{2}^{(M)})_{i,i+1}=1$ for each $i=1,...,M$.
However, if $(\hat{A}_{2}^{(M)})_{i,i+1}=0$, this transformation cannot be achieved for finite values of $d_{i}$.
Of course, in actual optimization it is not expected that $(\hat{A}_{2}^{(M)})_{i,i+1}=0$ holds exactly.
Nevertheless, even when $(\hat{A}_{2}^{(M)})_{i,i+1}$ is merely sufficiently close to zero, numerical instability can still arise.
For this reason, in the present work we keep $\hat{A}_{2}^{(M)}$ as a general complex matrix and leave the residual gauge freedom unfixed.

\begin{figure}[t] 
\centering
\includegraphics[width=1.0\textwidth]{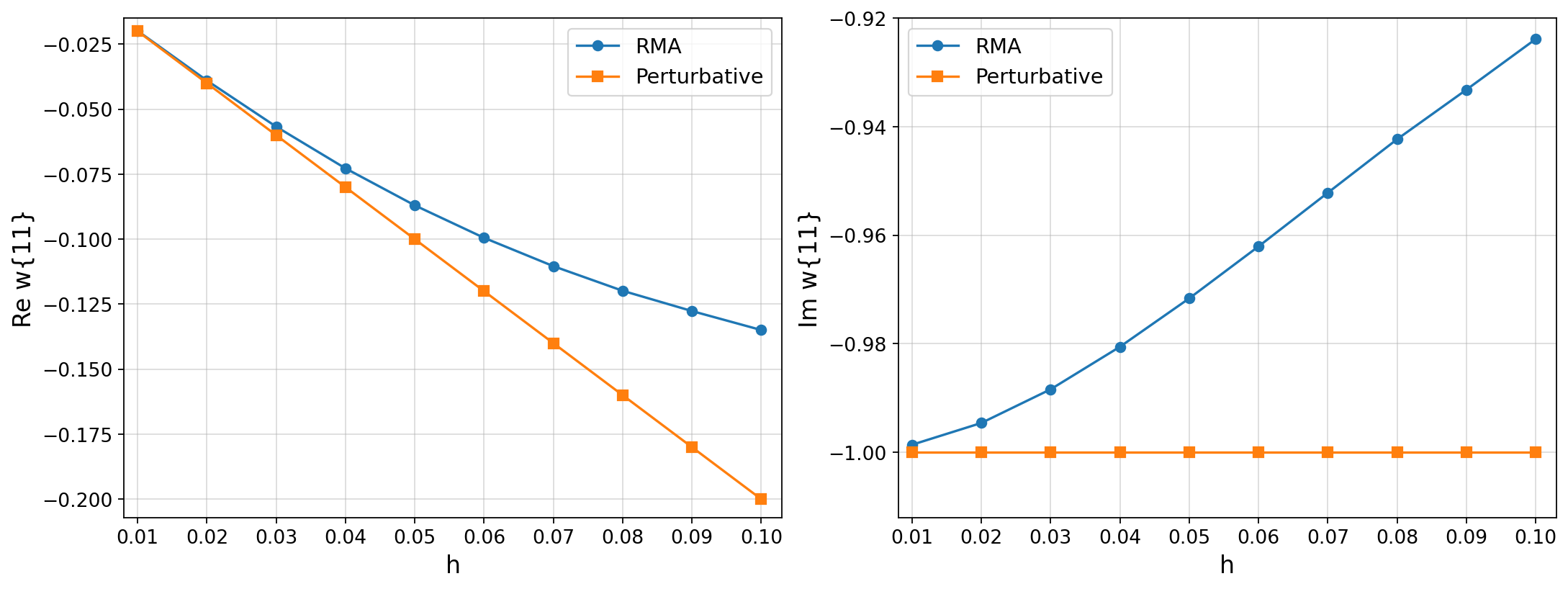}
\caption{
Comparison between the perturbative expansion of $w\{11\}$ up to first order in $h$ and the RMA results for $g=0$ and $|h|\ll1$, shown separately for the real and imaginary parts.
We consider ten values $h=0.01,0.02,...,0.1$.
The left panel shows the real part, and the right panel shows the imaginary part, where the blue solid lines represent the RMA results and the orange solid lines represent the perturbative results.
}
\label{fig:2-MM_perturbative_RMA_compare}
\end{figure}

We first present in Figure \ref{fig:2-MM_perturbative_RMA_compare} a comparison between perturbation theory and the RMA results.
Here we fix $M=8$ and $X=90$, and examine the ten values $h=0.01,0.02,...,0.1$.
Among the $X=90$ loop equations, the number of independent equations, determined from the singular values of the Jacobian, is 63, which is smaller than $M^{2}+1=65$.
Therefore, the present system $F^{(8)}=0$ is underdetermined.
This choice was made because for $X\ge91$, the optimization failed to achieve $F^{(8)}\sim10^{-30}$ regardless of the initial conditions.
In other words, $X=90$ was chosen as the largest value for which $F^{(8)}\sim10^{-30}$ could still be achieved.

As seen in the one-matrix example, even when $F^{(M)}=0$ forms an underdetermined system, achieving $F^{(M)}\sim10^{-30}$ may still be regarded as one indication that the optimization is functioning properly, although the possibility of false convergence nevertheless remains.
Indeed, Figure \ref{fig:2-MM_perturbative_RMA_compare} shows that the RMA method reproduces the perturbative results for $g=0$ and $|h|\ll1$ quite well.
Moreover, as expected, the agreement becomes increasingly accurate as the value of $h$ decreases.
We therefore conclude that the optimization is successful for all cases considered here, namely for $g=0$ and $h=0.01,0.02,...,0.1$.

Let us emphasize one point here.
As explained in Appendix \ref{app:appendixA}, this perturbative expansion can be derived from the loop equations.
On the other hand, $\hat{A}_{\mu}^{(M)}$ is chosen so that the loop equations are satisfied as well as possible.
One might therefore think that it is almost automatic that such $\hat{A}_{\mu}^{(M)}$ reproduces the perturbative result.
However, this is a misunderstanding.
As noted in Section \ref{subsubsec:Some_Remarks_on_the_RMA}, the loop equations, regarded as a system of algebraic equations, contain many unphysical solutions, and such unphysical solutions are incompatible with the perturbative result.
The reason why the perturbative expansion can be reproduced from the loop equations is precisely that we impose the physical assumption that the moments $w\{\mu_{1}...\mu_{n}\}$ admit power-series expansions in the coupling constants.
In contrast, in RMA we assume only the existence of a master field and make no assumptions whatsoever about the coupling constants or perturbative expansion.
The fact that RMA nevertheless reproduces the perturbative result strongly suggests that the existence of the master field plays a role in eliminating unphysical solutions.

\begin{figure}[t] 
\centering
\includegraphics[width=0.6\textwidth]{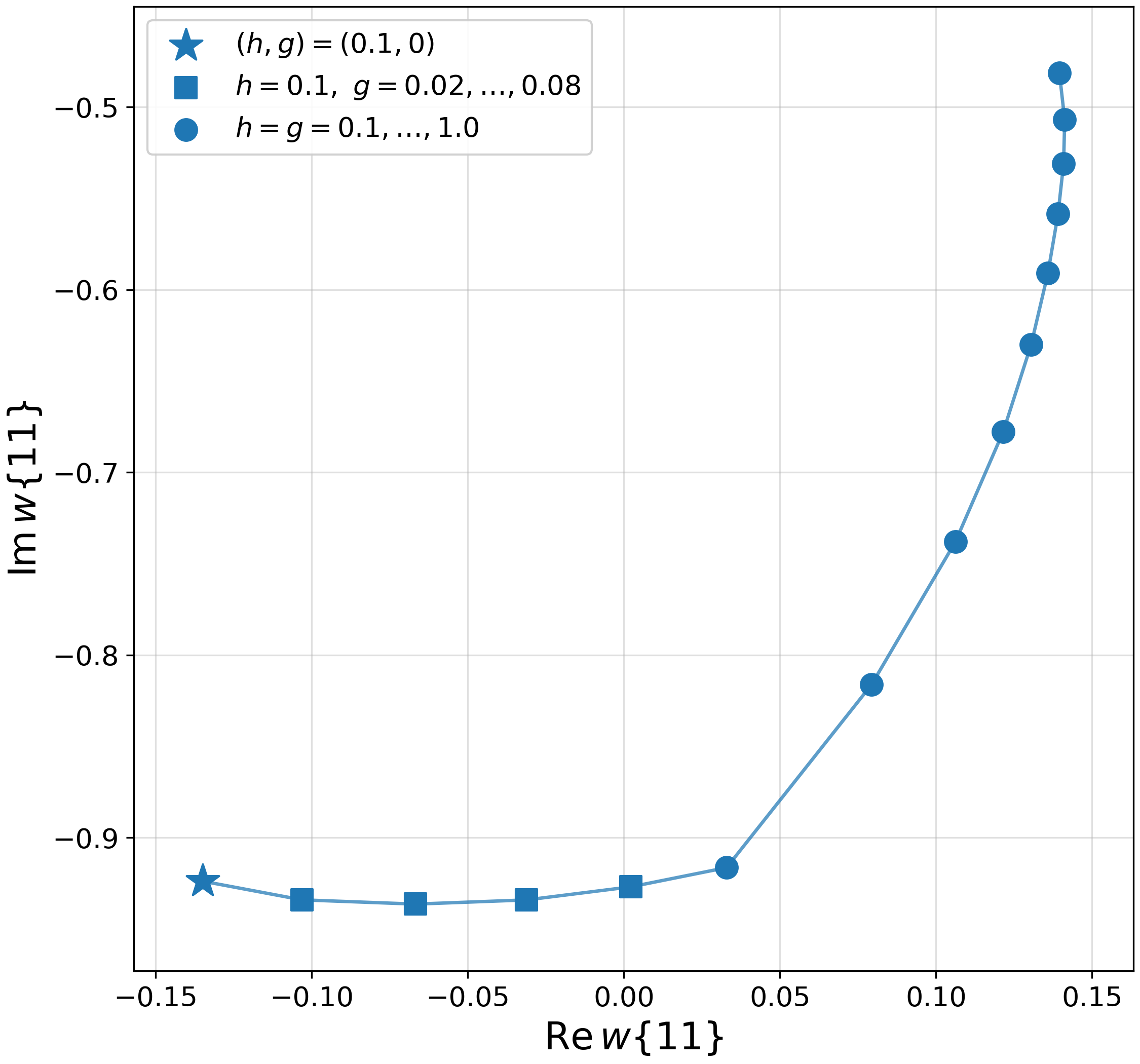}
\caption{
Values of $w^{(M)}\{11\}$ computed for different values of $h$ and $g$ with $M=8$.
The star-shaped marker corresponds to the case $g=0$ and $h=0.1$.
The square and circular markers represent the values of $w^{(M)}\{11\}$ obtained for $h=0.1$ with $g=0.02,...,0.08$, and for $h=g=0.1,...,1.0$, respectively.
We use $X=90$ for the case $g=0$ and $h=0.1$, while $X=86$ is used for all other cases.
The circular marker located at the upper right corresponds to the case $h=g=1.0$.
}
\label{fig:w11_various_h_g}
\end{figure}

Next, using the optimized configuration $\hat{A}_{\mu}^{(8)}$ obtained for $g=0$ and $h=0.1$ as the initial condition, we computed the case $g=0.02$ and $h=0.1$.
The resulting $\hat{A}_{\mu}^{(8)}$ was then reused as the initial condition for the computation with $g=0.04$ and $h=0.1$.
This procedure was continued until reaching $g=h=0.1$, after which the parameters were increased as $g=h=0.2,0.3,...,1.0$.
The resulting values are shown in Figure \ref{fig:w11_various_h_g}.
Here, except for the case $(g=0,\ h=0.1)$, we set $M=8$ and $X=86$ throughout.
In this case, the number of independent loop equations is $65$ for all $(g,h)$, so that $F^{(M)}=0$ forms a square system here.
The increase in the number of independent loop equations despite the decrease in $X$ is presumably due to the lifting of degeneracies among the equations once $g$ is turned on to a nonzero value.
Figure \ref{fig:w11_various_h_g} shows that the value of $w^{(M)}\{11\}$ changes smoothly from $w^{(M)}\{11\}\sim-0.13-0.92i$ at $g=0$ and $h=0.1$ to $w^{(M)}\{11\}\sim0.14-0.48i$ at $g=h=1.0$.

\begin{table}[t]
\centering
\begin{tabular}{ccccc}
\hline
$X$ & seed 0 & seed 1 & seed 2 & seed 3 \\
\hline
82 & $0.14444 - 0.47582i$ & $-0.47305 + 0.01844i$ & $0.12577 - 0.13037i$ & $0.13615 - 0.47964i$ \\
83 & $0.14484 - 0.47360i$ & $-0.48634 + 0.07667i$ & $0.12496 - 0.14606i$ & $0.13566 - 0.47926i$ \\
84 & $0.14524 - 0.47206i$ & $-0.47557 + 0.08019i$ & $0.11542 - 0.14959i$ & $0.13449 - 0.47988i$ \\
85 & $0.14606 - 0.47413i$ & $-0.45708 + 0.07803i$ & $0.12169 - 0.16485i$ & $0.12800 - 0.47952i$ \\
86 & $0.14980 - 0.47303i$ & $-0.52919 + 0.13278i$ & $0.11048 - 0.15774i$ & $0.13364 - 0.48626i$ \\
87 & $0.14964 - 0.46932i$ & $-0.54534 + 0.08209i$ & $0.13685 - 0.17973i$ & $0.13328 - 0.48760i$ \\
\hline
\end{tabular}
\caption{
Values of $w^{(8)}\{11\}$ obtained by the RMA for $g=h=1.0$, $M=8$, and $X=82,\ldots,87$.
The labels seed 0--3 denote different initial conditions at $X=82$.
For each computation at $X+1$, the corresponding $\hat{A}_{\mu}^{(M)}$ obtained at $X$ was used as the initial condition.
}
\label{tab:minkowski_two_matrix_M8_A82_87_w11}
\end{table}

To exclude false convergence arising from the dependence on initial conditions, it is important to examine the validity of the results from various perspectives.
In the following, we investigate the correctness of the numerical results for $g=h=1.0$.
Since perturbation theory cannot be trusted in this case, we first examine the stability of the approximate values against different initial conditions, in the same manner as for the one-matrix model in Section \ref{subsubsec:Euclidean-type_One-Matrix_Model}.
The values of $w^{(8)}\{11\}$ for $M=8$ and $X=82,\ldots,87$ are summarized in Table \ref{tab:minkowski_two_matrix_M8_A82_87_w11}.
Again, for each computation at $X+1$, the corresponding $\hat{A}_{\mu}^{(M)}$ obtained at $X$ was used as the initial condition.

In all four computations, $F^{(8)}$ decreases to the order of $10^{-31}$ at $X=86$ and then rises to the order of $10^{-10}$ at $X=87$.
Thus, from the viewpoint of the convergence of the objective function, there is essentially no difference among these four cases.
On the other hand, a clear difference is observed in the stability of the values of $w^{(8)}\{11\}$:
seed 0 and seed 3, which yield values close to $w^{(8)}\{11\}\sim0.14-0.48i$, give approximately consistent estimates for different values of $X$;
by contrast, seed 1 and seed 2, which yield values far from this region, show unstable variations in $w^{(8)}\{11\}$ as $X$ is varied.
This suggests that $w^{(8)}\{11\}\sim0.14-0.48i$ is close to a ``physical'' solution of the loop equations and that its value is therefore effectively protected.
We also note a subtle point:
although $F^{(8)}=0$ forms a square system at $X=86$, the approximate values obtained from seed 0 and seed 3 still differ slightly.
While this does not impair the practical usefulness of the approximation, the origin of this curious phenomenon should be clarified in future work.

\begin{figure}[t] 
\centering
\includegraphics[width=0.95\textwidth]{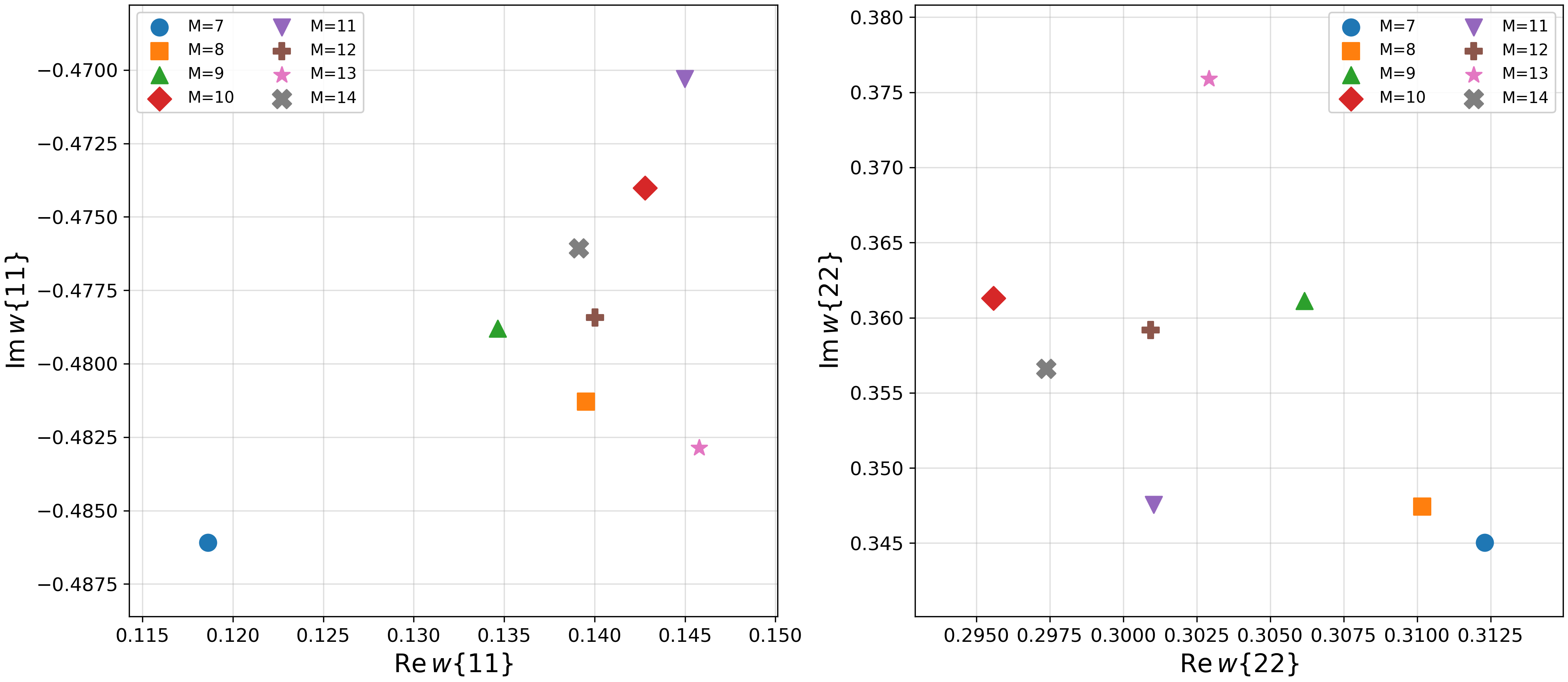}
\caption{
Values of $w^{(M)}\{11\}$ and $w^{(M)}\{22\}$ computed for $h=g=1.0$ with $M=7,...,14$.
The left panel corresponds to $w^{(M)}\{11\}$, while the right panel corresponds to $w^{(M)}\{22\}$.
For each value of $M$, we choose the largest value of $X$ for which $F^{(M)}\sim10^{-30}$ can still be achieved.
}
\label{fig:hg1_M7_M14_w11_w22_complex_plane_combined}
\end{figure}

\begin{table}[t]
\centering
\begin{tabular}{cccccc}
\hline
$M$ & $w^{(M)}\{11\}$ & $w^{(M)}\{22\}$ & $X$ & rank & $M^2+1$ \\
\hline
7 & $0.1186 - 0.4861i$ & $0.3123 + 0.3450i$ & 71 & 50 & 50 \\
8 & $0.1395 - 0.4813i$ & $0.3102 + 0.3474i$ & 86 & 65 & 65 \\
9 & $0.1346 - 0.4788i$ & $0.3062 + 0.3611i$ & 133 & 82 & 82 \\
10 & $0.1428 - 0.4740i$ & $0.2956 + 0.3613i$ & 152 & 101 & 101 \\
11 & $0.1450 - 0.4703i$ & $0.3010 + 0.3476i$ & 170 & 119 & 122 \\
12 & $0.1400 - 0.4784i$ & $0.3009 + 0.3592i$ & 266 & 145 & 145 \\
13 & $0.1458 - 0.4829i$ & $0.3029 + 0.3759i$ & 291 & 170 & 170 \\
14 & $0.1391 - 0.4761i$ & $0.2974 + 0.3566i$ & 317 & 195 & 197 \\
\hline
\end{tabular}
\caption{RMA results for the Minkowski-type two-matrix model at $h=g=1$. Here $X$ denotes the number of loop equations included in the objective function, ``rank'' denotes the number of nonzero singular values of the Jacobian matrix evaluated at the optimized RMA solution, and $M^2+1$ denotes the effective number of variables remaining after complete gauge fixing.}
\label{tab:minkowski_two_matrix_hg1_rma}
\end{table}

As a second test, we repeated the same procedure not only for $M=8$ but also for $M=7,\ldots,14$.
The resulting RMA values at $g=h=1.0$ are summarized in Figure \ref{fig:hg1_M7_M14_w11_w22_complex_plane_combined} and Table \ref{tab:minkowski_two_matrix_hg1_rma}.
For each value of $M$, the corresponding value of $X$ was chosen as the largest one for which $F^{(M)}\sim10^{-30}$ could still be achieved.
As an overall tendency, the moments clearly converge toward
\begin{equation} \label{eq:w11_w22_approximation}
\begin{aligned}
w^{(M)}\{11\} & \sim0.14-0.48i,\\
w^{(M)}\{22\} & \sim0.30+0.36i.
\end{aligned}
\end{equation}
Compared with the Euclidean two-matrix model, the scatter is somewhat larger.
We believe that this originates from the increased number of degrees of freedom that must be adjusted numerically when $\hat{A}_{\mu}$ is taken to be a complex matrix, as well as from the smaller number of loop equations included in the analysis.
Reducing the scatter while maintaining the reliability of the results remains an important problem for future work.
We also note that the values of $w^{(M)}\{11\}$ and $w^{(M)}\{22\}$ obtained from seed 0 and seed 3 in Table \ref{tab:minkowski_two_matrix_M8_A82_87_w11} are not plotted explicitly; nevertheless, they likewise lie within the ranges covered by the two plots in Figure \ref{fig:hg1_M7_M14_w11_w22_complex_plane_combined}.

Moreover, except for the cases $M=11,14$, the maximal value of $X$ was found to make the number of unknown variables equal to the number of constraints, namely the system $F^{(M)}=0$ becomes square.
Therefore, the corresponding matrix configurations $\hat{A}_{\mu}^{(M)}$ are likely to reproduce exact solutions of $F^{(M)}=0$ correctly.
For none of the values of $M$ considered here, however, was it possible to achieve $F^{(M)}\sim10^{-30}$ in an overdetermined system.
In any case, for the case $g=h=1.0$, there are several pieces of circumstantial evidence suggesting that \eqref{eq:w11_w22_approximation} reproduces a physical solution of the loop equations:
\begin{itemize}
\item the numerical values agree with the perturbative prediction for $g=0$ and $|h|\ll1$, and then vary smoothly as $g$ and $h$ are gradually increased toward $g=h=1.0$;

\item similar values are obtained even when the matrix size $M$ is varied;

\item the values remain stable under changes in the number $X$ of loop equations;

\item $F^{(M)}\sim10^{-30}$ is achieved multiple times for combinations of $(M,X)$ in which $F^{(M)}=0$ forms a square system.
\end{itemize}

\begin{figure}[t] 
\centering
\includegraphics[width=1.02\textwidth]{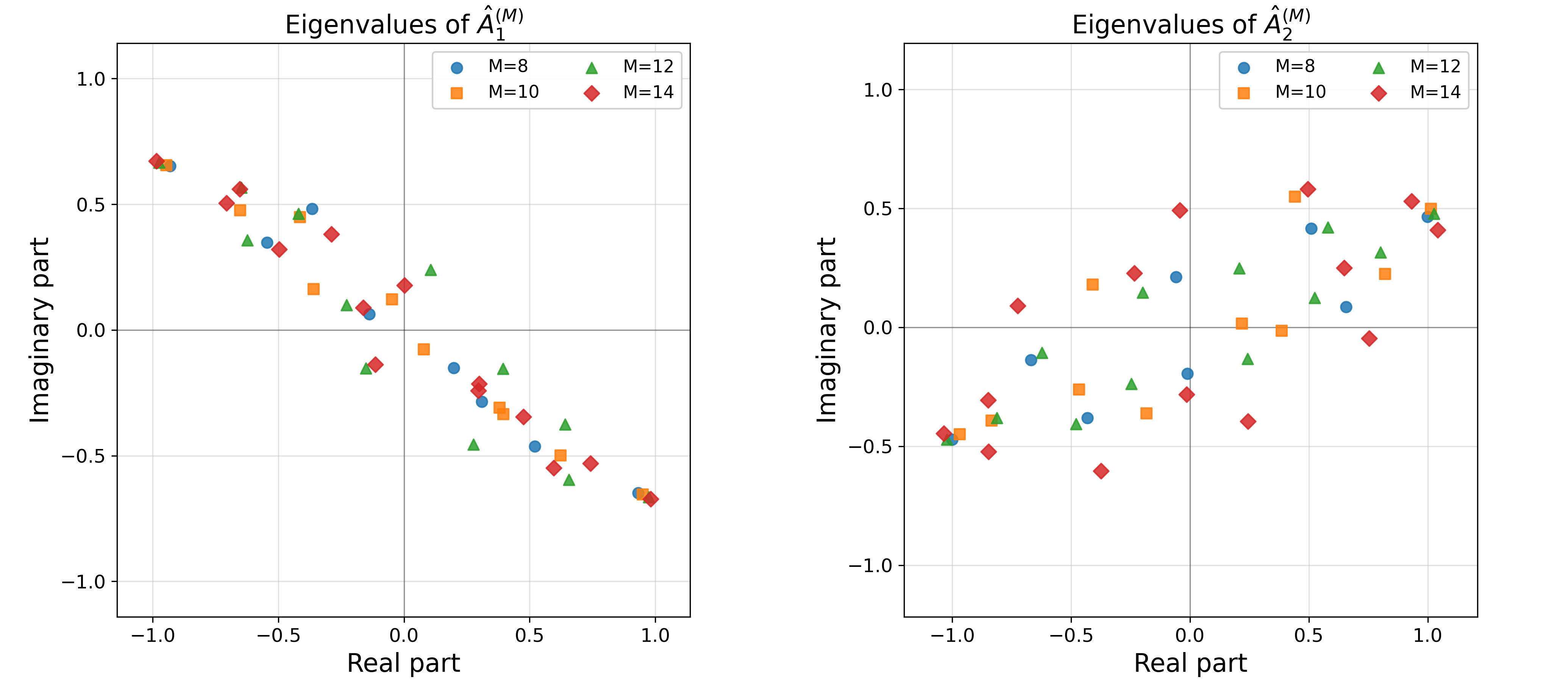}
\caption{
Eigenvalues of $\hat{A}_{1}^{(M)}$ and $\hat{A}_{2}^{(M)}$ obtained by applying the RMA method with $g=h=1$ and $M=8,10,12,14$, plotted on the complex plane.
The values of $M=8,10,12,14$ are distinguished by the colors and shapes of the points.
}
\label{fig:hg1_evenM_A1_A2_eigenvalues}
\end{figure}

Figure \ref{fig:hg1_evenM_A1_A2_eigenvalues} shows the eigenvalues of the regularized master fields $\hat{A}_{1}^{(M)}$ and $\hat{A}_{2}^{(M)}$ for $g=h=1$ and $M=8,...,14$.
The two plots appear completely different, which originates from the fact that the action $S_{M}$ does not treat $A_{1}$ and $A_{2}$ symmetrically.
Compared with the one-matrix model, the eigenvalue distributions in the Minkowski-type two-matrix model are somewhat irregular.
At present, except for the cases $M=11,14$, the systems are square and all satisfy $F^{(M)}\sim10^{-30}$.
Therefore, the resulting matrices $\hat{A}_{\mu}^{(M)}$ should provide exact solutions of $F^{(M)}=0$, namely all loop equations included in the analysis are expected to be satisfied.
The irregularity of the resulting eigenvalue distributions may instead indicate that the number of loop equations being considered is still insufficient.

Even so, as already observed in the one-matrix model, the RMA method is not particularly suited for reproducing the closed curve $C_{\text{const.}}$ on which the eigenvalues are distributed.
Rather, the quantities that can be approximated accurately are the various moments and the endpoints $\pm\alpha$ of the resolvent cut.
Regarding the latter, for each of $\hat{A}_{1}^{(M)}$ and $\hat{A}_{2}^{(M)}$, the eigenvalues appear to indicate an approximately common pair of endpoints as $M$ is varied.
These results suggest that, also in the Minkowski-type two-matrix model, the resolvents associated with \(A_1\) and \(A_2\) have a one-cut structure.

\section{Summary and Discussion} \label{sec:Summary_and_Discussion}

In this work, we propose an approximation method, the regularized master-field approximation (RMA), and apply it to Hermitian one- and two-matrix models.
In this method, we assume the existence of the master field and adopt the standpoint of regularizing it in a finite dimension $M$.
The matrix elements of the regularized master field $\hat{A}_{\mu}^{(M)}$ defined in this way are determined by imposing that they satisfy the loop equations as much as possible.
In practice, we compute them numerically using the least-squares method and explicitly obtain $\hat{A}_{\mu}^{(M)}$ for both Euclidean- and Minkowski-type cases.
The numerical results for the Euclidean-type case agree well with existing numerical results, and in the Minkowski-type case they also reproduce perturbative results accurately.
Furthermore, among the resulting approximate values, only those expected to reproduce physical solutions are found to be stable and insensitive to the details of the regularization scheme.
These results suggest that, at least for certain matrix models and under specific large-$N$ (large-$M$) limits, an object analogous to the large-$N$ master field may indeed exist.

Although the existence of the master field had long been conjectured, it had not been established in general (except for the eigenvalue distribution of the one-matrix model).
Moreover, even if such an object were to exist, it seems to have generally been regarded as an abstract operator associated only with the $N\to\infty$ limit, and as a concept with little practical utility apart from its theoretical interest.
One of the very few exceptions is the large-$N$ master-field optimization \cite{Jevicki:1982jj, Jevicki:1983hb, Koch:2021yeb, Mathaba:2023non}, of which RMA may be regarded as a variant.
The most striking result of this line of research is that, at least for certain matrix models and certain large-$N$ limits, an object resembling the master field actually exists and can be regularized by a finite-dimensional matrix of relatively small size $M\sim10$, accurately reproducing the known results.
As a consequence, for at least some classes of matrix models, numerical calculations are essentially reduced to finding the regularized master field, which can be carried out very efficiently.

Unlike the matrix bootstrap method that combines positivity constraints with semidefinite programming (SDP), this method does not yield rigorous upper or lower bounds on physical observables.
To compensate for this drawback, it is essential to perform independent checks of the validity of the approximation.
In this work, in addition to comparisons with perturbative calculations, we checked whether the values of the regularized moments $w^{(M)}\{\mu_{1}\ldots\mu_{n}\}$ remain stable when the set of loop equations included in the objective function is varied.
If $w^{(M)}\{\mu_{1}\ldots\mu_{n}\}$ is close to the true value, then it is expected to reproduce a ``physical'' solution of the loop equations, and hence should be insensitive to details of the regularization such as the matrix size and the number of loop equations.
This criterion was indeed found to likely function properly in both the one- and two-matrix models.
Moreover, in the one- and two-matrix examples studied here, the results suggest that the corresponding resolvents possess a one-cut structure even in the Minkowski case.
If this is indeed the case, then the method proposed in \cite{Li:2024ggr} for approximating physical observables is justified.
Since that approach is constructed from a viewpoint different from that of the master field, it may also provide an additional consistency check.
At present, the validity of the approximation is assessed through a combination of several methods, but in the future it would be desirable to develop consistency checks that can be implemented in a more systematic and broadly applicable manner.

As a next step of this work, it would be desirable to apply this method to higher-dimensional large-$N$ reduced gauge theories containing fermions.
More generally, many aspects of real-time dynamics in gauge theories, and in quantum field theory more broadly, remain difficult to understand from first principles.
From a phenomenological viewpoint, there are many important problems in which real-time dynamics and nonequilibrium quantum field theory play essential roles, such as the origin of the baryon asymmetry in the early universe, thermalization during reheating, and bubble nucleation associated with false vacuum decay.
These are all notoriously challenging problems for numerical approaches.
Although one would naturally like to analyze such real-time dynamics directly, efficient first-principles simulations based on Monte Carlo methods are obstructed by the sign problem.
By contrast, while the applicability of RMA is limited to large-$N$ theories and their reduced descriptions, it has the important advantage that it can be applied directly to Minkowski-type matrix models.
The results obtained in this work suggest that RMA may provide a complementary numerical approach to theories that are difficult to access by conventional methods.

From a different perspective, it is also of interest to apply this method to the IKKT matrix model, which is regarded as a nonperturbative formulation of superstring theory.
In recent years, it has been shown that a holographic dual description exists for the (polarized) IKKT matrix model \cite{Hartnoll:2024csr, Komatsu:2024bop, Komatsu:2024ydh, Ciceri:2025maa}, and it has also been proposed that gravity emerges based on the large-$N$ reduction scenario \cite{Ho:2025htr}, indicating that theoretical understanding has been steadily advancing, and hence numerical analysis is becoming increasingly important.
Regarding numerical studies of the IKKT matrix model, extensive work has been carried out using numerical integration methods including the complex Langevin method and the generalized Lefschetz thimble method \cite{Nishimura:2019qal, Anagnostopoulos:2022dak, Chou:2025moy, Anagnostopoulos:2026qvz,  Anagnostopoulos:2026utg}.
In particular, for the Lorentzian IKKT matrix model, the emergence of a (1+3)-dimensional spacetime has been consistently reported.

An important advantage of the RMA in this context is that it may allow direct numerical studies of the IKKT matrix model without introducing any deformation.
This possibility is particularly significant in light of several intriguing recent observations:
it has been shown, through analytic and numerical studies using the polarized IKKT matrix model, that introducing a mass term and then taking the massless limit after performing the matrix integral leads to a theory different from the original massless IKKT matrix model \cite{Komatsu:2024ydh, Chou:2025rwy}.
Direct numerical evaluation of the undeformed IKKT matrix integral is known to be extremely difficult, and most numerical studies to date have instead considered mass-deformed versions of the model (see, for example, \cite{Anagnostopoulos:2026qvz, Anagnostopoulos:2026utg}.)
Therefore, if the undeformed IKKT matrix integral can be analyzed numerically using the RMA, this may provide new insights into the model.

As a first step toward the numerical analysis of the IKKT matrix model, it is necessary to demonstrate that RMA can be applied to matrix models containing fermions.
To this end, it is natural to perform RMA calculations for models whose properties are already well understood.
Fortunately, the polarised IKKT matrix model has been studied extensively both analytically and numerically, and lower-dimensional mass-deformed SYM matrix models are also well suited for numerical investigations.
In particular, for the $D=4$ model, the Pfaffian is real, allowing direct comparisons with Monte Carlo results \cite{Martina:2025kwc}.
Applying RMA to these models therefore constitutes a natural first step toward the eventual analysis of the IKKT matrix model.
Specifically, the loop equations of matrix models containing fermions involve moments $\langle\text{tr}\,{\cal O}[A,\Psi]\rangle$ that depend on both bosonic and fermionic matrices.
These moments can be rewritten as
\begin{equation}
\begin{aligned}
\langle\text{tr}\,{\cal O}[A,\Psi]\rangle
&=
\int dA\,d\Psi\,
\text{tr}\,{\cal O}[A,\Psi]\,
e^{-S_{B}[A]-S_{F}[A,\Psi]}\\
&=
\int dA\,
\text{tr}\,{\cal O}_{B}[A]\,
e^{-S_{\mathrm{eff}}[A]},\\
S_{\mathrm{eff}}[A]
&\equiv
S_{B}[A]-\log\mathrm{Pf}({\cal M}[A]).
\end{aligned}
\end{equation}
Here $S_{B}[A]$ and $S_{F}[A,\Psi]$ denote the bosonic and fermionic parts of the action, respectively.
If the latter is quadratic in $\Psi$,\footnote{More precisely, after expanding $\Psi$ in terms of the generators $t^{a}$ of $U(N)$ as $\Psi=\sum_{a}\Psi^{a}t^{a}$, $S_{F}[A,\Psi]$ must be quadratic in the coefficients $\Psi^{a}$.} the fermionic path integral yields a Pfaffian, $\int d\Psi\,e^{-S_{F}[A,\Psi]}=\mathrm{Pf}({\cal M}[A])$.
Moreover, ${\cal O}_{B}[A]$ is obtained from ${\cal O}[A,\Psi]$ by integrating out the fermions, namely by performing the Wick contractions of the fermionic variables.
As a result of this procedure, all observables can be expressed solely in terms of the bosonic matrices $A_{\mu}$, with their expectation values weighted by the bosonic effective action $S_{\mathrm{eff}}[A]$.
It is therefore expected that the RMA can be formulated, as before, entirely in terms of the bosonic master field $\hat{A}_{\mu}$.
We are currently preparing a separate paper presenting detailed analytical and numerical studies of such models.
In addition, there exists a series of works investigating the relation between the master field and spacetime emergence in the IKKT matrix model \cite{Klinkhamer:2020wct, Klinkhamer:2020hyx, Klinkhamer:2020jqx}.
Since RMA is precisely a method for approximately reconstructing such a master field, it is highly compatible with these studies.

Finally, we comment on some caveats regarding the large-$N$ limit.
As already noted in Section \ref{subsubsec:Some_Remarks_on_the_RMA}, the RMA is tailored to the planar large-$N$ limit and relies on a nontrivial assumption, namely that after regularizing the master field $\hat{A}_{\mu}$ as a finite-dimensional matrix $\hat{A}_{\mu}^{(M)}$, one can take the limit $M\to\infty$.
While this provides a clearer operational definition of the large-$N$ limit, it also has the limitation that certain important classes of solutions are implicitly excluded.
For example, Heisenberg-type commutation relations $[A_{\mu},A_{\nu}]=i\theta_{\mu\nu}$ (with $\theta_{\mu\nu}\in\mathbb{R}$) and unitary representations of noncompact Lie algebras cannot be realized within this framework and are therefore outside the scope of the RMA.
These solutions correspond to field theories on noncommutative spaces and are of considerable physical interest.
In particular, studies of gravity theories emerging from such noncommutative geometries, as well as their applications to cosmology, have continued to develop in close connection with the IKKT matrix model \cite{Steinacker:2010rh, Battista:2022hqn, Steinacker:2026qzk}.
It is desirable to develop a formulation that can properly incorporate them.

The double-scaling limit is also outside the scope of the RMA.
The key idea behind the RMA, as well as its predecessor based on eigenvalue bootstrap, is that matrix models in the planar limit should have a simplified structure, and that by exploiting quantities that capture this simplicity, efficient numerical computation becomes possible.
The eigenvalue distribution, and its natural generalization to the master field, precisely play this role.
However, in the double-scaling limit, it becomes essential to include the $1/N$ corrections that are neglected in the planar limit, and as a result large-$N$ factorization no longer holds.
Therefore, the master field, which is the central assumption of this method, does not exist, and the double-scaling limit is fundamentally incompatible with the RMA.
To study such theories numerically, one must either perform Monte Carlo simulations (while somehow avoiding the sign problem) or employ bootstrap methods applicable at finite $N$ \cite{Laliberte:2026qce}.

Despite these limitations, the RMA provides a method applicable to a wide class of large-$N$ matrix models.
Applying this method to more physically interesting models, such as the Eguchi--Kawai model and the IKKT matrix model, and exploring their properties constitutes an important direction for future work.

\acknowledgments

The author would like to thank Robert Brandenberger, Yue Lei, Takeshi Morita, Jun Nishimura, Julia Pasiecznik, Diego Allan Reyna, and Hiromasa Watanabe for helpful discussions.
The author is also grateful to the organizers of the workshop “Thermal Quantum Field Theory and Their Applications” for their hospitality.
The author is supported by JST SPRING, Grant Number JPMJSP2132.

\appendix

\section{Perturbative Expansion Using Loop Equations} \label{app:appendixA}

In Section \ref{subsubsec:Minkowski-type_two-Matrix_Model}, we approximated the values of the moments by perturbative expansion.
Here, taking the Minkowski one-matrix model as an example, we derive this perturbative expansion from the loop equations.
The loop equations of this theory are given, for $n=1,2,\ldots$, by
\begin{equation}
w_{n}=i\sum_{k=0}^{n-2}w_{n-k-2}w_{k}+gw_{n+2}.
\end{equation}
To recover perturbation theory from this equation, we expand the moments $w_{n}$ in positive powers of $g$:
\begin{equation}
w_{n}=\sum_{m=0}^{\infty}w_{n;m}g^{m}.
\end{equation}
Here, $w_{n;m}$ denotes the expansion coefficient of $g^{m}$.
Since the loop equations hold for arbitrary $g$, when this power-series expansion is substituted into the original loop equations, the two sides must agree order by order in $g^{0},g^{1},\ldots$.
Extracting the coefficient of $g^{m}$, we find
\begin{equation} \label{eq:Recusion_Relation_from_Loopeq}
w_{n;m}=i\sum_{l=0}^{m}\sum_{k=0}^{n-2}w_{n-k-2;m-l}w_{k;l}+w_{n+2;m-1}.
\end{equation}
The important point is that this equation gives a recursion relation with respect to the two indices $(m,n)$ ordered lexicographically.
Indeed, for the left-hand side with indices $(m,n)$, the index pairs appearing on the right-hand side are $(m-l,n-k-2)$, $(l,k)$, and $(m-1,n+2)$, all of which are smaller than $(m,n)$ in the lexicographic order.
Thus, applying the loop equation in this form to $w_{n;m}$ always gives ``smaller'' coefficients $w_{n';m'}$.
Moreover, since $w_{n;m}$ is the coefficient in the expansion of $w_{n}$ in powers of $g$, and since $w_{0}=\langle\text{tr}1\rangle=1$ for $n=0$, the coefficients $w_{0;m}$ are independent of $g$ and satisfy
\begin{equation}
w_{0;m}=
\begin{cases}
1 & m=0,\\
0 & m>0.
\end{cases}
\end{equation}
This specifies the initial term and boundary condition for the recursion relation \eqref{eq:Recusion_Relation_from_Loopeq} for $w_{n;m}$.

Let us now use this relation to compute the power-series expansion of $w_{2}$.
First, the zeroth-order term in $g$, $w_{2;0}$, is immediately obtained as
\begin{equation}
w_{2;0}=iw_{0;0}w_{0;0}=i.
\end{equation}
Here, since we assume that $w_{n}$ is expanded in positive powers of $g$, $w_{n;-1}$ is always zero.
Next, considering $w_{2;1}$, we find
\begin{equation}
w_{2;1}=i\sum_{l=0}^{1}w_{0;1-l}w_{0;l}+w_{4;0}=w_{4;0}.
\end{equation}
In this case, the splitting term always contains $w_{0;1}$ and therefore vanishes, leaving only the second term, $w_{4;0}$.
Applying the recursion relation to $w_{4;0}$, we then obtain
\begin{equation}
w_{4;0}=i\sum_{k=0}^{2}w_{2-k;0}w_{k;0}=2iw_{2;0}+iw_{1;0}w_{1;0}=-2.
\end{equation}
Here, $w_{1;0}=0$ also follows immediately from the recursion relation.
We have therefore found $w_{2;1}=-2$.
This procedure can be continued to arbitrary order, yielding the power-series expansion
\begin{equation}
w_{2}=i-2g-9ig^{2}+\cdots
\end{equation}
from the loop equations.
The situation is essentially identical for the two-matrix model discussed in Section \ref{subsec:Overall_Setup_two-matrix_model}.
Here, for simplicity, we set $g=0$ and assume $h\ll1$, so that $h$ can be used as the expansion parameter in the perturbative expansion.
The corresponding loop equations are given by
\begin{equation}
\begin{aligned}
\eta^{\mu\nu}w_{m}\{\mu_{1}...\mu_{n}\nu\} & =-i\sum_{l=0}^{m}\sum_{k=1}^{n}\delta_{\mu_{k},\mu}w_{m-l}\{\mu_{1}...\mu_{k-1}\}w_{l}\{\mu_{k+1}...\mu_{n}\}\\
& \quad -\sum_{\nu=1}^{2}w_{m-1}\{\mu_{1}...\mu_{n}[\nu,[\nu,\mu]]\}.
\end{aligned}
\end{equation}
Here, $w_{m}\{\mu_{1}...\mu_{n}\}$ denotes the coefficient in the expansion of this quantity in powers of $h$.
We first consider the zeroth-order term in $h$ for the case $n=1$ and $\mu_{1}=\mu=1$.
Since $\eta^{11}=+1$, we obtain
\begin{equation}
w_{0}\{11\}=-iw_{0}\{\}.
\end{equation}
Here, $w\{\}\equiv\langle\mathrm{tr}\,1\rangle=1$.
By the same argument, one immediately obtains $w_{0}\{22\}=i$.
Next, let us consider the first-order term in $h$ for $\mu_{1}=\mu=1$:
\begin{equation}
w_{1}\{11\}=-iw_{1}\{\}-w_{0}\{1221\}-w_{0}\{1122\}+2w_{0}\{1212\}.
\end{equation}
Since $w_{1}\{\}=0$, the left-hand side is determined once the remaining three terms are evaluated.
These are given by
\begin{equation}
\begin{aligned}
w_{0}\{1221\} & =-iw_{0}\{22\}=1,\\
w_{0}\{1122\} & =-iw_{0}\{22\}=1,\\
w_{0}\{1212\} & =-iw_{0}\{1\}w_{0}\{1\}=0,
\end{aligned}
\end{equation}
where we assumed the $\mathbb{Z}_{2}$ symmetry so that $w\{12\}=w\{1\}=w\{2\}=0$.
We therefore obtain $w_{1}\{11\}=-2$, which leads to the expansion
\begin{equation}
w\{11\}=-i-2h+\cdots,
\end{equation}
corresponding to the expansion formula given in \eqref{eq:w11_perturbative}.

\section{Numerical Calculation of the One-Matrix Model Using Real Eigenvalues} \label{app:appendixB}

\begin{table}[t]
\centering
\begin{tabular}{cccc}
\hline
\(r\)
&
\(w_2^{(M)}\)
&
\(\left|w_2^{(M)}-w_2^{\rm exact}\right|\)
&
\(F_E^{(M)}\)
\\
\hline
1.0
&
0.5253
&
\(9.1748\times 10^{-3}\)
&
\(2.1666\times 10^{-3}\)
\\
0.8
&
0.5206
&
\(4.4799\times 10^{-3}\)
&
\(3.6959\times 10^{-4}\)
\\
0.6
&
0.5183
&
\(2.1126\times 10^{-3}\)
&
\(3.3439\times 10^{-5}\)
\\
0.4
&
0.5170
&
\(8.9153\times 10^{-4}\)
&
\(9.5458\times 10^{-7}\)
\\
0.2
&
0.5165
&
\(3.5545\times 10^{-4}\)
&
\(2.0259\times 10^{-9}\)
\\
\hline
\end{tabular}
\caption{
Numerical results obtained by applying RMA with $g=-1$, $M=\Lambda=10$, and $r=1.0,0.8,0.6,0.4,0.2$.
}
\label{tab:one_matrix_euclidean_r_dependence}
\end{table}

In Section \ref{subsubsec:Euclidean-type_One-Matrix_Model}, we treated the regularized master field $\hat{\phi}^{(M)}$ as a complex matrix, diagonalized it, and chose its complex eigenvalues $\hat{z}_{1}^{(M)},\ldots,\hat{z}_{M}^{(M)}$ as the unknown variables.
On the other hand, the master field $\hat{\phi}$ of the Euclidean one-matrix model should originally be Hermitian, and if it is regularized and diagonalized as it is, one obtains real eigenvalues $\hat{\lambda}_{1}^{(M)},\ldots,\hat{\lambda}_{M}^{(M)}$.
Naively, using these as the unknown variables in the RMA is the most natural procedure, and it is in fact possible.

However, the RMA based on $\hat{\lambda}_{1}^{(M)},\ldots,\hat{\lambda}_{M}^{(M)}$ does not optimize as well as the one based on $\hat{z}_{1}^{(M)},\ldots,\hat{z}_{M}^{(M)}$.
More specifically, the objective function $F_{E}^{(M)}$ does not become sufficiently small, and the values of the regularized moments $w_{n}^{(M)}$ deviate from the exact values $w_{n}^{\text{exact}}$.
Unfortunately, the cause of this phenomenon is not understood at present.

As an ad hoc remedy, however, we have confirmed that the values of $w_{n}^{(M)}$ approach $w_{n}^{\text{exact}}$ by choosing a smaller value of the parameter $r$ in the objective function $F_{E}^{(M)}=\sum_{n}r^{n}|{\cal L}_{n}|^{2}$.
Details of the calculation for $g=-1$ and $M=10$ are shown in Table \ref{tab:one_matrix_euclidean_r_dependence}.
One can see that a good approximation is obtained for values around $r\sim0.2$.
This situation appears peculiar when one recalls that the eigenvalue distribution $\rho_{E}(\lambda)$ of the Euclidean one-matrix model is given by
\begin{equation}
\begin{aligned}
\rho_{E}(\lambda) & =\frac{\left(1-\frac{\alpha^{2}g}{2}-g\lambda^{2}\right)\sqrt{(\alpha-\lambda)(\alpha+\lambda)}}{2\pi},\\
\alpha & =\sqrt{\frac{2(1-\sqrt{1-12g})}{3g}}.
\end{aligned}
\end{equation}
According to this expression, for $g=-1$, the endpoints of $\rho_{E}(\lambda)$ are located at approximately $\alpha\sim\pm1.3$.
As explained in Section \ref{subsec:Overall_Setup_one-matrix_model}, the moments $w_{n}$ behave as $w_{n}\sim\alpha^{n}$ for $n\gg1$, and therefore it should be reasonable to choose $r\sim\alpha^{-1}$ in order to stabilize the numerical calculation.
Indeed, in the case of the complex eigenvalues $\hat{z}_{1}^{(M)},\ldots,\hat{z}_{M}^{(M)}$, we chose $r=0.8$, and the optimization worked as expected with this value.
Compared with this, the value $r\sim0.2$ is clearly too small.
Nevertheless, for the real eigenvalues $\hat{\lambda}_{1}^{(M)},\ldots,\hat{\lambda}_{M}^{(M)}$, it gives the best approximation, whereas the approximation does not work so well for $r\sim0.8$.
For all values of $r$, the optimization terminates by the xtol condition, which suggests that $\hat{\lambda}_{1}^{(M)},...,\hat{\lambda}_{M}^{(M)}$ indeed yield values close to the minimum of $F^{(M)}$, although it is not currently understood why such small values of $r$ are required.

\begin{figure}[t] 
\centering
\includegraphics[width=0.9\textwidth]{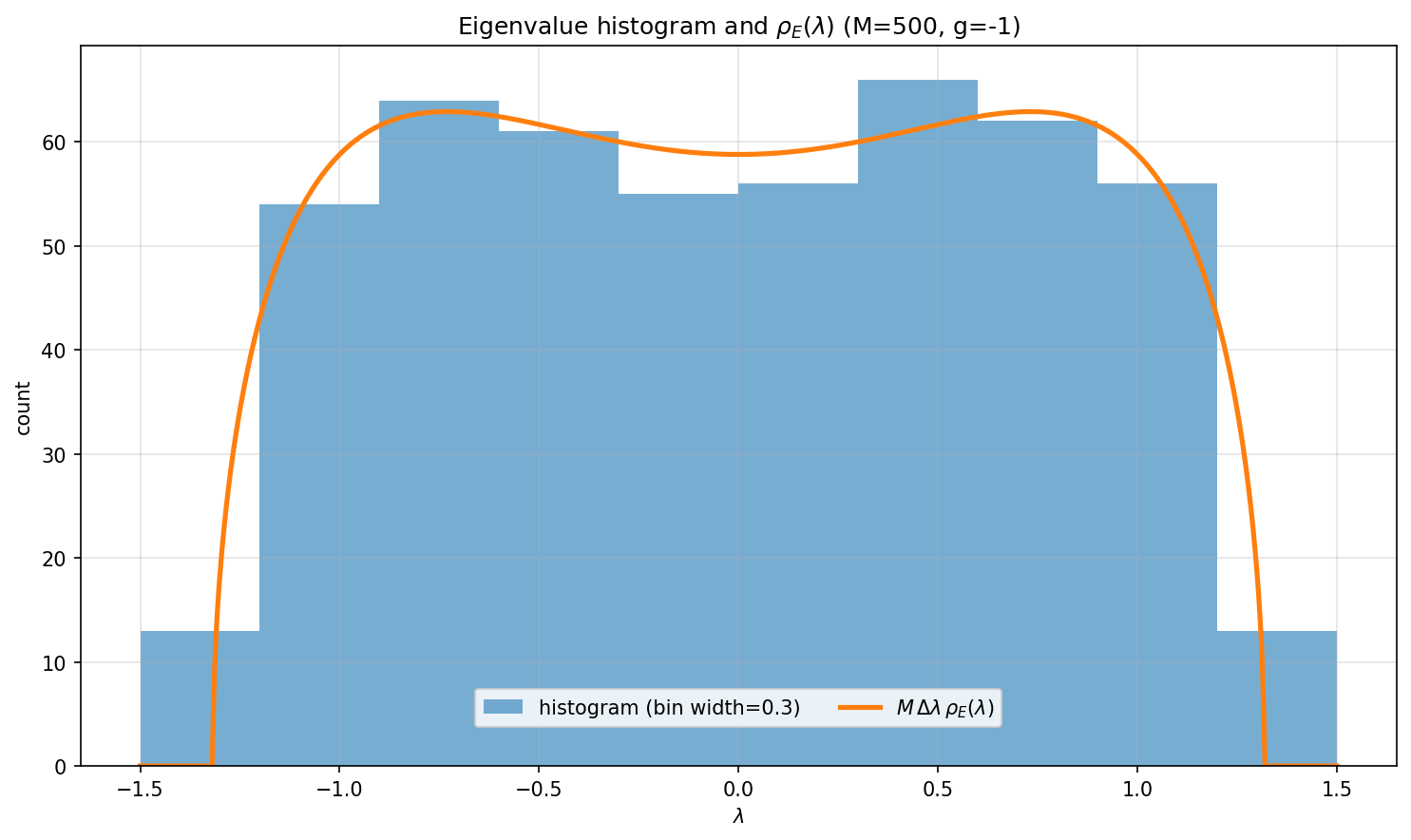}
\caption{
Histogram of $\hat{\lambda}_{1}^{(500)},\ldots,\hat{\lambda}_{500}^{(500)}$ obtained by the RMA with $g=-1$, $M=\Lambda=500$, and $r=0.3$, using bin width $\Delta\lambda=0.3$.
The orange solid curve represents the eigenvalue distribution $\rho_{E}(\lambda)$ for $g=-1$.
}
\label{fig:E-type_M=500_width=0.3_with_EVD}
\end{figure}

Leaving the issue of small $r$ for future work, let us numerically check here the relation between the real eigenvalues $\hat{\lambda}_{1},\ldots,\hat{\lambda}_{N}$ and the eigenvalue distribution $\rho_{E}(\lambda)$.
As explained in Section \ref{subsubsec:Master_Field_as_a_Generalization_of_the_Eigenvalue_Distribution}, the infinitely many eigenvalues $\hat{\lambda}_{1},\ldots,\hat{\lambda}_{N}$ of the master field $\hat{\phi}$ are expected to be distributed according to $\rho_{E}(\lambda)$.
This can be checked numerically by taking a sufficiently large value of $M$ and computing $\hat{\lambda}_{1}^{(M)},\ldots,\hat{\lambda}_{M}^{(M)}$.
Here, we set $g=-1$ and computed $\hat{\lambda}_{1}^{(500)},\ldots,\hat{\lambda}_{500}^{(500)}$ by the RMA with $r=0.3$ and $M=\Lambda=500$.
The resulting histogram is shown in Figure \ref{fig:E-type_M=500_width=0.3_with_EVD}.
It is constructed with bin width $\Delta\lambda=0.3$, and the orange solid curve represents $M\,\Delta\lambda\,\rho_{E}(\lambda)$ evaluated at $g=-1$.
As expected, the numerically obtained $\hat{\lambda}_{1}^{(500)},\ldots,\hat{\lambda}_{500}^{(500)}$ agree well with the eigenvalue distribution.
Comparing the computed value of $w_{2}^{(M)}$ with the exact value $w_{2}^{\text{exact}}$, we find
\begin{equation}
\begin{aligned}
w_{2}^{(M)} & =0.516151216\cdots,\\
w_{2}^{\text{exact}} & =0.516151232\cdots.
\end{aligned}
\end{equation}
Thus, the two values agree up to the seventh decimal place.
Although this is not a bad accuracy, it is honestly somewhat unsatisfactory, given that the approximation uses an extremely large matrix dimension, $M=500$, and that the one-matrix model itself has a simple structure.
This is probably because the optimization has not been fully completed.
For example, the histogram is slightly asymmetric, whereas the $\mathbb{Z}_{2}$ symmetry of the theory implies that it should be symmetric.

\section{Phase Transition in the Euclidean-type One-Matrix Model}

\begin{table}[t]
\centering
\begin{tabular}{ccccc}
\hline
\(g\)
&
\(\Lambda\)
&
\(w_2^{(20)}\)
&
\(\left|w_2^{(20)}-w_2^{\rm exact}\right|\)
&
\(F_E^{(20)}\)
\\
\hline
\(-1.0\)
&
20
&
0.5161486236
&
\(2.609\times 10^{-6}\)
&
\(3.230\times 10^{-31}\)
\\
\(-1.0\)
&
21
&
0.5161486236
&
\(2.609\times 10^{-6}\)
&
\(5.491\times 10^{-31}\)
\\
0.05
&
20
&
1.1332015741
&
\(2.311\times 10^{-9}\)
&
\(4.836\times 10^{-25}\)
\\
0.05
&
21
&
1.1332015741
&
\(2.311\times 10^{-9}\)
&
\(3.191\times 10^{-24}\)
\\
0.1
&
20
&
1.4920220549
&
-- 
&
\(4.040\times 10^{-21}\)
\\
0.1
&
21
&
1.3852381500
&
-- 
&
\(2.422\times 10^{-3}\)
\\
\hline
\end{tabular}
\caption{
Results of the RMA calculation using the complex eigenvalues $\hat{z}_{1}^{(20)},\ldots,\hat{z}_{20}^{(20)}$ with $M=\Lambda=20$ and $r=0.8$ for the three values of the coupling constant, $g=-1$, $0.05$, and $0.1$.
To examine whether the optimized configuration corresponds to a physical solution, we also list the results obtained by repeating the calculation with $M=20$, $\Lambda=21$, and $r=0.8$, using the optimized $\hat{z}_{1}^{(20)},\ldots,\hat{z}_{20}^{(20)}$ as the initial configuration.
Note that $w_{2}^{\mathrm{exact}}$ does not exist for $g=0.1$, since the corresponding matrix integral is not well defined.
}
\label{tab:one_matrix_euclidean_phase_transition}
\end{table}

The Euclidean-type one-matrix model is known to exhibit a phase transition as the coupling constant $g$ is varied, and this transition can be quantitatively captured within the RMA framework.
According to the definition of the action in \eqref{eq:Action_of_One-matrix_Models}, the critical coupling is given by $g_{C}=\frac{1}{12}$.
We therefore perform RMA calculations for three representative values of the coupling constant: $g=-1.0$, $0.05$, and $0.1$.
The value $g=-1.0$ corresponds to a well-defined matrix integral, $g=0.05$ corresponds to a regime in which the action $S$ is not bounded from below but the matrix integral remains well defined in the $N\to\infty$ limit, and $g=0.1>g_{C}$ corresponds to a regime in which the matrix integral is no longer well defined even in the $N\to\infty$ limit.
Here we set $M=\Lambda=20$ and perform the numerical calculation using the complex eigenvalues $\hat{z}_{1}^{(20)},\ldots,\hat{z}_{20}^{(20)}$.
Almost all of the remaining numerical settings are identical to those used in Section \ref{subsec:Numerical_Results_one-matrix_model}, except that the standard deviation of the initial conditions is changed to $\sigma=0.5$.
As will be discussed below, this modification is necessary because otherwise the optimization fails at $g=0.1$.
The numerical results are summarized in Table \ref{tab:one_matrix_euclidean_phase_transition}.

The optimization is successful for $g=-1.0$ and $0.05$.
In particular, the successful numerical calculation at $g=0.05$ is noteworthy.
As discussed above, the action $S$ is not bounded from below at $g=0.05$, and therefore the corresponding matrix integral is not well defined in general.
The large-$N$ limit, however, provides an exceptional case.
There, tunneling away from the local stationary point at $\phi=0$ is suppressed by large-$N$ effects, and the theory consequently remains well defined.

Turning now to RMA, the master field, which should originally be infinite dimensional, is regularized by a finite-dimensional matrix of size $M$.
This procedure may give the impression that the method is effectively performing numerical calculations for a finite-$N$ theory, namely the theory with $N=M$.
This interpretation, however, is incorrect.
The parameter $M$ serves only as a regularization parameter, whereas the theory under consideration is consistently the theory in the $N\to\infty$ limit.
The successful numerical calculation for the one-matrix model at $g=0.05$ can therefore be regarded as strong evidence supporting this interpretation.

As for the remaining case, $g=0.1$, clear signs of the breakdown of the approximation are observed.
First, as mentioned above, the optimization failed to achieve $F^{(20)}\sim10^{-30}$ for smaller values of $\sigma$.
After increasing the standard deviation to $\sigma=0.5$, the objective function decreased to approximately $F^{(20)}\sim10^{-20}$, indicating that the optimization itself was successful.
However, when the optimized $\hat{z}_{1}^{(20)},\ldots,\hat{z}_{20}^{(20)}$ were subsequently used as the initial configuration for a new calculation with $\Lambda=21$, the value of the moment $w_{2}^{(20)}$ changed substantially.
As explained in Sections \ref{subsubsec:Euclidean-type_One-Matrix_Model} and \ref{subsubsec:Minkowski-type_two-Matrix_Model}, this strongly suggests that the corresponding $\hat{z}_{1}^{(20)},\ldots,\hat{z}_{20}^{(20)}$ represent an unphysical solution.
This instability of $w_{2}^{(20)}$ persisted even after varying both $\sigma$ and the random seed, and no configuration yielding a stable value of $w_{2}^{(20)}$ could be found.
This is presumably because no physical solution exists for $g=0.1$.


 \bibliographystyle{JHEP}
 \bibliography{biblio}

@article{Berenstein:2025itw,
    author = "Berenstein, David and Rodriguez, Victor A.",
    title = "{Goldilocks and the bootstrap}",
    eprint = "2503.00104",
    archivePrefix = "arXiv",
    primaryClass = "hep-th",
    doi = "10.1007/JHEP09(2025)109",
    journal = "JHEP",
    volume = "09",
    pages = "109",
    year = "2025"
}

@article{Berenstein:2026wky,
    author = "Berenstein, David and Rodrigues, Jo{\~a}o and Rodriguez, Victor A.",
    title = "{Asymptotic bootstrap for unitary matrix integrals at complex coupling}",
    eprint = "2602.18559",
    archivePrefix = "arXiv",
    primaryClass = "hep-th",
    month = "2",
    year = "2026"
}

@article{Maeta:2026oku,
    author = "Maeta, Reishi",
    title = "{Matrix Bootstrap Approximation without Positivity Constraint}",
    eprint = "2601.16099",
    archivePrefix = "arXiv",
    primaryClass = "hep-th",
    month = "1",
    year = "2026"
}

@article{Li:2024ggr,
    author = "Li, Wenliang",
    title = "{Analytic trajectory bootstrap for matrix models}",
    eprint = "2407.08593",
    archivePrefix = "arXiv",
    primaryClass = "hep-th",
    doi = "10.1007/JHEP02(2025)098",
    journal = "JHEP",
    volume = "02",
    pages = "098",
    year = "2025"
}

@article{Anagnostopoulos:2026qvz,
    author = "Anagnostopoulos, Konstantinos N. and Azuma, Takehiro and Hirasawa, Mitsuaki and Nishimura, Jun and Papadoudis, Stratos and Tsuchiya, Asato",
    title = "{The emergence of (3+1)-dimensional expanding spacetime from complex Langevin simulations of the Lorentzian type IIB matrix model with deformations}",
    eprint = "2604.19836",
    archivePrefix = "arXiv",
    primaryClass = "hep-th",
    reportNumber = "KEK-TH-2826",
    month = "4",
    year = "2026"
}

@inproceedings{Anagnostopoulos:2026utg,
    author = "Anagnostopoulos, Konstantinos N. and Azuma, Takehiro and Hirasawa, Mitsuaki and Nishimura, Jun and Tsuchiya, Asato and Yamamori, Naoyuki",
    title = "{Impact of supersymmetry on the dynamical emergence of the spacetime in the type IIB matrix model with the Lorentz symmetry ''gauge fixed''}",
    eprint = "2604.25564",
    archivePrefix = "arXiv",
    primaryClass = "hep-lat",
    reportNumber = "KEK-TH 2806",
    month = "4",
    year = "2026"
}

@article{Gross:1980he,
    author = "Gross, D. J. and Witten, Edward",
    title = "{Possible Third Order Phase Transition in the Large N Lattice Gauge Theory}",
    doi = "10.1103/PhysRevD.21.446",
    journal = "Phys. Rev. D",
    volume = "21",
    pages = "446--453",
    year = "1980"
}

@article{Wadia:1980cp,
    author = "Wadia, Spenta R.",
    title = "{$N$ = Infinity Phase Transition in a Class of Exactly Soluble Model Lattice Gauge Theories}",
    reportNumber = "EFI-80/15-CHICAGO",
    doi = "10.1016/0370-2693(80)90353-6",
    journal = "Phys. Lett. B",
    volume = "93",
    pages = "403--410",
    year = "1980"
}

@article{Ho:2025htr,
    author = "Ho, Pei-Ming and Kawai, Hikaru and Steinacker, Harold C.",
    title = "{General Relativity in IIB matrix model}",
    eprint = "2509.06646",
    archivePrefix = "arXiv",
    primaryClass = "hep-th",
    reportNumber = "UWThPh 2025-17, NITEP 258",
    doi = "10.1007/JHEP02(2026)070",
    journal = "JHEP",
    volume = "02",
    pages = "070",
    year = "2026"
}

@article{tHooft:1993dmi,
    author = "'t Hooft, Gerard",
    title = "{Dimensional reduction in quantum gravity}",
    eprint = "gr-qc/9310026",
    archivePrefix = "arXiv",
    reportNumber = "THU-93-26",
    journal = "Conf. Proc. C",
    volume = "930308",
    pages = "284--296",
    year = "1993"
}

@article{Susskind:1994vu,
    author = "Susskind, Leonard",
    title = "{The World as a hologram}",
    eprint = "hep-th/9409089",
    archivePrefix = "arXiv",
    reportNumber = "SU-ITP-94-33",
    doi = "10.1063/1.531249",
    journal = "J. Math. Phys.",
    volume = "36",
    pages = "6377--6396",
    year = "1995"
}

@article{Maldacena:1997re,
    author = "Maldacena, Juan Martin",
    title = "{The Large $N$ limit of superconformal field theories and supergravity}",
    eprint = "hep-th/9711200",
    archivePrefix = "arXiv",
    reportNumber = "HUTP-97-A097, HUTP-98-A097",
    doi = "10.4310/ATMP.1998.v2.n2.a1",
    journal = "Adv. Theor. Math. Phys.",
    volume = "2",
    pages = "231--252",
    year = "1998"
}

@article{tHooft:1973alw,
    author = "'t Hooft, Gerard",
    editor = "Taylor, J. C.",
    title = "{A Planar Diagram Theory for Strong Interactions}",
    reportNumber = "CERN-TH-1786",
    doi = "10.1016/0550-3213(74)90154-0",
    journal = "Nucl. Phys. B",
    volume = "72",
    pages = "461",
    year = "1974"
}

@article{Dijkgraaf:1997vv,
    author = "Dijkgraaf, Robbert and Verlinde, Erik P. and Verlinde, Herman L.",
    title = "{Matrix string theory}",
    eprint = "hep-th/9703030",
    archivePrefix = "arXiv",
    reportNumber = "CERN-TH-97-034, CERN-TH-97-34, THU-97-06, UTFA-97-06",
    doi = "10.1016/S0550-3213(97)00326-X",
    journal = "Nucl. Phys. B",
    volume = "500",
    pages = "43--61",
    year = "1997"
}

@article{Liao:2025yfb,
    author = "Liao, Henry and Maeta, Reishi",
    title = "{A New Type of Saddle in Euclidean IKKT Matrix Model and Its Effective Geometry}",
    eprint = "2512.03161",
    archivePrefix = "arXiv",
    primaryClass = "hep-th",
    month = "12",
    year = "2025"
}

@article{Eguchi:1982nm,
    author = "Eguchi, Tohru and Kawai, Hikaru",
    title = "{Reduction of Dynamical Degrees of Freedom in the Large N Gauge Theory}",
    reportNumber = "UT-378-TOKYO",
    doi = "10.1103/PhysRevLett.48.1063",
    journal = "Phys. Rev. Lett.",
    volume = "48",
    pages = "1063",
    year = "1982"
}

@article{Gonzalez-Arroyo:1982hwr,
    author = "Gonzalez-Arroyo, Antonio and Okawa, M.",
    title = "{A Twisted Model for Large $N$ Lattice Gauge Theory}",
    reportNumber = "BNL-31689",
    doi = "10.1016/0370-2693(83)90647-0",
    journal = "Phys. Lett. B",
    volume = "120",
    pages = "174--178",
    year = "1983"
}

@article{Ishibashi:1996xs,
    author = "Ishibashi, N. and Kawai, H. and Kitazawa, Y. and Tsuchiya, A.",
    title = "{A Large N reduced model as superstring}",
    eprint = "hep-th/9612115",
    archivePrefix = "arXiv",
    reportNumber = "KEK-TH-503",
    doi = "10.1016/S0550-3213(97)00290-3",
    journal = "Nucl. Phys. B",
    volume = "498",
    pages = "467--491",
    year = "1997"
}

@article{Banks:1996vh,
    author = "Banks, Tom and Fischler, W. and Shenker, S. H. and Susskind, Leonard",
    title = "{M theory as a matrix model: A conjecture}",
    eprint = "hep-th/9610043",
    archivePrefix = "arXiv",
    reportNumber = "RU-96-95, SU-ITP-96-12, UTTG-13-96",
    doi = "10.1201/9781482268737-37",
    journal = "Phys. Rev. D",
    volume = "55",
    pages = "5112--5128",
    year = "1997"
}

@article{Jevicki:1982jj,
    author = "Jevicki, A. and Karim, O. and Rodrigues, J. P. and Levine, H.",
    title = "{Loop Space Hamiltonians and Numerical Methods for Large $N$ Gauge Theories}",
    reportNumber = "BROWN-HET-485",
    doi = "10.1016/0550-3213(83)90180-3",
    journal = "Nucl. Phys. B",
    volume = "213",
    pages = "169--188",
    year = "1983"
}

@article{Jevicki:1983hb,
    author = "Jevicki, Antal and Rodrigues, Joao P.",
    title = "{Master Variables and Spectrum Equations in Large $N$ Theories}",
    reportNumber = "BROWN-HET-506",
    doi = "10.1016/0550-3213(84)90216-5",
    journal = "Nucl. Phys. B",
    volume = "230",
    pages = "317--335",
    year = "1984"
}

@article{Koch:2021yeb,
    author = "Koch, Robert de Mello and Jevicki, Antal and Liu, Xianlong and Mathaba, Kagiso and Rodrigues, Jo{\~a}o P.",
    title = "{Large N optimization for multi-matrix systems}",
    eprint = "2108.08803",
    archivePrefix = "arXiv",
    primaryClass = "hep-th",
    doi = "10.1007/JHEP01(2022)168",
    journal = "JHEP",
    volume = "01",
    pages = "168",
    year = "2022"
}

@article{Mathaba:2023non,
    author = "Mathaba, Kagiso and Mulokwe, Mbavhalelo and Rodrigues, Jo{\~a}o P.",
    title = "{Large N master field optimization: the quantum mechanics of two Yang-Mills coupled matrices}",
    eprint = "2306.00935",
    archivePrefix = "arXiv",
    primaryClass = "hep-th",
    doi = "10.1007/JHEP02(2024)054",
    journal = "JHEP",
    volume = "02",
    pages = "054",
    year = "2024"
}

@article{Han:2020bkb,
    author = "Han, Xizhi and Hartnoll, Sean A. and Kruthoff, Jorrit",
    title = "{Bootstrapping Matrix Quantum Mechanics}",
    eprint = "2004.10212",
    archivePrefix = "arXiv",
    primaryClass = "hep-th",
    doi = "10.1103/PhysRevLett.125.041601",
    journal = "Phys. Rev. Lett.",
    volume = "125",
    number = "4",
    pages = "041601",
    year = "2020"
}

@article{Lin:2023owt,
    author = "Lin, Henry W.",
    title = "{Bootstrap bounds on D0-brane quantum mechanics}",
    eprint = "2302.04416",
    archivePrefix = "arXiv",
    primaryClass = "hep-th",
    doi = "10.1007/JHEP06(2023)038",
    journal = "JHEP",
    volume = "06",
    pages = "038",
    year = "2023"
}

@article{Lin:2024vvg,
    author = "Lin, Henry W. and Zheng, Zechuan",
    title = "{Bootstrapping ground state correlators in matrix theory. Part I}",
    eprint = "2410.14647",
    archivePrefix = "arXiv",
    primaryClass = "hep-th",
    doi = "10.1007/JHEP01(2025)190",
    journal = "JHEP",
    volume = "01",
    pages = "190",
    year = "2025"
}

@article{Lawrence:2024mnj,
    author = "Lawrence, Scott and McPeak, Brian and Neill, Duff",
    title = "{Bootstrapping time-evolution in quantum mechanics}",
    eprint = "2412.08721",
    archivePrefix = "arXiv",
    primaryClass = "hep-th",
    reportNumber = "LA-UR-24-33001",
    month = "12",
    year = "2024"
}

@article{Kazakov:2021lel,
    author = "Kazakov, Vladimir and Zheng, Zechuan",
    title = "{Analytic and numerical bootstrap for one-matrix model and {\textquotedblleft}unsolvable{\textquotedblright} two-matrix model}",
    eprint = "2108.04830",
    archivePrefix = "arXiv",
    primaryClass = "hep-th",
    doi = "10.1007/JHEP06(2022)030",
    journal = "JHEP",
    volume = "06",
    pages = "030",
    year = "2022"
}

@article{Lin:2020mme,
    author = "Lin, Henry W.",
    title = "{Bootstraps to strings: solving random matrix models with positivity}",
    eprint = "2002.08387",
    archivePrefix = "arXiv",
    primaryClass = "hep-th",
    doi = "10.1007/JHEP06(2020)090",
    journal = "JHEP",
    volume = "06",
    pages = "090",
    year = "2020"
}

@article{DiFrancesco:1993cyw,
    author = "Di Francesco, P. and Ginsparg, Paul H. and Zinn-Justin, Jean",
    title = "{2-D Gravity and random matrices}",
    eprint = "hep-th/9306153",
    archivePrefix = "arXiv",
    reportNumber = "LA-UR-93-1722, SACLAY-SPH-T-93-061",
    doi = "10.1016/0370-1573(94)00084-G",
    journal = "Phys. Rept.",
    volume = "254",
    pages = "1--133",
    year = "1995"
}

@article{Ambjorn:2000dx,
    author = "Ambjorn, Jan and Anagnostopoulos, K. N. and Bietenholz, Wolfgang and Hotta, T. and Nishimura, J.",
    title = "{Monte Carlo studies of the IIB matrix model at large N}",
    eprint = "hep-th/0005147",
    archivePrefix = "arXiv",
    reportNumber = "NBI-HE-00-24, NORDITA-2000-50-HE",
    doi = "10.1088/1126-6708/2000/07/011",
    journal = "JHEP",
    volume = "07",
    pages = "011",
    year = "2000"
}

@article{Martin:2004un,
    author = "Martin, Xavier",
    title = "{A Matrix phase for the phi**4 scalar field on the fuzzy sphere}",
    eprint = "hep-th/0402230",
    archivePrefix = "arXiv",
    doi = "10.1088/1126-6708/2004/04/077",
    journal = "JHEP",
    volume = "04",
    pages = "077",
    year = "2004"
}

@article{Azuma:2004zq,
    author = "Azuma, Takehiro and Bal, Subrata and Nagao, Keiichi and Nishimura, Jun",
    title = "{Nonperturbative studies of fuzzy spheres in a matrix model with the Chern-Simons term}",
    eprint = "hep-th/0401038",
    archivePrefix = "arXiv",
    reportNumber = "KUNS-1883, KEK-TH-929",
    doi = "10.1088/1126-6708/2004/05/005",
    journal = "JHEP",
    volume = "05",
    pages = "005",
    year = "2004"
}

@article{Panero:2006bx,
    author = "Panero, Marco",
    title = "{Numerical simulations of a non-commutative theory: The Scalar model on the fuzzy sphere}",
    eprint = "hep-th/0608202",
    archivePrefix = "arXiv",
    reportNumber = "DIAS-STP-06-11",
    doi = "10.1088/1126-6708/2007/05/082",
    journal = "JHEP",
    volume = "05",
    pages = "082",
    year = "2007"
}

@article{Hanada:2008gy,
    author = "Hanada, Masanori and Miwa, Akitsugu and Nishimura, Jun and Takeuchi, Shingo",
    title = "{Schwarzschild radius from Monte Carlo calculation of the Wilson loop in supersymmetric matrix quantum mechanics}",
    eprint = "0811.2081",
    archivePrefix = "arXiv",
    primaryClass = "hep-th",
    doi = "10.1103/PhysRevLett.102.181602",
    journal = "Phys. Rev. Lett.",
    volume = "102",
    pages = "181602",
    year = "2009"
}

@article{Gonzalez-Arroyo:2010omx,
    author = "Gonzalez-Arroyo, Antonio and Okawa, Masanori",
    title = "{Large $N$ reduction with the Twisted Eguchi-Kawai model}",
    eprint = "1005.1981",
    archivePrefix = "arXiv",
    primaryClass = "hep-th",
    reportNumber = "IFT-UAM-CSIC-10-32, FTUAM-2010-9, HUPD-1002",
    doi = "10.1007/JHEP07(2010)043",
    journal = "JHEP",
    volume = "07",
    pages = "043",
    year = "2010"
}

@article{Hanada:2011fq,
    author = "Hanada, Masanori and Nishimura, Jun and Sekino, Yasuhiro and Yoneya, Tamiaki",
    title = "{Direct test of the gauge-gravity correspondence for Matrix theory correlation functions}",
    eprint = "1108.5153",
    archivePrefix = "arXiv",
    primaryClass = "hep-th",
    reportNumber = "KEK-TH-1490, KEK{\_}TH-1490",
    doi = "10.1007/JHEP12(2011)020",
    journal = "JHEP",
    volume = "12",
    pages = "020",
    year = "2011"
}

@article{Kim:2011cr,
    author = "Kim, Sang-Woo and Nishimura, Jun and Tsuchiya, Asato",
    title = "{Expanding (3+1)-dimensional universe from a Lorentzian matrix model for superstring theory in (9+1)-dimensions}",
    eprint = "1108.1540",
    archivePrefix = "arXiv",
    primaryClass = "hep-th",
    reportNumber = "KEK-TH-1484, OU-HET-720-2011",
    doi = "10.1103/PhysRevLett.108.011601",
    journal = "Phys. Rev. Lett.",
    volume = "108",
    pages = "011601",
    year = "2012"
}

@article{Filev:2015hia,
    author = "Filev, Veselin G. and O'Connor, Denjoe",
    title = "{The BFSS model on the lattice}",
    eprint = "1506.01366",
    archivePrefix = "arXiv",
    primaryClass = "hep-th",
    reportNumber = "DIAS-STP-15-09",
    doi = "10.1007/JHEP05(2016)167",
    journal = "JHEP",
    volume = "05",
    pages = "167",
    year = "2016"
}

@article{Kabat:2000zv,
    author = "Kabat, Daniel N. and Lifschytz, Gilad and Lowe, David A.",
    editor = "Duff, Michael J. and Liu, J. T. and Lu, J.",
    title = "{Black hole thermodynamics from calculations in strongly coupled gauge theory}",
    eprint = "hep-th/0007051",
    archivePrefix = "arXiv",
    reportNumber = "BROWN-HET-1233, CU-TP-981, IASSNS-99-114, PUPT-1942",
    doi = "10.1103/PhysRevLett.86.1426",
    journal = "Int. J. Mod. Phys. A",
    volume = "16",
    pages = "856--865",
    year = "2001"
}

@article{Gonzalez-Arroyo:2014dua,
    author = "Gonzalez-Arroyo, Antonio and Okawa, Masanori",
    title = "{Testing volume independence of SU(N) pure gauge theories at large N}",
    eprint = "1410.6405",
    archivePrefix = "arXiv",
    primaryClass = "hep-lat",
    reportNumber = "IFT-UAM-CSIC-14-102, FTUAM-14-39, HUPD-1406",
    doi = "10.1007/JHEP12(2014)106",
    journal = "JHEP",
    volume = "12",
    pages = "106",
    year = "2014"
}

@article{Parisi:1982gp,
    author = "Parisi, Giorgio",
    title = "{A Simple Expression for Planar Field Theories}",
    reportNumber = "Print-82-0193 (FRASCATI)",
    doi = "10.1016/0370-2693(82)90849-8",
    journal = "Phys. Lett. B",
    volume = "112",
    pages = "463--464",
    year = "1982"
}

@article{GROSS1982440,
author = {Gross, David J and Kitazawa, Yoshihisa},
title = {A quenched momentum prescription for large-N theories},
journal = {Nuclear Physics B},
volume = {206},
number = {3},
pages = {440-472},
year = {1982},
issn = {0550-3213},
doi = {https://doi.org/10.1016/0550-3213(82)90278-4}
}

@article{Parisi:1983mgm,
    author = "Parisi, G.",
    title = "{ON COMPLEX PROBABILITIES}",
    doi = "10.1016/0370-2693(83)90525-7",
    journal = "Phys. Lett. B",
    volume = "131",
    pages = "393--395",
    year = "1983"
}

@article{Klauder:1983sp,
    author = "Klauder, John R.",
    title = "{Coherent State Langevin Equations for Canonical Quantum Systems With Applications to the Quantized Hall Effect}",
    reportNumber = "Print-83-0902 (BTL)",
    doi = "10.1103/PhysRevA.29.2036",
    journal = "Phys. Rev. A",
    volume = "29",
    pages = "2036--2047",
    year = "1984"
}

@article{Aarts:2009uq,
    author = "Aarts, Gert and Seiler, Erhard and Stamatescu, Ion-Olimpiu",
    title = "{The Complex Langevin method: When can it be trusted?}",
    eprint = "0912.3360",
    archivePrefix = "arXiv",
    primaryClass = "hep-lat",
    doi = "10.1103/PhysRevD.81.054508",
    journal = "Phys. Rev. D",
    volume = "81",
    pages = "054508",
    year = "2010"
}

@article{Cristoforetti:2012su,
    author = "Cristoforetti, Marco and Di Renzo, Francesco and Scorzato, Luigi",
    collaboration = "AuroraScience",
    title = "{New approach to the sign problem in quantum field theories: High density QCD on a Lefschetz thimble}",
    eprint = "1205.3996",
    archivePrefix = "arXiv",
    primaryClass = "hep-lat",
    doi = "10.1103/PhysRevD.86.074506",
    journal = "Phys. Rev. D",
    volume = "86",
    pages = "074506",
    year = "2012"
}

@article{Alexandru:2015sua,
    author = "Alexandru, Andrei and Basar, Gokce and Bedaque, Paulo F. and Ridgway, Gregory W. and Warrington, Neill C.",
    title = "{Sign problem and Monte Carlo calculations beyond Lefschetz thimbles}",
    eprint = "1512.08764",
    archivePrefix = "arXiv",
    primaryClass = "hep-lat",
    doi = "10.1007/JHEP05(2016)053",
    journal = "JHEP",
    volume = "05",
    pages = "053",
    year = "2016"
}

@article{Nishimura:2019qal,
    author = "Nishimura, Jun and Tsuchiya, Asato",
    title = "{Complex Langevin analysis of the space-time structure in the Lorentzian type IIB matrix model}",
    eprint = "1904.05919",
    archivePrefix = "arXiv",
    primaryClass = "hep-th",
    reportNumber = "KEK-TH-2119",
    doi = "10.1007/JHEP06(2019)077",
    journal = "JHEP",
    volume = "06",
    pages = "077",
    year = "2019"
}

@article{Chou:2025moy,
    author = "Chou, Chien-Yu and Nishimura, Jun and Tripathi, Ashutosh",
    title = "{Inequivalence between the Euclidean and Lorentzian Versions of the Type IIB Matrix Model from Lefschetz Thimble Calculations}",
    eprint = "2501.17798",
    archivePrefix = "arXiv",
    primaryClass = "hep-th",
    reportNumber = "KEK-TH-2686",
    doi = "10.1103/PhysRevLett.134.211601",
    journal = "Phys. Rev. Lett.",
    volume = "134",
    number = "21",
    pages = "211601",
    year = "2025"
}

@article{Anagnostopoulos:2022dak,
    author = "Anagnostopoulos, Konstantinos N. and Azuma, Takehiro and Hatakeyama, Kohta and Hirasawa, Mitsuaki and Ito, Yuta and Nishimura, Jun and Papadoudis, Stratos Kovalkov and Tsuchiya, Asato",
    title = "{Progress in the numerical studies of the type IIB matrix model}",
    eprint = "2210.17537",
    archivePrefix = "arXiv",
    primaryClass = "hep-th",
    reportNumber = "KEK-TH-2470",
    doi = "10.1140/epjs/s11734-023-00849-x",
    journal = "Eur. Phys. J. ST",
    volume = "232",
    number = "23-24",
    pages = "3681--3695",
    year = "2023"
}

@article{Witten:1979pi,
    author = "Witten, Edward",
    editor = "'t Hooft, Gerard and Itzykson, C. and Jaffe, A. and Lehmann, H. and Mitter, P. K. and Singer, I. M. and Stora, R.",
    title = "{THE 1 / N EXPANSION IN ATOMIC AND PARTICLE PHYSICS}",
    reportNumber = "HUTP-79/A078",
    doi = "10.1007/978-1-4684-7571-5_21",
    journal = "NATO Sci. Ser. B",
    volume = "59",
    pages = "403--419",
    year = "1980"
}

@article{virtanen2020scipy,
  title={SciPy 1.0: Fundamental Algorithms for Scientific Computing in Python},
  author={Virtanen, Pauli and Gommers, Ralf and Oliphant, Travis E and Haberland, Matt and Reddy, Tyler and Cournapeau, David and others},
  journal={Nature Methods},
  volume={17},
  pages={261--272},
  year={2020},
  publisher={Nature Publishing Group}
}

@article{Hartnoll:2024csr,
    author = "Hartnoll, Sean A. and Liu, Jun",
    title = "{The polarised IKKT matrix model}",
    eprint = "2409.18706",
    archivePrefix = "arXiv",
    primaryClass = "hep-th",
    doi = "10.1007/JHEP03(2025)060",
    journal = "JHEP",
    volume = "03",
    pages = "060",
    year = "2025"
}

@article{Komatsu:2024bop,
    author = "Komatsu, Shota and Martina, Adrien and Penedones, Jo{\~a}o and Vuignier, Antoine and Zhao, Xiang",
    title = "{Einstein gravity from a matrix integral -- Part I}",
    eprint = "2410.18173",
    archivePrefix = "arXiv",
    primaryClass = "hep-th",
    month = "10",
    year = "2024"
}

@article{Komatsu:2024ydh,
    author = "Komatsu, Shota and Martina, Adrien and Penedones, Joao and Vuignier, Antoine and Zhao, Xiang",
    title = "{Einstein gravity from a matrix integral -- Part II}",
    eprint = "2411.18678",
    archivePrefix = "arXiv",
    primaryClass = "hep-th",
    month = "11",
    year = "2024"
}

@article{Ciceri:2025maa,
    author = "Ciceri, Franz and Samtleben, Henning",
    title = "{Holography for the Ishibashi-Kawai-Kitazawa-Tsuchiya Matrix Model}",
    eprint = "2503.08771",
    archivePrefix = "arXiv",
    primaryClass = "hep-th",
    doi = "10.1103/fb8g-b8fd",
    journal = "Phys. Rev. Lett.",
    volume = "135",
    number = "6",
    pages = "061601",
    year = "2025"
}

@article{Chou:2025rwy,
    author = "Chou, Chien-Yu and Nishimura, Jun and Wang, Cheng-Tsung",
    title = "{Monte~Carlo Studies of the Emergent Spacetime in the Polarized Type IIB Matrix Model}",
    eprint = "2507.18472",
    archivePrefix = "arXiv",
    primaryClass = "hep-th",
    reportNumber = "KEK-TH-2740",
    doi = "10.1103/y1rm-n85b",
    journal = "Phys. Rev. Lett.",
    volume = "135",
    number = "22",
    pages = "221601",
    year = "2025"
}

@article{Martina:2025kwc,
    author = "Martina, Adrien",
    title = "{Massive deformations of supersymmetric Yang-Mills matrix models}",
    eprint = "2507.17813",
    archivePrefix = "arXiv",
    primaryClass = "hep-th",
    doi = "10.1007/JHEP03(2026)250",
    journal = "JHEP",
    volume = "03",
    pages = "250",
    year = "2026"
}

@article{Klinkhamer:2020wct,
    author = "Klinkhamer, F. R.",
    title = "{IIB matrix model: Emergent spacetime from the master field}",
    eprint = "2007.08485",
    archivePrefix = "arXiv",
    primaryClass = "hep-th",
    reportNumber = "KA-TP-08-2020",
    doi = "10.1093/ptep/ptaa168",
    journal = "PTEP",
    volume = "2021",
    number = "1",
    pages = "013B04",
    year = "2021"
}

@article{Klinkhamer:2020hyx,
    author = "Klinkhamer, F. R.",
    title = "{IIB matrix model: Extracting the spacetime points}",
    eprint = "2008.01058",
    archivePrefix = "arXiv",
    primaryClass = "hep-th",
    reportNumber = "KA-TP-09-2020",
    month = "8",
    year = "2020"
}

@article{Klinkhamer:2020jqx,
    author = "Klinkhamer, F. R.",
    title = "{IIB matrix model: Extracting the spacetime metric}",
    eprint = "2008.11699",
    archivePrefix = "arXiv",
    primaryClass = "hep-th",
    reportNumber = "KA-TP-10-2020",
    month = "8",
    year = "2020"
}

@article{Laliberte:2026qce,
    author = "Laliberte, Samuel and Toriumi, Reiko",
    title = "{Finite-$N$ Bootstrap Constraints in Matrix and Tensor Models}",
    eprint = "2603.17364",
    archivePrefix = "arXiv",
    primaryClass = "hep-th",
    month = "3",
    year = "2026"
}

@article{Chatzistavrakidis:2011su,
    author = "Chatzistavrakidis, Athanasios",
    title = "{On Lie-algebraic solutions of the type IIB matrix model}",
    eprint = "1108.1107",
    archivePrefix = "arXiv",
    primaryClass = "hep-th",
    doi = "10.1103/PhysRevD.84.106010",
    journal = "Phys. Rev. D",
    volume = "84",
    pages = "106010",
    year = "2011"
}

@article{Manta:2025tcl,
    author = "Manta, Alessandro and Steinacker, Harold C.",
    title = "{Dynamical covariant quantum spacetime with fuzzy extra dimensions in the IKKT model}",
    eprint = "2509.24753",
    archivePrefix = "arXiv",
    primaryClass = "hep-th",
    doi = "10.1007/JHEP02(2026)062",
    journal = "JHEP",
    volume = "02",
    pages = "062",
    year = "2026"
}

@article{Steinacker:2010rh,
    author = "Steinacker, Harold",
    title = "{Emergent Geometry and Gravity from Matrix Models: an Introduction}",
    eprint = "1003.4134",
    archivePrefix = "arXiv",
    primaryClass = "hep-th",
    reportNumber = "UWTHPH-2010-4",
    doi = "10.1088/0264-9381/27/13/133001",
    journal = "Class. Quant. Grav.",
    volume = "27",
    pages = "133001",
    year = "2010"
}

@article{Battista:2022hqn,
    author = "Battista, Emmanuele and Steinacker, Harold C.",
    title = "{On the propagation across the big bounce in an open quantum FLRW cosmology}",
    eprint = "2207.01295",
    archivePrefix = "arXiv",
    primaryClass = "gr-qc",
    reportNumber = "UWThPh-2022-10",
    doi = "10.1140/epjc/s10052-022-10874-0",
    journal = "Eur. Phys. J. C",
    volume = "82",
    number = "10",
    pages = "909",
    year = "2022"
}

@article{Steinacker:2026qzk,
    author = "Steinacker, Harold C.",
    title = "{Modified gravity at large scales on quantum spacetime in the IKKT model}",
    eprint = "2601.08031",
    archivePrefix = "arXiv",
    primaryClass = "hep-th",
    doi = "10.1007/JHEP04(2026)044",
    journal = "JHEP",
    volume = "04",
    pages = "044",
    year = "2026"
}


\end{document}